\documentclass[11pt]{article}
\usepackage[nonatbib,preprint]{neurips_2024}
\pdfoutput=1

\usepackage{amsmath}
\usepackage{bm} 
\usepackage{graphicx}
\usepackage{booktabs}
\usepackage{tabularx}
\usepackage{hyperref}
\usepackage{upgreek}
\usepackage{multirow}

\usepackage[ justification=justified,format=plain]{caption} 
\graphicspath{{./all_images/}}
\usepackage{comment}
\usepackage{sidecap}
\sidecaptionvpos{figure}{c}
\usepackage[title]{appendix}
\usepackage{enumerate}
\usepackage[font={footnotesize}]{caption}
\usepackage[]{todo}
\usepackage{tabularx}
\usepackage{threeparttable}
\usepackage[table,xcdraw]{xcolor}
\usepackage{booktabs}
\usepackage{lscape}
\usepackage{pdfpages}
\usepackage{ulem}
\usepackage{authblk}
\usepackage[section]{placeins}
\usepackage{textcomp}
\usepackage{pgfplots}
\usepackage{tikz}
\usetikzlibrary{arrows}
\usetikzlibrary{calc}
\usepackage[
backend=bibtex,
style=numeric-comp,
sorting=none,
url=false,
isbn=false,
date=year,
maxbibnames=20
]{biblatex}
\AtEveryBibitem{%
  \clearfield{note}%
  \clearfield{Publisher}
  \clearfield{pages}
  \clearfield{volume}
  \clearfield{number}
  \clearlist{language}
  
}
\newcommand{\np}{\par \vspace{0.25 cm} \noindent}
\renewcommand{\v}[1]{\mathbf{#1}}
\newcommand{\vs}[1]{\boldsymbol{#1}}
\newcommand{\M}[1]{{\rm\mathbf{#1}}}
\newcommand{\Realn}[1]{\mathbb{R}^{#1}}
\DeclareMathOperator*{\argmin}{arg\,min}

\newcommand{\rev}[1]{{\color{black}#1}}
\newcommand{\citeauthoryear}[1]{\citeauthor{#1}~(\citeyear{#1})~\cite{#1}}
\usepackage{lineno}
\usepackage{setspace} 
\usepackage{bm}

\DeclareUnicodeCharacter{2009}{\,}

\title{A compact model of \textit{Escherichia coli} core and biosynthetic metabolism}

\author{%
    \textbf{Marco Corrao}$^{1,*}$, \textbf{Hai He}$^{2}$, \textbf{Wolfram Liebermeister}$^{3}$, \textbf{Elad Noor}$^{4}$, \textbf{Arren Bar-Even}$^{5,\mp}$ \\
    \begin{small}
    $^{1}$ Department of Engineering Science, University of Oxford, Parks Road, OX1 3PJ Oxford, UK \\
    $^{2}$  Max Planck Institute for Terrestrial Microbiology, Karl-von-Frisch-Str.~10, 35043 Marburg, Germany \\
    $^{3}$  Universit\'e Paris-Saclay, INRAE, MaIAGE, 78350 Jouy-en-Josas, France \\
    $^{4}$  Department of Plant and Environmental Sciences, Weizmann Institute of Science, 7610001 Rehovot, Israel \\
    $^{5}$ Max Planck Institute of Molecular Plant Physiology, Am Mühlenberg 1, 14476 Potsdam-Golm, Germany \\
    $^{*}$ Corresponding author: Marco Corrao; email: marco.corrao@eng.ox.ac.uk \\
    ${\mp}$ Deceased in 2020
    \end{small}
}

\begin{document}

\maketitle

\begin{abstract}
Metabolic models condense biochemical knowledge about organisms in a structured and standardised way. As large-scale network reconstructions are readily available for many organisms, genome-scale models are being widely used among modellers and engineers. However, these large models can be difficult to analyse and visualise and occasionally generate predictions that are hard to interpret or even biologically unrealistic. Of the thousands of enzymatic reactions in a typical bacterial metabolism, only a few hundred form the metabolic pathways essential to produce energy carriers and biosynthetic precursors. These pathways carry relatively high flux, are central to maintaining and reproducing the cell, and provide precursors and energy to engineered metabolic pathways. Focusing on these central metabolic subsystems, we present \rev{\textit{i}CH360}, a manually curated medium-scale model of energy and biosynthesis metabolism for the well-studied bacterium \textit{Escherichia coli} K-12 MG1655. The model is a sub-network of the most recent genome-scale reconstruction, \textit{i}ML1515, and comes with an updated layer of database annotations and with a range of metabolic maps for visualisation. We enriched the stoichiometric network with extensive biological information and quantitative data, enhancing the scope and applicability of the model. In addition, we assess the properties of this model in comparison to its genome-scale parent and demonstrate the use of the network and supporting data in various scenarios, including enzyme-constrained flux balance analysis, elementary flux mode analysis, and thermodynamic analysis. Overall, we believe this model holds the potential to become a reference medium-scale metabolic model for \textit{E.~coli}.
\end{abstract}


\textbf{Abbreviations}: EFM: Elementary Flux Mode; FBA: Flux Balance Analysis; GEM: Genome-scale models; MDF: Max-Min Driving Force; PMO: Probabilistic Metabolic Optimisation; RMSE: Root Mean Squared Error; 3GP: glycerol 3-phosphate; 3hmrsACP: (R)-3-Hydroxy\-tetra\-decanoyl-ACP; 3MOB: 3-Methyl-2-oxo\-butanoate; Ala: L-alanine; Asp: L-aspartate; ATP: adenosine triphosphate; F6P: fructose 6-phosphate; G3P: glycer\-aldehyde 3-phosphate; HdeACP: cis-hexadec-9-enoyl-ACP; palmACP: palmitoyl-ACP; PEP: phospho\-enol\-pyruvate; Ru5P: D-ribulose 5-phosphate.

\section{Introduction}
Metabolic models are a valuable tool for biologists and biotechnologists who want to elucidate and engineer cell metabolism \cite{sarkar_engineering_2019,gu_current_2019,lu_silico_2023}. In their simplest form, such models encode basic biochemical knowledge, such as network structure, reaction stoichiometries, or known reaction directionalities, in a structured and standardised format. However, the scope of these models is often wider, including information on catalysing enzymes and kinetic parameters, as well as annotations that link model elements to external databases.
Rapid development of high-throughput experimental and computational pipelines has led to genome-scale metabolic network reconstructions now existing for a wide range of microorganisms \cite{gu_current_2019, fang_reconstructing_2020}. One of them, \textit{Escherichia coli}, has been the most studied prokaryotic organism and, as such, its metabolism has been the subject of extensive modelling efforts spanning over three decades \cite{reed_thirteen_2003, feist_genomescale_2007,orth_comprehensive_2011,monk_iml1515_2017}. In particular, for the common laboratory strain \textit{E.~coli} K-12 MG1655, the most recent genome-scale reconstruction, \textit{i}ML1515, accounts for 2712 enzyme-catalysed reactions mapped in detail to 1515 genes \cite{monk_iml1515_2017}.

Genome-scale metabolic network models (GEMs) provide a comprehensive picture of cell metabolism, and constraint-based modelling algorithms that use these models have shown remarkable predictive power, for example when predicting gene essentiality in bacteria \cite{bernstein_evaluating_2023}. However, working with such large models comes with some disadvantages. In the absence of sufficient constraining or parameterisation, simulations based on large networks can easily lead to biologically unrealistic solutions. For example, when designing and testing gene knockout strategies, genome-scale networks often wrongly predict unphysiological metabolic bypasses that must be manually inspected and filtered out \cite{he_optimized_2020,satanowski_awakening_2020} (see Supplementary Table \ref{examples_of_unrealsistic_GEM_behaviour} for some examples). Another issue is that, owing to their size and complexity, analysis of genome-scale networks is often limited to relatively simple modelling frameworks, such as flux balance analysis (FBA), that can only answer a limited range of questions. More complex methods, including sampling of metabolic flux distributions, elementary flux mode (EFM) analysis \cite{wortel_metabolic_2018}, or kinetic modelling can be used to gain additional insight into the governing principles and constraints of microbial metabolism, but are difficult to apply to large models. Finally, genome-scale models are often hard to visualise comprehensively, which can make the interpretation of computed flux distributions cumbersome and unintuitive.

For all these reasons, small-scale models of \textit{E.~coli} metabolism are commonly used instead, both for strain design and for the development of novel modelling frameworks. Among these, the \textit{E.~coli core model} (ECC) developed by \citeauthor*{orth_reconstruction_2010} \cite{orth_reconstruction_2010} has been widely used in the literature. Although popular as an educational and benchmark tool, ECC is limited in scope: it lacks, among others, most biosynthesis pathways, which would be relevant to many metabolic engineering applications. This limitation was addressed by \citeauthor{hadicke_ecolicore2_2017} \cite{hadicke_ecolicore2_2017}, who constructed a medium-scale model, \textit{E.~coli core 2} (ECC2), as a subnetwork of \textit{i}JO1366, the most up-to-date GEM available at the time \cite{orth_comprehensive_2011}. ECC2 was obtained through an algorithmic reduction \cite{erdrich_algorithm_2015} that iteratively prunes reactions from a template model while retaining user-specified structural and phenotypic features, such as the ability to grow under a defined set of conditions. However, to enforce the desired phenotypes, the algorithm relied only on steady-state stoichiometric modelling and did not account for other important factors, such as thermodynamics, kinetics, or regulatory effects, which are relevant under physiological conditions. Therefore, while the resulting submodel satisfies the stoichiometric constraints imposed by construction, further manual curation is often needed depending on the application at hand. 

Here, we introduce \textit{i}CH360 (named, according to convention, by the initials of the authors followed by the number of genes covered by the model), a manually curated ``Goldilocks-sized'' model of \textit{E.~coli} K-12 MG1655 energy and biosynthesis metabolism. The model was derived from the most recent genome-scale reconstruction (\textit{i}ML1515 \cite{monk_iml1515_2017}) and includes all pathways required for energy production and for the biosynthesis of the main biomass building blocks, such as amino acids, nucleotides, and fatty acids, while the conversion of these precursors into more complex biomass components is described by an effective biomass-producing reaction. We extended the coverage of annotations that point to external databases from \textit{i}ML1515, and built custom metabolic maps to facilitate visualisation of the model and its subsystems \cite{jahan_development_2016}. We complemented the stoichiometric network structure with a curated layer of biological information on catalytic function, protein complex composition, and small molecule regulation. Finally, we enriched the model with useful quantitative data, including thermodynamic and kinetic constants. Thanks to all these extra layers of information, our model can support a wide range of modelling methods beyond simple stoichiometric ones.

In the following, we present the model and demonstrate several use cases across various modelling scenarios, including enzyme-allocation predictions, EFM analysis, and thermodynamic analysis. The model is freely available in the standard formats SBML and JSON and can be used directly with popular metabolic modelling tools such as the COBRA toolbox \cite{ebrahim_cobrapy_2013}.

\section{Results}
\subsection{A compact model of \textit{Escherichia coli} energy and biosynthesis metabolism}

To assemble the \textit{i}CH360 model, we started from the core metabolic reactions present in ECC and extended them with a curated set of pathways required for the biosynthesis of the main biomass building blocks, including the twenty amino acids, the four nucleotides, and both saturated and unsaturated fatty acids 
(Fig.~\ref{pfba_solution_escher}, Table \ref{model_subsystems}, and Supplementary Figures \ref{aa_biosynthesis}-\ref{c1-metabolism}). \rev{On the other hand, we specifically omitted from our model the pathways required for the biosynthesis of complex biomass components and polymers, most degradation pathways, the pathways involved in the \textit{de novo} biosynthesis of cofactors, and the reactions involved in the uptake of metals and ions. } In addition, while not performing a comprehensive review of \textit{i}ML1515, we applied a small number of corrections to the original reactions based on the knowledge from the literature (Supplementary Information \ref{GEM_corrections}).
\begin{figure}[t!]
	\centering
    {\textsf{\text{Metabolic model  \textit{i}CH360}}}\\[5mm]
	\includegraphics[width=\textwidth]{all_images/map_w_fluxes_and_labels_ECC_shading.pdf}
	\caption{Metabolic map of the \textit{i}CH360 model, built with the metabolic visualisation tool Escher \cite{king_escher_2015} and showing the metabolic subsystems included in the model. Shaded areas denote metabolic subsystems already present in the ECC model \cite{orth_reconstruction_2010}. Reaction and metabolite names were omitted from the plot for clarity. Overlaid onto the map is a flux distribution for aerobic growth on glucose, computed via parsimonious FBA.}
 \label{pfba_solution_escher}
\end{figure}
\begin{table}[t!]
	 \caption{The main metabolic subsystems covered by \textit{i}CH360.}\ \\
     
  \begin{tabularx}{\textwidth}{
			|>{\hsize=0.55\hsize}X
			|>{\hsize=1\hsize}X
            |>{\hsize=0.45\hsize}X|
		}
		\hline
		Subsystem &  Description & Metabolic map\\
        \hline
		Carbon uptake and transport &
		Reactions required for the uptake and assimilation of the following carbon sources: glucose, fructose, ribose, xylose, lactate, acetate, gluconate, pyruvate, glycerol, glycerate, succinate, 2-ketoglutarate & 
        Figure \ref{pfba_solution_escher}\\
		\hline
		Central carbon metabolism & 
		Glycolysis, pentose phosphate pathway, pyruvate fermentation, TCA cycle, oxidative phosphorylation & 
        Figure \ref{pfba_solution_escher}\\	
		\hline 
		Amino acids biosynthesis &
		Biosynthesis of all 20 amino acids from core metabolism precursors & 
        Supplementary Figure \ref{aa_biosynthesis}\\
		\hline
		Nucleotide biosynthesis &
		Biosynthesis of purine and pyrimidine nucleotides (and deoxynucleotides) from core and amino acid metabolism &
        Supplementary Figure \ref{nt_biosynthesis}\\
		\hline	
		Fatty-acids biosynthesis &
		Biosynthesis of saturated and unsaturated fatty acids present in {\textit{E.~coli}} from acetyl-CoA & 
        Supplementary Figure \ref{fa_biosynthesis}\\
		\hline
		C1 metabolism &
		One-carbon metabolism &
        Supplementary Figure \ref{c1-metabolism}\\
		\hline
	\end{tabularx}
     \label{model_subsystems}
\end{table}

The final assembled model (Figure \ref{pfba_solution_escher}) comprises \(304\) compartment-specific metabolites \rev{(\(254\) chemically unique compounds)} and \(323\) metabolic reactions mapped to \(360\) genes, thus qualifying as a medium-scale model ranging in between ECC and \textit{i}ML1515 (Supplementary Figure \ref{model_size_comparison}). Although similar in scale, our model and ECC2 present a fundamental structural difference. ECC2 was built by systematically removing reactions from its genome-scale parent (\textit{i}JO1366) \cite{hadicke_ecolicore2_2017}. Thus, its metabolic space spans the production of all compounds consumed in the \textit{i}JO1336 biomass reaction. In contrast, the metabolic space of \textit{i}CH360 only reaches biomass building blocks, without explicitly considering peripheral pathways such as the assembly of cell-membrane components, \textit{de novo} synthesis of cofactors, and active transport of ions in the cell. To make the model comparable to its parent model \textit{i}ML1515, we constructed an equivalent biomass reaction, in which all biomass requirements not present in our model are summarised by an equivalent metabolic cost in terms of precursors, based on manually curated bioproduction routes (Table \ref{biomass_precursors}, Supplementary Information \ref{equivalent_biomass_computation}, and Supplementary File S1).

\begin{table}[t!]
\begin{threeparttable}
    \caption{Biosynthesis pathways outside of the \textit{i}CH360 model that were considered to construct a biomass reaction equivalent to the biomass reaction in the \textit{i}ML1515 model. The right column shows the main precursors present in \textit{i}CH360. Note that only the main precursors are shown here, but the equivalent biomass reaction computed also accounts for any net production or consumption of metabolites in the reduced model.\ \\}
    
	\begin{tabularx}{\textwidth}{
			|>{\hsize=1.1\hsize}X|
			>{\hsize=0.9\hsize}X|
		}
		\hline
		Pathway & Precursor in \textit{i}CH360\\
		\hline
		Biosynthesis of phosphatidylethanolamine (C16:0 and C16:1) & 
		3GP, PalmACP (C16:0), HdeACP (C16:1)  \\	
		\hline 
		Biosynthesis of KDO2-Lipid-A &
		F6P, Ru5P, 3hmrsACP\\
		\hline
		Murein Biosynthesis &
		G3P, PEP, F6P, Ala \\
		\hline	
		NAD/NADP de novo biosynthesis &
		Asp, DHAP  \\
		\hline
		FAD de novo biosynthesis &
		GTP, Ru5P\\
		\hline
		CoA de novo biosynthesis &
		3MOB, Asp\\
		\hline
		Active transport of ions&
		ATP		\\
		\hline
	\end{tabularx}
     \label{biomass_precursors}
\end{threeparttable}
\end{table}

\subsection{Range of metabolic conversions described by production envelopes}
\label{sec:ProductionEnvelopes}

To check how its significantly smaller size and complexity affect the solution space of our model, we first looked at the maximum achievable biomass production flux under a range of growth conditions (Supplementary Figure \ref{model_comparison_growth}).  Across most conditions considered, the model achieves biomass fluxes comparable to those of its genome-scale parent. The main differences exist in anaerobic growth on fumarate, alpha-ketoglutarate (AKG), and glycerol, where our model predicts zero growth, while the GEM achieves some (albeit small) biomass production rate. In practice, fermentation in these scenarios is biologically unrealistic (e.g.~\cite{boecker_enabling_2022}). The reduced model is thus a good basis for metabolic simulation frameworks with few constraints such as FBA, since the reduction procedure, at least in this test, limits the original solution space of the GEM to more physiologically relevant regions. 

\begin{figure}[]
	\centering
	\includegraphics[width=.8\textwidth]{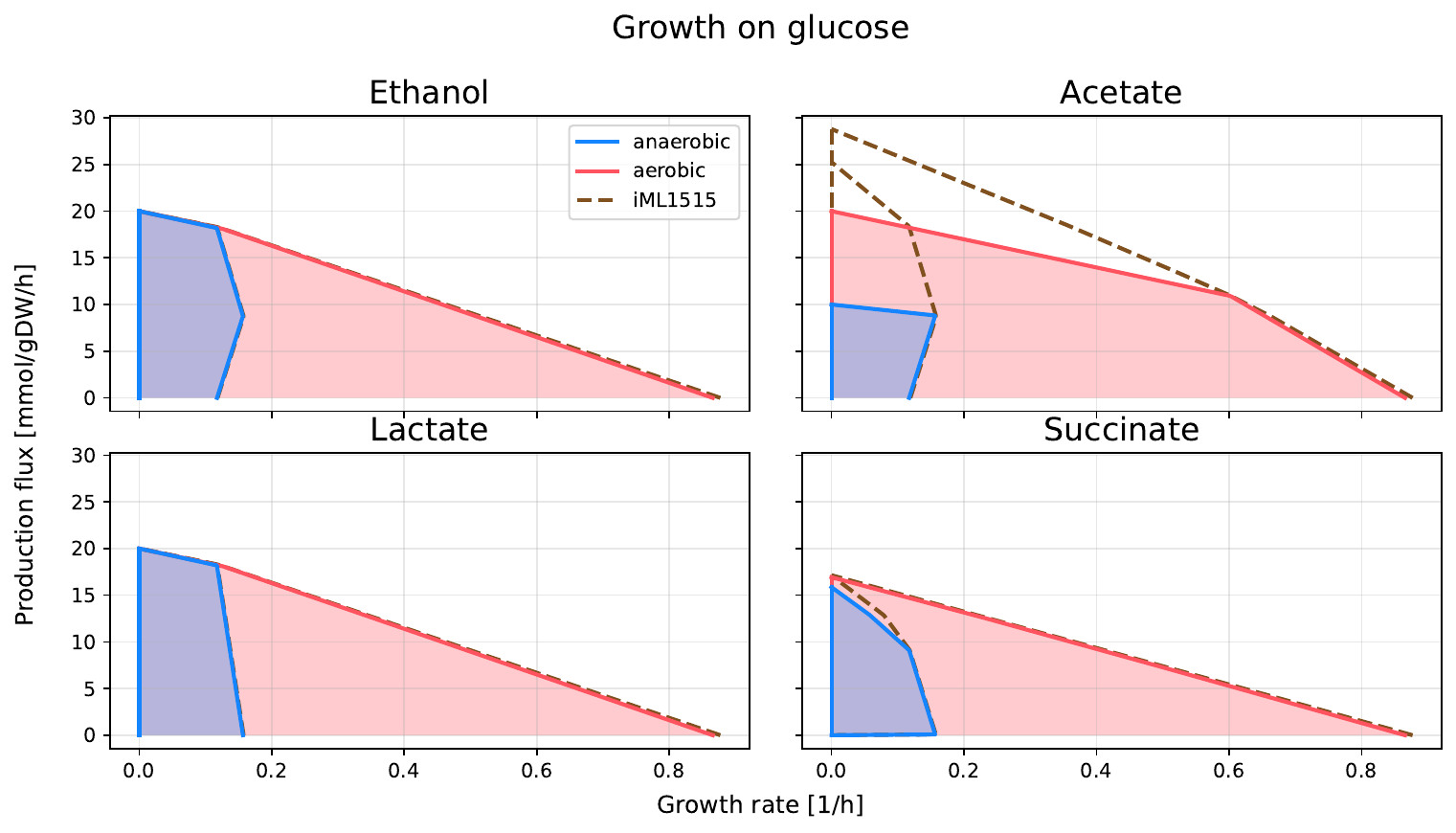}
	\caption{The \textit{i}CH360 model shows similar, but more precise metabolic capabilities than \textit{i}ML1515. Considering glucose as a feedstock and studying ethanol, lactate, and succinate production, a production envelope analysis yields similar results in the two models (note that the dashed line representing the production envelope of \textit{i}ML1515 is sometimes hidden behind the coloured lines). Growth rate and production fluxes were computed by limiting the glucose uptake rate to a maximum of 10 mmol/gDW/h. In the scenario of acetate production (top right panel) \textit{i}CH360 avoids an unrealistically high production flux \cite{hadicke_ecolicore2_2017} as predicted by \textit{i}ML1515.}
 \label{production_envelopes}
\end{figure}

To further investigate the metabolic capabilities of our model, we looked at production envelopes (projections of the model solution space onto a smaller set of dimensions, also known as phenotypic phase planes) describing production rates for biomass and a range of metabolites (Methods). Figure \ref{production_envelopes} shows the resulting envelopes for aerobic growth on glucose. The reduced model has production envelopes comparable to \textit{i}ML1515, except for the production of acetate, where the genome-scale model can achieve considerably higher yields, both aerobically and anaerobically.
\rev{
In order to understand the cause of these differences, we investigated optimal acetate production routes in both models using Flux Balance Analysis. Under aerobic conditions, we found that the differences can be traced to different production abilities for acetyl-CoA, a precursor of acetate. In \textit{i}CH360, acetyl-CoA is produced entirely from pyruvate oxidation via either the Pyruvate Dehydrogenase (PDH) or the Pyruvate-Formate-Lyase (PFL) reactions (Supplementary Figure \ref{aerobic_acetate_production_iCH360_vs_iML1515}A, left). On the other hand, \textit{i}ML1515 can produce acetyl-Coa via a number of additional pathways not included in our model (Supplementary Figure \ref{aerobic_acetate_production_iCH360_vs_iML1515}A, right). These include: the degradation of threonine either directly to acetyl-CoA (THRD, GLYAT), or indirectly via acetaldehyde (THRA, THRD, ATHRDHr, THRA2), which is then converted to acetyl-CoA; the degradation of 2-deoxy-D-ribose 5-phosphate into acetaldehyde (DRPA), followed by conversion into acetyl-CoA; the degradation of autoinducer 2 into acetyl-CoA (AI2K, PAI2I, PAI2T). Indeed, we found that, in the region of the production envelope where the two models diverge significantly, up to \(50 \%\) of acetyl-CoA production in the genome-scale model is accounted for by these degradation pathways. This is in disagreement with the understanding of the existing literature that under aerobic growth on glucose, most of the acetyl-CoA is produced from oxidation of pyruvate \cite{krivoruchko_microbial_2015, zhang_metabolic_2019}. Confirming these findings, the simultaneous deletion of four reactions blocking these degradation pathways (DRPA, PAI2T, THRA, THRD) in \textit{i}ML1515 brings aerobic acetate production to levels comparable to \textit{i}CH360 (Supplementary Figure \ref{aerobic_acetate_production_iCH360_vs_iML1515}C).

In the anaerobic scenario, we found that the differences between the two models are exacerbated by the ability of the genome-scale model to produce more pyruvate (which in turn results in higher acetyl-CoA production) than \textit{i}CH360. Investigating the cause of this difference, we found that, under anaerobic conditions, the \textit{i}ML1515 solution involves the uptake of external CO\textsubscript{2} and its use as a sink for electrons produced in glycolysis (Supplementary Figure \ref{aerobic_acetate_production_iCH360_vs_iML1515}C), which is thermodynamically unrealistic under ambient CO\textsubscript{2} levels. Blocking CO\textsubscript{2} uptake reduces the maximal anaerobic pyruvate yield in the genome-scale model, but does not fully close the gap with the production capabilities of our model (Supplementary Figure \ref{aerobic_acetate_production_iCH360_vs_iML1515}D), implying the existence of additional routes for anaerobic production of pyruvate that are not included in \textit{i}CH360.  A similar pattern was found when comparing production envelopes between the two models under different growth conditions (Supplementary Figure \ref{additional_production_envelopes_lactate_glycerol}). 

Overall, the higher maximal yields achievable by \textit{i}ML1515 for acetyl-CoA (aerobically) and pyruvate (anaerobically) appear unrealistic. Hence, in agreement with what was previously done with ECC2 \cite{hadicke_ecolicore2_2017}, we chose not to include in \textit{i}CH360 any additional reactions that would allow for higher acetate yields.} 

\rev{Production envelopes generated with \textit{i}CH360 were also comparable, albeit not identical, with those computed on other existing medium-scale models, namely ECC and ECC2 (Supplementary Figures \ref{production_envelopes_ich360_vs_ecc_ecc2_glucose}). Particularly, the maximal yields achievable for each product were nearly identical for the products considered, both aerobically and anaerobically. Some differences between the envelopes can be observed for solutions with higher predicted growth rates. However, since the three models are equipped with different biomass reactions sourced from different parental models (and hence predict different biomass yields), these differences are expected.}

\subsection{Connecting reactions to their catalysing enzymes and the enzymes' protein components}
\label{annotation}

\begin{figure}[t!]
	\centering
	\includegraphics[width=\textwidth]{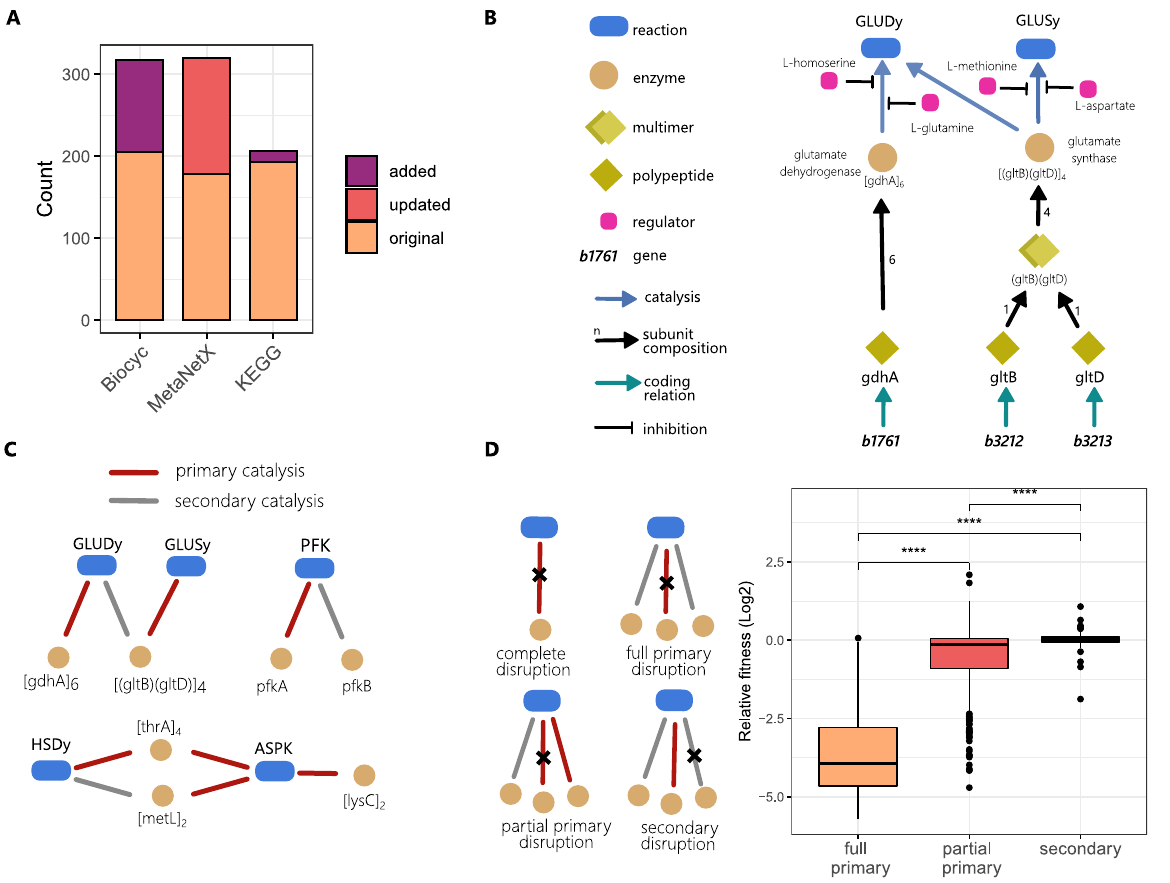}
	\caption{Layers of annotation and biological knowledge supporting the stoichiometric model  in \textit{i}CH360. \textbf{A}: Annotations for the model reactions point to the BioCyc, metaNetX, and KEGG databases. Bars show the numbers of annotations, highlighting the share of annotations that were added to or updated from the parent model \textit{i}ML1515. \textbf{B}: Some of the biological knowledge parsed from EcoCyc (and manually curated) included in the model-supporting functional annotation graph. The graph captures catalytic relationships between reactions and enzymes, protein subunit compositions, protein-gene mappings, and small-molecule regulation interactions, among others. Shown here as an example are the branches of the graph corresponding to the Glutamate Dehydrogenase (GLUDy) and Glutamate Synthase (GLUSy) reactions. \textbf{C}: Examples of catalytic relationships functionally annotated as either primary or secondary in the graph. Note that all catalytic relationships were classified as primary by default, unless sufficient evidence was found to annotate them as secondary. \textbf{D}: Functional annotation of catalytic edges as primary or secondary supports can be used to improve phenotypic predictions. Left: Classification of catalytic edge disruptions in the network resulting from simulated knockout of genes associated with essential reactions in the model across \(9\) growth conditions (see text for a description of each disruption class). Right: comparison of predicted disruption outcomes against a large data set of mutant fitness data \cite{price_mutant_2018} shows that the different types of disruption tend to lead to significantly different fitness changes.}
 \label{annotation_graph}
\end{figure}

Metabolic models often contain annotations that connect model elements to entries in biological databases, such as BioCyc \cite{karp_biocyc_2019}, KEGG \cite{kanehisa_kegg_2017} and MetaNetX \cite{moretti_metanetxmnxref_2021}. However, even for the subset of reactions included in \textit{i}CH360, the annotations present in \textit{i}ML1515 were incomplete and, in part, outdated. To fill these gaps, we extended and corrected the original annotations through a mixture of automated querying and manual curation (Figure \ref{annotation_graph}A). \rev{ Notably, the annotations pointing to the BioCyc knowledgebase \cite{karp_biocyc_2019}, are nearly complete: Out of \(321\) enzymatic reactions in the model, \(317\) are mapped to BioCyc with a single ID (for the remaining four unannotated reactions, involved in the biosynthesis of unsaturated fatty acids,  a match in the database could not be found for the specific use of NADPH as a redox cofactor). Further, nearly all of these BioCyc annotations (\(315\) / \(317\)) are in the ECOLI namespace and therefore point to the organism-specific EcoCyc database, a widely used and extensive reference for \textit{E. coli} molecular biology \cite{keseler_ecocyc_2017,EcoCyc2023}. The remaining two reactions map instead to the broader MetaCyc database, also part of the the BioCyc ecosystem, via the META namespace.} Aditionally, we found \(134\) deprecated annotations pointing to the MetaNetX database, which were consequently updated with the most up-to-date IDs.

Using the estensive mapping to the EcoCyc database, we parsed, assembled, and manually curated a knowledge graph that enhances the stoichiometric model with a detailed layer of information about enzymes, polypeptides, and genes related to the network (Methods). This data structure takes the form of a weighted graph, where nodes represent biological entities (reactions, proteins, genes, or compounds) and edges represent (potentially quantitative) functional relationships between them (Methods, Figure \ref{annotation_graph}B, Supplementary Tables \ref{node_description_table} and \ref{edge_description_table}), such as catalysis, regulation, protein modification, and protein subunit composition. This graph contains the information collected in a unified form, allowing users to perform a number of tasks that occur in metabolic modelling applications. For example, by explicitly mapping reactions to their catalysing enzymes rather than to the associated genes (which may additionally include protein activators or cofactors), it simplifies the definition of meaningful enzyme capacity constraints in the model. Similarly, since the graph topology implicitly defines associations between reactions and genes, boolean gene-protein-reaction (GPR) rules needed for \textit{in silico} knockout studies can be generated that account for catalytic and noncatalytic requirements for each reaction (Supplementary Information \ref{graph_based_computations}). Crucially, these GPR rules can be regenerated as needed whenever the graph is updated with new nodes or edges. Finally, we used it to estimate the abundance of protein complexes included in the model from the measured polypeptide abundances using a simple automated procedure (Supplementary Information \ref{complex_abundance_estimation}, Results section \ref{EC-iCH360}).

Through this annotation graph, \(318\) metabolic reactions in the model are linked to \(289\) catalysing enzymes, with more than \(25\%\) of the reactions being catalysed by multiple enzymes (isozymes). Such enzymatic redundancy plays an important role, for example, when designing metabolic engineering strategies to prevent flux through a pathway. However, the different isoenzymes of a reaction need not all be equally important, and treating them as completely equivalent may generate inaccuracies in some phenotypic predictions. 
For example, while phosphofructokinase activity in {\textit{E.~coli}} (reaction PFK) is known to be carried by two isozymes, \textit{pfkA} and \textit{pfkB}, the latter is known to be responsible only for minor activity under normal physiological conditions \cite{babul_phosphofructokinases_1978}.
If these differences are not taken into account, metabolic models can overpredict redundancy in the network, for example, when predicting phenotypes after gene knockouts \cite{monk_iml1515_2017}. 

To address this issue and make the model usable in a wider range of modelling scenarios, we classified the catalytic edges in the graph as either \textit{primary} or \textit{secondary catalysis} (Methods, Figure \ref{annotation_graph}C). The catalytic relationship between a reaction and an enzyme was annotated as secondary whenever the enzyme, according to experiments, accounts only for negligible activity for the reaction in the wild-type strain. As experimental evidence we considered \textit{in vitro} and \textit{in vivo} complementation studies. Through this curation process, a total of \(72\) catalytic edges were functionally annotated as secondary. 

To test how this functional annotation can support phenotypic predictions, we investigated the disruption of each essential reaction in the model, resulting from the knockout of each of its associated genes (Methods). We classified the outcome of each possible knockout as: (i) complete disruption, if a reaction loses all of its catalytic edges; (ii) full primary disruption, if a reaction loses all of its primary catalysis edges, but secondary ones are left; (iii) partial primary disruption, if the reaction loses some, but not all, of its primary catalysis edges, or (iv) secondary disruption, if some or all secondary edges are disrupted, but none of the primary ones. We classified the effect of \textit{in silico} knockouts for aerobic growth on \(9\) different carbon sources and compared the results against a large data set of mutant fitness data, obtained via competitive fitness assays, from~\citeauthoryear{price_mutant_2018}. 

Based on this analysis, we found that disruptions of the primary edges as a result of a knockout were significantly associated with greater fitness losses than disruptions of secondary edges (Wilcoxon rank-sum test, \(p<10^{-6}\), Figure \ref{annotation_graph}D). In addition, primary disruptions that were only partial were associated with more contained fitness losses. Finally, minor fitness gains could be identified when mutants were associated with secondary and partial primary disruptions, but not when complete primary disruptions occurred. We did not find any significant differences between complete disruptions and full primary disruptions (Supplementary Figure \ref{complete_vs_full_primary_fig}), supporting the idea that catalytic relationships annotated as secondary are unlikely to be strong enough to stand-in for disrupted primary ones under normal physiological conditions.

\subsection{Enzyme-constrained Flux Balance Analysis with EC-\textit{i}CH360}
\label{EC-iCH360}

To make \textit{i}CH360 applicable for enzyme-constrained flux simulations, we constructed a version of the model containing all necessary extra information, which we denote as EC-\textit{i}CH360.  We constructed EC-\textit{i}CH360 in the sMOMENT \cite{bekiaris_automatic_2020} format (Methods). The sMOMENT framework is inherently simple and generates the same solution space as more complex model formats, such as GECKO, unless reaction-specific capacity constraints are specified in the latter~\cite{sanchez_improving_2017,bekiaris_automatic_2020}. Since it requires unique reaction-enzyme mappings, we used the knowledge graph to remove all secondary catalytic relationships in the model (Section \ref{annotation}). 

We first parametrised the model by defining a flux-specific enzyme cost for each reaction (in units of grams of enzyme per unit flux), using as values the estimated \textit{in vivo} turnover numbers from~\citeauthoryear{heckmann_kinetic_2020}. 
Using this enzyme-constrained model, we then predicted enzyme abundances and compared them to experimental enzyme abundances, estimated from a data set of proteomic measurements for aerobic growth on eight different carbon sources \cite{schmidt_quantitative_2016} (Methods). This analysis led to predictions with root mean squared error (RMSE, computed for log\textsubscript{10}-transformed enzyme abundances) ranging from \(0.53\) to \(0.62\) (Supplementary Figure \ref{proteome_allocation_predictions_all_conditions}). To assess the nature of residuals between measurements and predictions, we investigated the geometric mean of enzyme abundances across all eight conditions (Supplementary Figure \ref{proteome_allocation_predictions_all_conditions}, bottom right panel). \rev{If the mismatch between measurements and predictions were due to each enzyme operating at different saturation levels in each condition, one would expect that averaging would reduce these differences.} However, averaging the predicted and measured enzyme abundances, respectively, across conditions did not significantly improve the RMSE, indicating that the abundances of individual enzymes were systematically over- or under-predicted between conditions. 

\begin{figure}[t!]
	\centering	\includegraphics[width=0.95\textwidth]{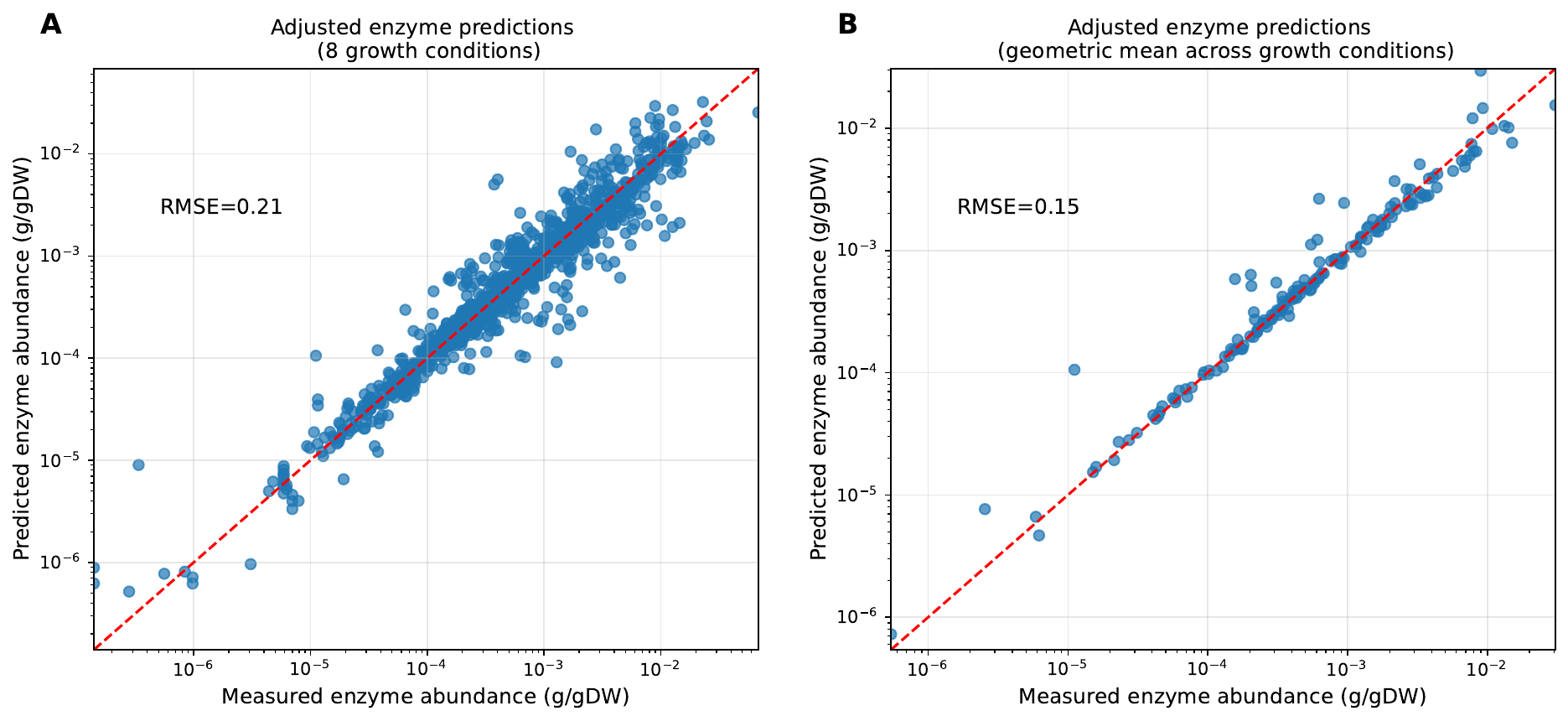}
	\caption{Enzyme allocation predictions obtained with the model variant EC-\textit{i}CH360 after adjusting the turnover parameters. \textbf{A}: Predicted vs measured enzyme abundances for aerobic growth on eight different carbon sources. Each data point represents an enzyme-condition pair. A total of \(325\) data points corresponding to zero predictions (enzymes associated with zero-flux in the enzyme-constrained FBA solution for a given condition) were omitted from the plot. \textbf{B}: Geometric mean across conditions of predicted vs measured enzyme abundances. For each enzyme, the geometric mean was computed across the conditions with non-zero predicted abundance. A total of \(27\) data points, corresponding to enzymes with zero predictions across all conditions, were omitted from the plot.  }
 \label{proteome_predictions}
 \end{figure}

To increase the predictive capacity of the model, we adjusted the turnover numbers by fitting them to experimental measurements through a custom heuristic (Methods, Supplementary information \ref{kcat_adjustment_SI}). By \textit{simultaneously} fitting all available conditions \rev{with a single set of parameters}, we ensured that our adjustment procedure is robust to condition-specific biases. Furthermore, by introducing regularisation within our adjustment scheme, we penalised large deviations of parameters from the original data set, increasing the robustness of the procedure to overfitting. Our adjusted parameter set shows a mean absolute deviation (computed for log\textsubscript{10}-transformed turnover values) from the original parameter set of $\approx$ 0.22 (Supplementary Figure \ref{original_vs_adjusted_kcats}) and results in significantly better agreement with experimental measurements across conditions (Figure \ref{proteome_predictions}A). Further, mean enzyme abundances across conditions are very well predicted with the adjusted parameter set (Figure \ref{proteome_predictions}B), implying that residuals between measurements and predictions are now to be attributed to \rev{variability in enzyme saturation} across conditions (which simple frameworks such as enzyme-constrained FBA, which use a constant enzyme cost per unit flux, cannot account for), rather than systematic over- or under-predictions. Thus, each adjusted turnover parameter can be thought of as a ``typical'' apparent \(k_{\rm cat}\) value, incorporating average saturation trends for an enzyme across growth conditions. \rev{To further confirm this aspect, we performed a leave-one-out cross-validation analysis (Methods), where each condition was in turn excluded from the model fitting dataset, but used to evaluate predictions. Results (Supplementary Figure \ref{leave_one_out_cross_validation}) show that the predictions of enzyme abundances in a given condition are considerably improved even when data from that condition is not used for fitting, confirming that our parameter fitting heuristics is capturing global trends and not condition-specific effects. Training the model on the full dataset, as we did to compute the final parameter set, further improves predictions, even if to a lesser extent. This can be explained by noting that, by design, our procedure cannot adjust the turnover parameters for enzymes associated with zero flux in the reference flux distribution used for fitting (Supplementary information \ref{kcat_adjustment_SI}). Hence, including all conditions in the training set allows for the turnover number of highly condition-specific enzymes (those associated to nonzero flux in only one of the reference flux distributions used by the procedure) to also be adjusted, further improving the overall prediction metrics. } 

\subsection{Elementary Flux Modes in the reduced model variant \textit{i}CH360\textsubscript{red}}
\label{efm_enumeration}

Despite the small size of \textit{i}CH360, we found the explicit enumeration of its elementary flux modes (EFMs) to be intractable. This is not necessarily surprising, since EFM count is crucially dependent on the topology of a metabolic network rather than its sheer size. Metabolic networks can possess different types of redundancy, such as the presence of alternative pathways for the production of the same metabolite, the use of alternative cofactors for the same catalytic step in a pathway, or the presence of alternative transporters for the uptake/excretion of a compound. Although knowledge about these redundancies is often valuable, including them in the model can increase the number of EFMs exponentially, hampering or even preventing EFM-based analyses.

To address this issue, we identified and removed a small set of \rev{alternative metabolic routes} in \textit{i}CH360, using available information from the literature, whenever possible, to ensure that the most physiologically relevant alternative was maintained (Supplementary Table \ref{metabolic_redundancies}). This resulted in a metabolic submodel of \textit{i}CH360, a model variant that we denote by \textit{i}CH360red. \textit{i}CH360red contains \(305\) metabolic reactions (\(18\) less than \textit{i}CH360) and shows the same production envelopes as its parent model for a number of metabolites of interest (Supplementary Figure \ref{ich360_vs_ich360red_production_envelopes_glucose}). While the number of EFMs in \textit{i}CH360red is still relatively large (\(\approx 13.5\) millions for aerobic growth on glucose, see Supplementary Table \ref{EFM_count}), it is not prohibitive for most types of EFM-based analysis, and their explicit enumeration does not require high-performance computing (Methods). 

\begin{figure}[]
	\centering	\includegraphics[width=0.8\textwidth]{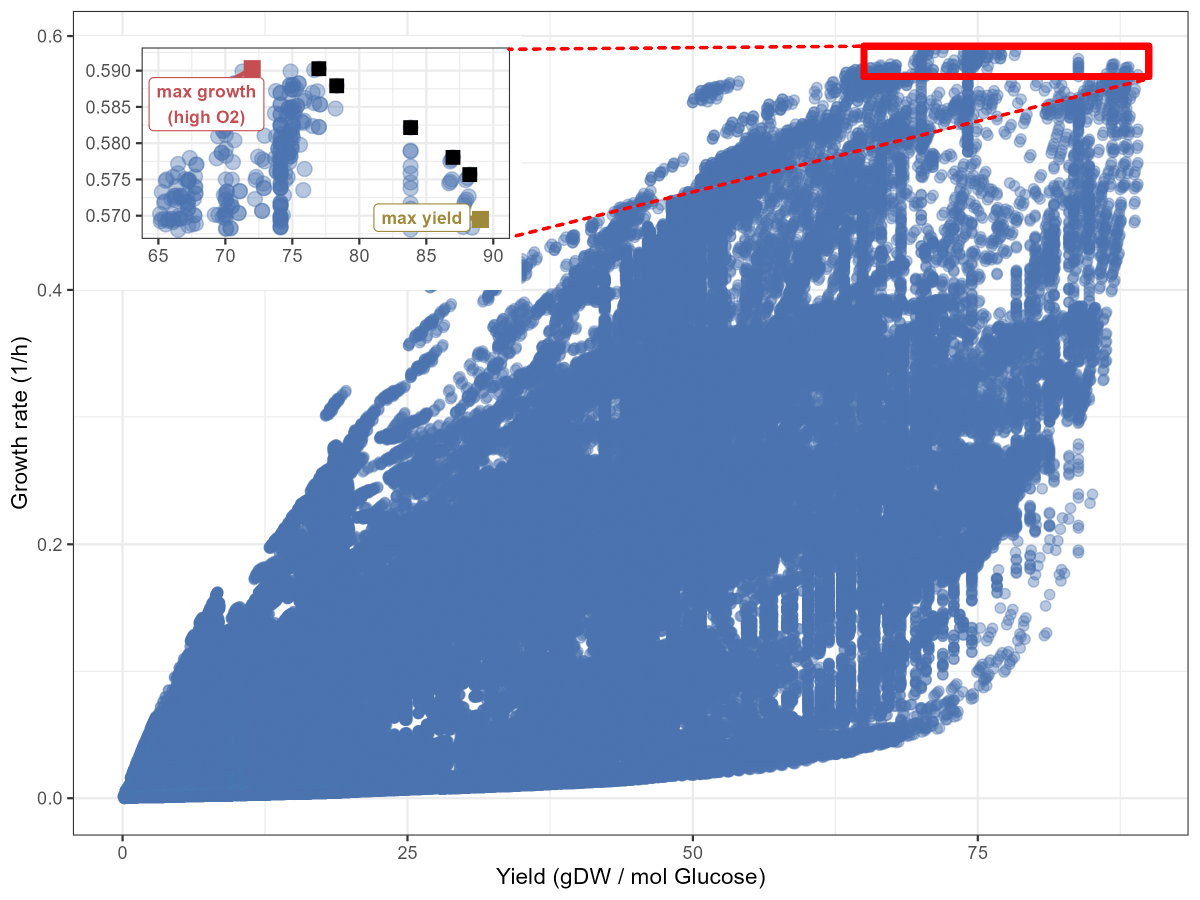}
	\caption{ Growth rates and biomass yields achieved by different elementary flux modes of \textit{i}CH360$_{\rm red}$  for growth on glucose. The inset on the top left (corresponding to the area of the plot enclosed by the red rectangle) highlights the front of Pareto-optimal EFMs (squares), with the maximum-growth and maximum yield modes laying at the extremes of the front. The growth rate of each mode was estimated by assuming that the metabolic enzymes in the model occupy, by mass, a constant fraction of the cell's dry weight (see Methods).}
 \label{EFM_yield_cost_tradeoff}
\end{figure}

We used the EFMs of \textit{i}CH360red to study the possible combinations of achievable growth rates and yields in the network \cite{wortel_metabolic_2018}. To this end, we considered growth on glucose as a scenario and computed, for each EFM from \textit{i}CH360red, its yield, \rev{computed as the ratio of biomass flux and glucose uptake}, and its achievable cell growth rate, which we estimated based on the enzyme costs defined for the enzyme-constrained model (Methods). Based on this analysis, we identified a front of Pareto-optimal EFMs, along which any increase in the growth rate will necessarily lead to a reduction in yield (Figure \ref{EFM_yield_cost_tradeoff}). Along the Pareto front, we observe a transition from a purely respiratory mode at maximum yield (Supplementary Figure \ref{max_yield_efm_escher_map}) to a mixed respiratory-fermentative mode at maximum growth (Supplementary Figure \ref{max_growth_efm_escher_map}). Quantitatively, the extent of this trade-off was rather modest: the EFM with maximal yield reaches almost the maximal growth rate, so only minor gains in growth rate can be achieved by using other,  fermentative modes along the Pareto front. However, it is worth noting that this analysis was performed using a simple capacity-based enzyme cost function, \rev{which ignores metabolite concentrations by assuming a constant enzyme cost per unit flux for each reaction. Repeating the analysis with a more complete enzyme-cost function, such as one that accounts for variable thermodynamic driving force and enzyme saturation,  could help elucidate the nature of this trade-off \cite{noor_protein_2016}.}
To demonstrate that the shape and size of the Pareto front depend strongly on growth conditions, we also simulated an environment with very low oxygen levels. We implemented this by increasing the flux-specific enzyme cost of the oxygen-dependent reaction in the respiratory chain, equivalent to assuming a lower enzyme efficiency due to a lower oxygen level (Methods). Results (Supplementary Figure \ref{pareto_front_low_o2}) show a much broader front of Pareto optimal EFMs, indicating that the nature of the observed trade-off is indeed condition-dependent. Notably, a similar dependence of the Pareto front on extracellular oxygen availability has been previously observed in a small-scale model of \textit{E.~coli} core metabolism \cite{wortel_metabolic_2018}.

\subsection{Saturation FBA and modelling of overflow metabolism}

\begin{figure}[b!]
	\centering	\includegraphics[width=\textwidth]{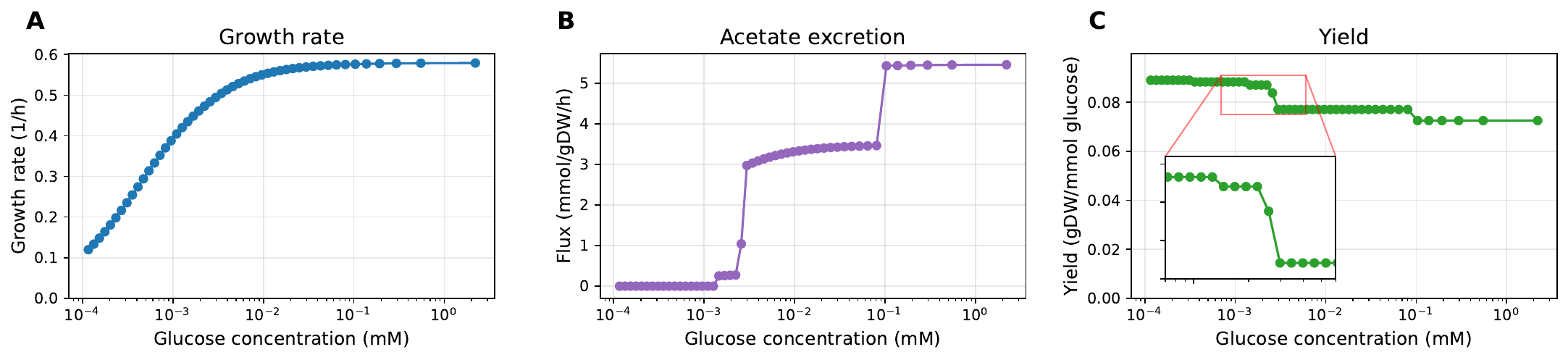}
	\caption{Saturation FBA enables the exploration of the optimal switching across elementary modes as a function of the growth environment. \textbf{A}: satFBA predictions for the growth rate as a function of external glucose concentration, showing a typical Monod curve. Note that satFBA computes the cell's growth rate by assuming a fixed total enzyme mass budget while varying the saturation of the substrate transporter as a function of external substrate concentration. Importantly, although the curve is continuous and smooth, it comprises many smaller sections, each dominated by a different elementary mode. \textbf{B}: satFBA predictions for the acetate excretion flux, showing progressively higher use of fermentative metabolism in the optimal solution as external glucose availability increases. \textbf{C}: The yield of the optimal satFBA solution progressively decreases in a step-like manner as external glucose availability increases. Each jump represents a switch in the optimal elementary flux mode.}
 \label{satFBA_results}
\end{figure}

In order to study the effect of external conditions on optimal metabolic strategies in more detail, we used another framework which does not require an enumeration of EFMs and allows for additional flux bounds, for example, to impose a minimum ATP consumption rate for cell maintenance.  The saturation FBA (satFBA) framework~\cite{muller_resource_2015} is a variant of enzyme-constrained FBA, wherein a fixed enzyme cost per flux is assumed for all metabolic reactions in a model, except for the substrate transporter, for which a complete kinetic law is used (Methods). Since the external substrate concentration is a simple parameter, screening this concentration is equivalent to screening the values of the transporter efficiency. Here we used satFBA to simulate how the growth-maximising solution of the network varies in response to changing extracellular glucose concentration. By solving the satFBA problem for a range of glucose concentrations, we predicted the dependence of the cell's growth rate on substrate concentration, resulting in the typically observed Monod curve (Figure \ref{satFBA_results}A). If the problem does not contain any further flux bounds (so that the magnitude of fluxes at the optimum is solely limited by the maximum enzyme availability) the solution of satFBA problems will be an elementary flux mode~\cite{muller_resource_2015, wortel_metabolic_2014}. Hence, in this case we can use satFBA to explore how a cell should switch across elementary modes as a function of the growth environment. 

At low glucose concentrations, the glucose transporter operates at low saturation and glucose uptake is enzymatically expensive, leading to a high-yield, purely respiratory metabolic mode at the optimum (Figure \ref{satFBA_results}B and C). As substrate availability is increased, the cost of substrate uptake decreases and higher growth rates are achieved by switching to lower yield, acetate-secreting modes~\cite{muller_resource_2015}. \rev{Since the yield of a flux distribution is, by definition, constant along an elementary flux mode, the yield varies in step-like manner as a function of external glucose concentration (Figure \ref{satFBA_results}C, inset), where each jump represents a change of optimal mode.} The satFBA formalism can also be used with additional flux bounds. For example, if a positive lower bound on ATP hydrolysis is added as a maintenance requirement, optimal solutions to the satFBA problem will no longer be elementary modes, and the yield of the optimal solution no longer follows a piecewise constant profile (Supplementary Figure \ref{satFBA_with_ATPM}C).


\subsection{Equilibrium constants, thermodynamic forces, and thermodynamically feasible states}

Living systems operate outside of thermodynamic equilibrium, and thermodynamics places strong constraints on the operation of metabolic systems. In any metabolic state, the flux directions must follow the signs of thermodynamic forces, which depend on metabolite concentrations and equilibrium constants. To provide these constants as parts of our model, we used the component contribution framework \cite{noor_consistent_2013} to estimate the standard Gibbs free energy (\(\Delta_rG'^\circ\)) of each reaction (Methods). These estimates account for compartment-specific chemical environment parameters, such as pH, pMg, and ionic strength, and were corrected to account for protons and charge translocation in multi-compartment reactions.

The resulting parameter set covers the vast majority of the model (over 97\% of metabolic reactions, with the remaining being not covered by the component contribution database used here) and accounts for the uncertainty in the estimates through a multivariate covariance matrix. Accounting for the correlations between different \(\Delta_rG'^\circ\) values becomes important when imposing thermodynamic constraints on the model. For example, the fatty acid biosynthesis subsystem in the model consists of repeated elongation cycles, where a short sequence of chemical transformations is repeatedly performed on a growing carbon chain. As a result, even if the \(\Delta_rG'^{\circ}\) for each reaction in the pathway is known with some uncertainty, this uncertainty is tightly correlated across the reactions, which constrains the set of achievable thermodynamic states in the network.

\begin{figure}[]
	\centering	\includegraphics[width=\textwidth]{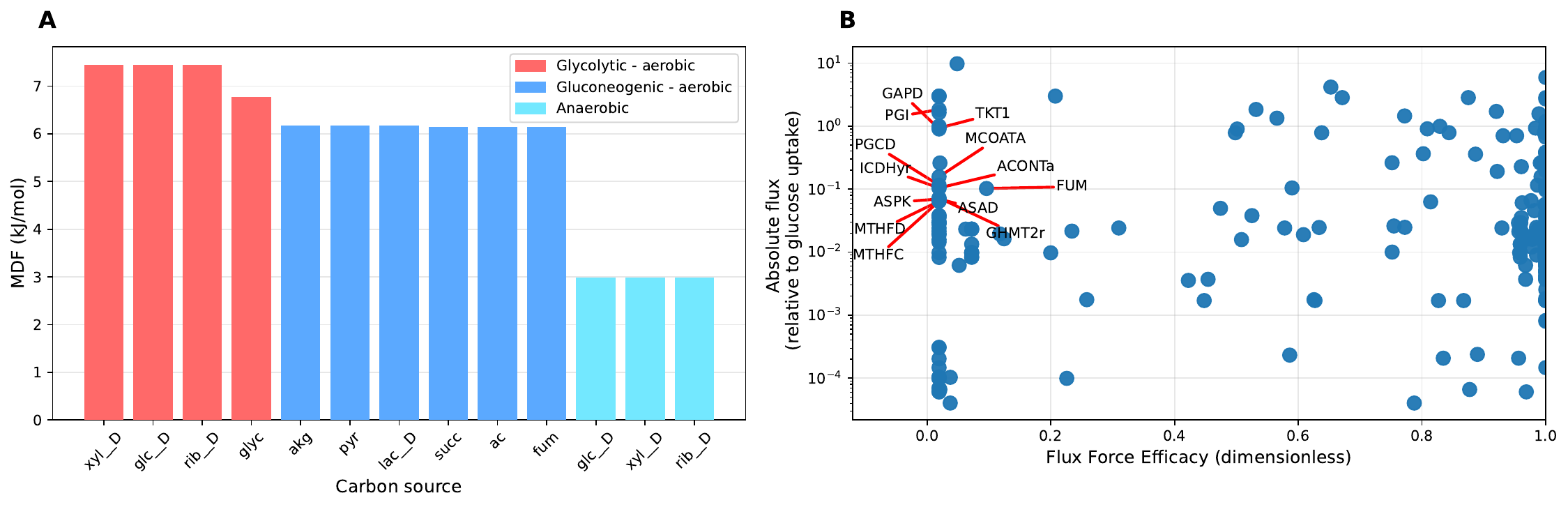}
	\caption{Thermodynamic analysis of the model via the curated thermodynamic parameter set. \textbf{A} Probabilistic max-min driving force (MDF) analysis of flux distributions obtained by parsimonious flux balance analysis for a total of \(12\) growth conditions. All flux distributions tested have a positive MDF, implying they are thermodynamically feasible under reasonable physiological metabolite concentration ranges. The computed MDF values cluster in three groups, corresponding to glycolytic aerobic, gluconeogenic aerobic, and anaerobic growth conditions. \textbf{B}: Fluxes relative to the glucose uptake flux (EX\_glc\_\_D\_e) and flux-force efficacies computed by probabilistic metabolic optimisation (PMO). The labelled data points represent examples of reactions (excluding transport and spontaneous reactions) with low predicted flux-force efficacy (here, below \(20 \%\)), but carrying high relative flux in the optimal solution (here, more than \(5\%\) of the glucose uptake flux). xyl\_\_D: D-xylose; glc\_\_D: D-glucose; rib\_\_D: D-ribose; glyc: glycerol; akg: alpha-keto-glutarate; ac: acetate; pyr: pyruvate; lac\_\_D: D-lactate; succ: succinate. fum: fumarate.}
 \label{MDF_PTA_combined}
\end{figure}

Using this set of thermodynamic constants, we first tested whether some typical flux distributions obtained from the model are thermodynamically feasible. To this end, we considered flux distributions generated by parsimonious Flux Balance Analysis (pFBA) across \(12\) growth conditions and computed their max-min driving force (MDF) \cite{noor_pathway_2014}, accounting for uncertainty in the estimates (Methods). We found a positive MDF for each of the flux distributions, indicating that all pFBA solutions tested are thermodynamically feasible. Notably, we found that the computed MDF values cluster very clearly in three groups, corresponding to aerobic growth on glycolytic substrates \rev{(high MDF)}, aerobic growth on gluconeogenic substrates \rev{(medium MDF)}, and anaerobic conditions (\rev{low MDF,} Figure \ref{MDF_PTA_combined}A).

Having confirmed that our reference FBA-derived flux distributions are thermodynamically feasible, we then employed an alternative flux prediction method that ensures thermodynamic feasibility by construction. The probabilistic metabolic optimisation (PMO) framework \cite{gollub_probabilistic_2021} uses a mixed-integer quadratic programming approach to compute a set of fluxes, metabolite concentrations, and reaction driving forces which is probabilistically most in agreement with experimentally measured metabolite concentrations (Methods). Using this framework, we computed a maximum-likelihood thermodynamic state for the model, simulating aerobic growth on glucose. In the thermodynamic state computed by PMO, we found that all metabolite concentrations lie within physiologically reasonable ranges (1 $\mu$M -- 1 mM). In addition, all ``anomalous concentrations'' identified by the framework (metabolite concentrations lying more than one standard deviation away from the mean experimental value) had been identified and explained previously \cite{gollub_probabilistic_2021}, and can potentially be addressed by lumping together those reactions for which substrate channelling is known to happen, as reported in literature. 

We used the PMO-derived thermodynamic state to identify candidate bottlenecks, in terms of enzyme demand, across the network. To this end, we first computed the flux-force efficacy of each reaction in the thermodynamic state~\cite{noor_pathway_2014} (Methods). The flux-force efficacy is a unitless quantity, ranging from \(0\) to \(1\), denoting the ratio between net flux (forward minus backward flux) and total flux (forward plus backward flux) of a reaction. Reactions operating at low flux-force efficacy have a lower net flux due to two reasons: (1) the forward flux is lower in absolute terms, and (2) the higher backward flux is counter-productive and subtracts from the forward flux. Therefore, to achieve a given required net flux, the cell has to invest more resources in maintaining a higher enzyme level. To identify potential thermodynamic bottlenecks that lead to high enzyme costs in specific reactions, we therefore screened all reactions for their predicted flux-force efficacy and flux (Figure \ref{MDF_PTA_combined}B)  and identified those predicted to operate at low efficacy, while carrying significant flux (Figure \ref{MDF_PTA_combined}B, labelled points). 

To assess to which extent the thermodynamic operating states predicted by \rev{PMO} are predictive of enzyme investment, we split reactions into two groups based on whether their predicted flux-force efficacy sits above or below \(50 \%\), and compared the distributions of measured enzyme abundances between the two groups (Methods, Supplementary Figure \ref{FFE_vs_enzyme_abundance_comparison}). While there are many determinants of enzyme abundance beyond thermodynamics, including enzyme turnover, affinity, or regulation, we observed a significant difference between the two distributions (\(p<0.001\), two-sided Wilcoxon rank-sum test), with the low efficacy having approximately a 3-fold higher median enzyme abundance than the high efficacy group. \rev{Establishing a causal link can be very difficult, but we can speculate that this negative correlation might be explained by metabolism evolving to compensate for the low efficacy of some reactions with a higher expression level of their catalysing enzyme \cite{noor_pathway_2014, khana_thermodynamics_2025}.}

\section{Discussion}

\begin{table}[t!]
\caption{A summary of knowledge captured by the \textit{i}CH360 model, as well as example simulations and analyses shown in this article.}
\footnotesize
\begin{center}
  \begin{tabularx}{\textwidth}{
        |>{\RaggedRight \hsize=0.3\hsize}X
		>{\hsize=0.5\hsize}X
        >{\hsize=0.2\hsize}X|}
		\hline
\textbf{Model structure} & Notes & Data source \\ \hline
Network (reaction stoichiometries)  &  Selected reactions from \textit{i}ML1515, hand-curated & \cite{monk_iml1515_2017}
\\
Annotations to external databases & Parsed from \textit{i}ML1515, extended, and updated &\cite{monk_iml1515_2017}, manual curation 
\\
Network graphics   & Escher maps of the full model and its subsystems & 
\\
Biological knowledge supporting the stoichiometric model   & Catalytic relationships, protein complex composition, small-molecule regulations, and others &\cite{keseler_ecocyc_2017} 
\\
\hline\hline
\textbf{Physico-chemical parameters mapped to model} & &\\\hline
Thermodynamic constants ($\Delta G^0$) &  Account for compartment-specific chemical environment (pH, pMg, ionic strength, and potential) and include corrections for reactions occurring across compartments. &\cite{noor_consistent_2013}, manual curation \\
$k_{\rm app}^{\rm max}$ values  & \textit{in vivo} estimates of catalytic turnover numbers & \cite{heckmann_kinetic_2020}\\
Typical $k_{\rm app}$ values  & Adjusted estimates of turnover numbers, fitted to proteomic data,  accounting for typical saturation levels across growth conditions & \cite{heckmann_kinetic_2020}~\cite{schmidt_quantitative_2016}\\
Protein molecular masses  &Molecular masses for all proteins/protein complexes covered by the model  & \cite{keseler_ecocyc_2017}
\\\hline\hline
\textbf{Cell-state data mapped to the model}   &  & 
\\ \hline
Protein abundances & Abundance across different growth conditions for proteins/protein complexes covered by the model, estimated from experimentally measured polypeptide abundances   & \cite{schmidt_quantitative_2016} 
\\ 
Metabolite concentrations & Measured metabolite concentrations across growth conditions   & \cite{gerosa_pseudo-transition_2015} 
\\ 
Metabolic fluxes & Measured metabolite fluxes for aerobic growth on glucose  & \cite{heckmann_kinetic_2020}, \cite{long_metabolic_2019} \\
\hline\hline
\textbf{Example applications shown} & &\\\hline
Production envelope analysis & See Figures \ref{production_envelopes} and \ref{additional_production_envelopes_lactate_glycerol} & \\
Enzyme-constrained FBA &With enzyme-constrained version of the model, EC-\textit{i}CH360, constructed in sMOMENT format \cite{bekiaris_automatic_2020} & \\
EFM analysis   & With reduced model variant \textit{i}CH360red. & \\
Saturation FBA &Predicts metabolic fluxes as a function of external substrate concentration & \\
Max-Min Driving force & Formulation from \citeauthoryear{noor_pathway_2014}, extended to account for correlated uncertainty in the themodynamic constants estimates.  & \\
Probabilistic Metabolic Optimisation &Prediction of thermodynamic states (metabolite concentrations and relative fluxes) maximally consistent with measured metabolite concentrations \cite{gollub_probabilistic_2021} &\\
\hline
\end{tabularx}
\end{center}
    \label{summary_table}
\end{table}

Here we presented \textit{i}CH360, a medium-scale metabolic model of \textit{E.~coli} covering central and biosynthesis metabolism, together with the associated data and metabolic maps and results from several analysed use cases. Similarly to previously constructed \textit{core} models \cite{orth_reconstruction_2010,hadicke_ecolicore2_2017}, this model trades metabolic coverage for usability, interpretability, and ease of visualisation. It is well suited whenever a relatively small, highly curated network is desired, when computationally demanding analyses are to be performed, or as an educational tool in the field of metabolic modelling. 
When comparing some key properties of this model with those of its parent genome-scale model, we observed only small differences in the achievable biomass and product yields across a range of growth conditions, validating that, despite its contained size, the model captures the most salient metabolic features of the genome-scale network. 

\rev{Further, we showed that the use of a well-curated, smaller-scale model can, in some instances, even correct unrealistic phenotypes predicted by its genome-scale parent. These unrealistic predictions from \textit{i}ML1515 are not the result of “errors” in the metabolic model. Rather, they are the result of applying simple stoichiometric methods, such as FBA, to a large network with many degrees of freedom. It is possible that the inclusion of additional constraints, such as thermodynamic feasibility under physiological conditions or proteome allocation bounds, would automatically render such solutions infeasible. However, these constraints, and the parameters required to implement them, are not always readily available. By assembling a smaller model and curating it with expert knowledge, we filter out many of these behaviours \textit{by construction}, providing users with a versatile and interpretable tool to investigate central and biosynthetic metabolism in \textit{E. coli}. Clearly, this comes at the cost of limited applicability in other scenarios, such as those where the metabolic subsystems not included in \textit{i}CH360 (e.g. degradation pathways) are crucial to explaining or modelling a given phenotype. In these cases, the use of a genome-scale model (or an \textit{ad hoc} reduction thereof) would still be an invaluable tool.}

To make this model easily usable in a variety of applications, we enriched the stoichiometric network structure with a curated layer of biological knowledge in the form of a knowledge graph.  This graph encodes information about biological entities in the network in a structured, ready-to-use format, including catalytic relationships between reactions and enzymes, the stoichiometric composition of protein complexes, and small-molecule regulation interactions. In addition, we mapped to the model a range of quantitative parameters, including \textit{in vivo} turnover number estimates and thermodynamic constants, extending the use of the model beyond a simple stoichiometric analysis. A summary of the biological knowledge captured by \textit{i}CH360 is shown in Table \ref{summary_table}.

Due to its medium size and the high level of curation, \textit{i}CH360 lends itself to a wide range of modelling methods. Here, we demonstrated some representative examples. These include the calculation of production envelopes, the modelling of metabolic proteome allocation via the enzyme-constraint model variant EC-\textit{i}CH360, enumerating and analysing elementary flux modes in the network via the reduced model variant \textit{i}CH360red, and performing thermodynamic-based analysis using the set of thermodynamic constants provided. \rev{Nevertheless, we believe that many other analyses are possible. For example, alternative definitions of elementary pathways, such as elementary conversion modes, would be valuable to explore as a more tractable alternative to elementary flux modes \cite{clement_unlocking_2021}. Similarly, our analyses of metabolic enzyme cost presented here have relied on simple, capacity-based definitions of enzyme cost, where a constant cost per unit flux is assumed for each reaction regardless of the condition studied. However, with additional kinetic parametrisation of the model, more complete enzyme cost functions, explicitly accounting for condition-specific metabolite concentrations and, consequently, enzyme costs, could be used to generate more realistic estimates of metabolic tradeoffs predicted by the model \cite{noor_protein_2016}.}

Indeed, while at this stage the parameterisation of the model is limited to turnover numbers and thermodynamic constants, we anticipate that additional parameter sets can easily be mapped to the model. Facilitated by extensive annotations present in the model and by the recent development of machine learning-enabled kinetic constant estimators \cite{kroll_deep_2021,li_deep_2022, kroll_turnover_2023,gollub_enkie_2023}, a complete kinetic parameterisation of \textit{i}CH360 is thus a valuable potential future development. In addition, probabilistic estimates of the kinetic parameters \cite{gollub_enkie_2023, lubitz_parameter_2019} can be combined with our existing thermodynamic parameterisation, making it possible to account for (potentially correlated) parameter uncertainty throughout the kinetic modelling process.

In light of the above results, we believe that \textit{i}CH360 has the potential to become a reference metabolic model for \textit{E.~coli}.

\section{Methods}

\subsection{Model assembly and curation}

All relevant pathways included in the model (the core metabolism reactions from ECC \cite{orth_reconstruction_2010} and the biosynthesis pathways shown in Supplementary Figures \ref{aa_biosynthesis}-\ref{c1-metabolism})  were assembled and curated based on information available in the EcoCyc \cite{keseler_ecocyc_2017} and KEGG \cite{kanehisa_kegg_2017} databases. The respective reactions were then extracted from \textit{i}ML1515 and parsed into a new model. To compute an equivalent biomass reaction, we first collected all the pathways required for the production of the components present in the \textit{i}ML1515 biomass reaction (with the exception of a small number of compounds present with very small stoichiometry, which were neglected from this analysis for simplicity), but not in our model. \rev{We used the "core" biomass reaction from \textit{i}ML1515 (\text{BIOMASS\_Ec\_iML1515\_core\_75p37M}), rather than alternative "WT" reaction (BIOMASS\_Ec\_iML1515\_WT\_75p37M), since its smaller number of requirements made it easier to manually curate the pathways for their production.} These additional pathways (available in the repository supporting this manuscript) were manually curated based on the available literature and database annotations to ensure they represent the biologically most relevant bioproduction route for each biomass component. By adding these pathways to \textit{i}CH360, we obtained an extended model that was able to predict growth rates directly through the original \textit{i}ML1515 biomass reaction. The equivalent biomass function was then computed based on a reference flux distribution computed on this extended model, as explained in Supplementary Information \ref{equivalent_biomass_computation}. \rev{Both growth-associated (the stoichiometry of intracellular ATP in the biomass reaction) and non-growth-associated (the lower bound on the maintenance reaction, ATPM) energy requirements were directly inherited from \textit{i}ML1515}. Model assembly, manipulation and validation were performed using the COBRA Toolbox \cite{ebrahim_cobrapy_2013}. The extension of database cross-annotation for the model reactions was performed through a mixture of automated database query and manual curation.

\subsection{Network graphics}

 \textit{i}CH360 can be visualised through a series of custom-built maps using the metabolic visualisation tool Escher \cite{king_escher_2015} (see Figure \ref{pfba_solution_escher}). There are three main ways to visualise the model or solutions thereof. First, a complete map of the model, including all of its reactions, can be used (Figure \ref{pfba_solution_escher}). In order to provide a more compact representation of the network, a compressed second variant of the same map was constructed (Supplementary Figure \ref{compressed_escher_map}). Here, long biosynthetic linear pathways were lumped into single pseudo-reactions, which only show the net production or consumption of metabolites by the pathway while omitting intermediates. Finally, individual maps are provided for each of the main subsystems in the model.
 \rev{
\subsection{Production Envelope Analysis}
All production envelopes shown in the main text and Supplementary Information were generated using the built-in production envelopes tools from the \texttt{ cobra.flux\_analysis.phenotype\_phase\_plane} module in the COBRApy package \cite{ebrahim_cobrapy_2013}. Briefly, the algorithm first computes the maximum and minimum production rates of the metabolite of interest, given the existing constraints in the model (including the specified bounds on the uptake of the carbon source and oxygen). The interval between the maximum and minimum achievable production rates is then discretised into an equally spaced grid of points. For each point in the interval, the production rate is fixed and the model's objective (here, the growth rate) is sequentially maximised and minimised, thus generating the boundary of the production envelope. All prediction envelopes were computed by specifying an upper bound on the uptake of the carbon source of \(10\) mmol/gDW/h, and blocking oxygen uptake for the anaerobic scenario.  For comparisons with ECC and ECC2, the maintenance  requirement (lower bound on the ATPM reaction) of these two models was set to the same value used in \textit{i}ML1515 and \textit{i}CH360 (\(6.86\) mmol/gDW/h).
}
\subsection{Knowledge graph for linking reactions to enzymes and proteins}\label{model_graph}

Information about the enzymes and proteins behind the stoichiometric model was collected in  a knowledge graph. To build this graph, all available data on reaction-protein association and subunit composition were retrieved by automatically querying the BioCyc database through its REST-based data retrieval API (\url{https://biocyc.org/web-services.shtml}). This information was then extended and curated based on a comparison with existing \textit{i}ML1515 GPR annotations. The resulting data were used to generate a directed graph in which nodes represent biological entities (such as reactions, proteins, genes, and metabolites) and edges represent functional dependencies across them, including catalysis, subunit composition, post-translational modifications, and others. A complete list of node and edge types in the graph is provided in Supplementary Tables \ref{node_description_table} and \ref{edge_description_table}, respectively. All polypeptide nodes were annotated with their molecular mass (parsed from EcoCyc), enabling recursive computation of the molecular mass of \textit{any} protein node in the graph (see Supplementary Information \ref{graph_based_computations}). All manipulation and analysis of the graph data structure were performed using the NetworkX Python package \cite{hagberg_exploring_2008}, and the final data structure is provided to the user in Cytoscape and GML formats.

\subsection{Primary and secondary catalytic edges}

Catalytic edges in the graph, i.e.~edges connecting reaction and enzyme nodes, were manually annotated as either primary or secondary, based on available evidence for the activity of each enzyme in the model with respect to its associated reactions (Supplementary File S2). More specifically, a catalytic edge between a reaction-enzyme pair was labelled as secondary whenever the enzyme was shown in the literature to
account for only minor catalytic activity for a reaction when compared with another isozyme. In this case, references to the relevant literature were included as metadata for that edge. Whenever sufficient information was not available, all isozyme edges for a reaction were conservatively treated as primary. 

\subsection{Catalytic disruption analysis}
For the catalytic disruption test, we identified condition-specific essential reactions in the following way: in the model, for each condition considered, we determined the reactions whose knockout led to the inability to produce biomass. A small number of known false positives, which are essential in \textit{i}CH360 only due to its lack of certain reactions or pathways, were excluded from the analysis. For each essential reaction, we considered all associated genes and investigated the removal of each of them individually from the graph. The effect of the simulated knockout was propagated across the graph by removing all nodes from the graph for which the gene is required according to the nodes' Boolean GPR rules (see Supplementary Information \ref{graph_based_computations}). Finally, the result of each simulated knockout was catalogued according to the reaction-level disruption it caused (see the main text). Whenever multiple reactions were disrupted by the knockout of a gene, the strongest disruption among them was assigned to the knockout, in the following precedence order: complete disruption, full primary disruption, partial primary disruption, secondary disruption. The analysis was repeated for a total of \(9\) growth conditions, and each condition-gene pair, labelled with the assigned disruption type, was compared with experimentally measured mutant relative fitness (averaged across available replicates for that condition-gene combination), using the data set from \citeauthoryear{price_mutant_2018}.

\subsection{Construction of the enzyme-constrained metabolic model}

In the sMOMENT formalism, the positive and negative fluxes in each reaction are formally described, respectively, as positive fluxes in separate ``forward'' and ``backward'' versions of the reaction.
To construct the enzyme-constrained model in the sMOMENT format, reversible reactions in the model were duplicated in the model to separately represent fluxes in forward or backward direction. Direction-specific turnover number estimates were parsed from \citeauthoryear{heckmann_kinetic_2020} and a default value of \(65~\rm{s}^{-1}\) was used for transporters as in \cite{heckmann_machine_2018}. To account for the fact that \cite{heckmann_kinetic_2020} reports values as turnover numbers \textit{per polypeptide}, the values were multiplied by the number of polypeptide subunits in each enzyme. For each reaction-enzyme pair, an enzyme cost per unit flux (in units of \({\rm g \cdot h \cdot mmol^{-1}}\)) was then defined as:
\begin{equation}\label{enzyme_cost_per_unit_flux}
    a_i=\frac{M_i}{k_{\text{cat},i} ~\sigma}
\end{equation}
where \(a_i\) is the enzyme cost of reaction-enzyme pair \(i\), \(k_{\text{cat},i} \) is the turnover rate estimate for the pair (here, in units of \(\rm h^{-1}\)), \(M_i\) is the molecular mass of the enzyme involved (in \({\rm kDa}\)), and \(\sigma\) is a unitless condition-specific scaling factor (typically interpreted as an average enzyme saturation value). Then, a unique enzyme was assigned to each reaction.  To this end, secondary catalytic relationships were first discarded and, for reactions with multiple annotated primary isoenzymes, the enzyme with the highest measured abundance in the integrated PAX Database \cite{huang_paxdb_2023} was heuristically chosen. Based on these costs per unit flux, an enzyme capacity constraint was introduced as:
\begin{equation}\label{proteome_constraint}
    \sum_i a_i~v_i \leq e_{\rm tot}
\end{equation}
where \(v_i\) denotes the flux in reaction \(i\) (in \({\rm mmol~h^{-1}~gDW^{-1}}\)) and \(e_{\rm tot}\) is a parameter denoting the total amount of enzyme mass (in g/gDW) that can be allocated to the flux mode. The constraint is enforced by augmenting the stoichiometric matrix of the model with an additional enzyme supply pseudoreaction (upper-bounded by \(e_{\rm tot}\)) and an enzyme pool pseudometabolite consumed in each reaction with stoichiometry \(a_i\) \cite{bekiaris_automatic_2020}.

\subsection{Adjustment of turnover numbers across conditions}\label{kcat_fitting_methods}
Turnover numbers from~\citeauthoryear{heckmann_kinetic_2020} were adjusted based on the nonlinear programming (NLP) formulation detailed in Supplementary Information \ref{kcat_adjustment_SI}). The reference flux distributions required by the procedure were obtained using the original (unadjusted) set of turnover numbers and the bounds on allowable adjustments (\(\v{u_{\rm min}}\) and \(\v{u_{\rm max}}\) in Eq. \ref{kcat_adjustment_problem}) set to \(\pm 2\) (corresponding to a maximal 100-fold increase or reduction of each parameter from the original value). The linear program used to obtain reference flux distributions was formulated and solved with GUROBI \cite{gurobi}, while the nonlinear program used for turnover adjustment was formulated and solved with the open-source optimisation package CasADi \cite{Andersson2019}. To investigate the effect of the ridge regularisation hyperparameter (\(\rho\) in equation (\ref{kcat_adjustment_problem})), we solved the adjustment problem for a broad range of values of this parameter and, each time, we computed the RMSE between measurements and predictions of enzyme abundance, as well as the mean absolute deviation of between original and adjusted turnover values, both computed for log\textsubscript{10}-transformed data (Supplementary Figure \ref{regulatisation_parameter_sweep}). Based on this information, a value of \(\rho=1\), after which any further decrease in the amount of regularisation results in marginal reduction of the RMSE, was heuristically chosen to compute the final set of adjusted turnover numbers. Upon reparametrisation into \textit{apparent} turnover numbers, (see Supplementary Information \ref{turnover_reparametrisation}), this set of adjusted parameters was used to parametrise the enzyme-constrained model variant EC-\textit{i}CH360. \rev{For the leave-one-out cross validation, the adjustment procedure was run multiple times, each time excluding abundance data from a condition from the training dataset. The resulting adjusted parameter set was then used to generate enzyme abundance predictions for the condition left out, which were then compared against measured values. }

\subsection{Enzyme allocation predictions}

To validate enzyme allocation predictions against experimental values, we first retrieved measured polypeptide abundances for each growth condition from \citeauthoryear{schmidt_quantitative_2016} and imputed missing values using, whenever available,  abundance values from the PAX Database \cite{huang_paxdb_2023}. Then we estimated enzyme counts across conditions from polypeptide counts using non-negative least-squares estimation (Supplementary Information \ref{complex_abundance_estimation}). Next, we converted them into mass abundances (in units of \({\rm g~gDW^{-1}}\)) based on the molecular mass of each complex (see Supplementary Information \ref{graph_based_computations}) and assuming a cell dry mass of \(2.8 \times 10^{-13} \rm{g}\) (BIONUMBER ID 103904 \cite{milo_bionumbers--database_2010}). 
Enzyme allocation predictions for each condition were then computed via EC-\textit{i}CH360 by fixing the growth rate to the experimentally measured one from \citeauthoryear{schmidt_quantitative_2016} and minimising the total enzyme cost, initially using a value of \(1\) for the average saturation parameter \(\sigma\) (Eq. (\ref{enzyme_cost_per_unit_flux})). For the predictions computed using the turnover number estimates from \citeauthoryear{heckmann_kinetic_2020}, the average saturation coefficient \(\sigma\) was then estimated from data, for each condition, as:
\begin{equation}
    \sigma=\frac{\sum_{i \in \mathcal{M}} e_i}{\bar{e}_{\rm{tot}}}
\end{equation}
where \(\mathcal{M}\) denotes the index set of model enzymes for which measurements are available, \(e_i\) is the predicted abundance for the \(i\)th enzyme,  and \(\bar{e}_{\rm{tot}}\) is the total measured model enzyme abundance for that condition. This value of \(\sigma\) was then used to scale the predicted enzyme abundances before comparing them with the ones measured experimentally. Note that this choice of \(\sigma\) ensures that the sum of predicted abundances matches that of measured ones. To compute predictions with the adjusted parameter set, the scaling factors for each condition were obtained as part of the fitting procedure (see Supplementary Information \ref{kcat_adjustment_SI}). All reported root mean squared errors were computed on log\textsubscript{10} transformed enzyme abundances, excluding enzymes with zero predicted abundance from the computation.

\subsection{Enumeration of Elementary Flux Modes}

Elementary Flux Modes (EFMs) for the submodel \textit{i}CH360red were enumerated, for each growth condition, using EFMtools \cite{terzer_large-scale_2008}. Filtered modes (Supplementary Table \ref{EFM_count}) were defined as those supporting nonzero biomass flux and, for aerobic modes, nonzero oxygen uptake. In addition, in the aerobic case, filtered modes exclude those carrying flux in either of three reactions -- Pyruvate-Formate Lyase (PFL), Fumarate Reductase (FRD2), and the menaquinone-dependent Dihydroorotate dehydrogenase (DHORD5) -- known to be physiologically active only under anaerobic conditions \cite{zhang_inactivation_2001,cecchini_succinate_2002,andrews_anaerobic_1977}.

\subsection{Growth/yield trade-off analysis}

To analyse trade-offs between growth rate and yield, we computed \rev{the yield and a predicted growth rate of each EFM, using the unit conventions common in stoichiometric metabolic models. The specific growth rate $\mu$ (1/h) of a cell or a cell population can be defined as the rate of biomass production per amount of biomass present. In stoichiometric metabolic models, units are chosen in a specific way such that the growth rate is identically given by the rate $v_{\rm BM}$ of the biomass reaction. In FBA models, normal reaction fluxes are given in units of mmol/gDW/h, so the yield of an EFM, computed as the ratio of its biomass flux $v_{\rm BM}$ to its glucose uptake flux, has a unit of gDW/mmol. The biomass flux, whose conventional unit is \(\rm{h}^{-1}\) was directly interpreted as the cell growth rate $\mu$.} 
To determine the cell growth rate allowed by a flux distribution, we first computed its absolute enzyme cost
\begin{equation}
    c_{\rm enz}=\sum_i{a_i~v_i}
\end{equation}
where \(c_{\rm enz}\) is the enzyme cost of the flux distribution (an enzyme mass, measured in g/gDW), \(v_i\) is the flux of reaction \(i\), and \(a_i\) is the enzyme cost per unit flux in reaction $i$, computed as per equation (\ref{enzyme_cost_per_unit_flux}) using the set of adjusted turnover numbers. \rev{Since an elementary flux mode can be scaled arbitrarily,  both \(v_{\rm BM}\) and \(c_{\rm enz}\) depend on the particular choice of scaling of the mode (though their ratio, \(v_{\rm BM}/c_{\rm enz}\), does not)}. In order to obtain an estimate for the achievable growth rate of a mode \rev{\(\mu\)} that is a unique property of each EFM, independent of its scaling, we thus normalised all modes to the same total enzyme cost \(f_{\rm enz}\) (in g/gDW), and looked at the resulting flux through the biomass reaction. More formally, the achievable growth rate \(\mu\) (in \(\rm h^{-1}\)) for an (arbitrarily scaled) EFM with biomass flux \(v_{\rm BM}\) and enzyme cost \(c_{\rm enz}\) was computed as:
\begin{equation}
    \mu=f_{\rm enz}~ \frac{v_{\rm BM}}{c_{\rm enz}}
\end{equation}
where $f_{\rm enz}$ denotes the total mass of enzyme available for the flux distribution, relative to the total dry mass of the cell.
For simplicity, we approximate $f_{\rm enz}$ by a constant value of \(f_{\rm enz}=0.285\) g/gDW, which we obtained by taking the minimum enzyme investment required by the enzyme-constrained model to support the experimentally measured growth rate reported in the proteomic data set \cite{schmidt_quantitative_2016} for the condition of interest in this analysis (aerobic growth on glucose).

To simulate low-oxygen conditions, the cost per unit flux of all oxygen-consuming reactions (CYTBO3\_4pp, CYTBDpp, CYTBD2pp) was increased by a \(1000\)-fold, mimicking the physiological state in which these reactions operate at low saturation with oxygen.

\subsection{Saturation FBA analysis}

Saturation FBA calculations were performed by optimising biomass production in the enzyme-constrained model, setting the saturation coefficient  (\(\sigma\) in 
Eq.~(\ref{enzyme_cost_per_unit_flux}) to the value fitted as part of the turnover number adjusting procedure (Section \ref{kcat_fitting_methods}) for all reactions except the glucose transporter (GLCptspp). The saturation of the glucose transporter, \(\sigma_{\rm up}\), was computed as a function of external glucose concentration and assuming irreversible Michaelis Mentens kinetics, so that:
\begin{equation}
    \sigma_{\rm up}=\frac{\rm [Glc]}{K_{\rm m}+{\rm [Glc]}}
\end{equation}
where \(\rm [Glc]\) is the external glucose concentration (in mM) and a value of \(0.116\) mM was used for the Michaelis constant \(K_{\rm m}\) \cite{wortel_metabolic_2018}.

\subsection{Component contribution estimates of thermodynamic constants}

Estimates of the free energies of reactions, and their uncertainties, were obtained using the component contribution framework previously described in \cite{noor_consistent_2013}. Several reactions in the model involve protein side groups, such as the acyl-carrying protein (ACP), or cofactors, such as glutaredoxin, for which a decomposition in terms of chemical groups is not available. As a result, thermodynamic constants for these reactions cannot be \textit{directly} estimated through database-based implementations of the component contribution method, such as eQuilibrator \cite{beber_equilibrator_2022}. However, since these non-decomposable protein groups are conserved in all reactions within our model, their net contribution to the reaction thermodynamics is, at least from a group contribution perspective, null. If we were only interested in computing standard free energies of reaction, $\Delta_rG^\circ$, we could simply treat protein groups as non-decomposable ``black-box'' units and add them to the group incidence matrix of the eQuilibrator database (see Section S1.1 in \cite{beber_equilibrator_2022}), enabling us to construct a group decomposition for compounds that contain them. However, for the computation of  \textit{transformed} standard free energies of reaction, 
$\Delta_r G'^\circ$, an exact chemical definition of each metabolite is required. 

To address this issue, protein groups were replaced by an appropriate chemical moiety that best approximates the metabolite's chemical environment. Specifically, ACP- groups were replaced by a phosphopantetheine group, the natural prosthetic group of acyl-carrier proteins, with a methyl group at the attachment site to the protein scaffold. Similarly, glutaredoxin was replaced by its Cys-Pro-Tyr-Cys active site, with the two cysteines being either free (for the reduced form of the cofactor) or linked by a disulfide bridge (for the oxidised state).  International Chemical Identifiers (InChI) were constructed for these ``replacement'' metabolites (Supplementary File S3) and used to extend the default compound cache of the eQuilibrator database. This custom-extended eQuilibrator compound cache is available in the repository supporting this manuscript.

Corrections for reactions that occur in different compartments were calculated as described in \cite{beber_equilibrator_2022}, using compartment-specific pH, pMg, ionic strength, and potentials from \citeauthoryear{gollub_enkie_2023}.

\subsection{Max-min driving force computation}

The max-min driving force (MDF) of each reference flux distribution was calculated by extending the original formulation described in \cite{noor_pathway_2014} to account for correlated uncertainty in the estimates of the thermodynamic constants. Let \(\Delta G_r'^\circ \in \Realn{N}\) be a random vector representing the (uncertain) standard Gibbs free energy of reaction for the reactions in the network. Importantly, this vector includes only balanced metabolic reactions and excludes pseudoreactions such as exchange reactions. To describe our uncertain knowledge, we assume that the vector \(\Delta G_r'^\circ\) follows a multivariate normal distribution:
\begin{equation}\label{standard_normal_transformation_of_drg0}
    \Delta G_r'^\circ \sim \mathcal{N}(\Delta \bar{G}_r'^\circ,\bm{\Sigma})
\end{equation}

where \(\Delta \bar{G}_r'^\circ\) and \({\bm{\Sigma}}\) are the mean vector and covariance matrix of the estimates obtained through the component contribution methods. This random vector can equivalently be expressed as:
\begin{equation}
    \Delta G_r'^\circ=\Delta \bar{G}_r'^\circ+\mathbf{Q}~\mathbf{z}
\end{equation}
where \(\mathbf{z}\) is a standard normal random vector in \(\Realn{q}\), with \(q=\text{rank}(\bm{\Sigma})\), and \(\mathbf{Q} \in \Realn{N \times q}\) is a square root of the covariance matrix, i.e.~it satisfies:
\begin{equation}
    \bm{\Sigma}=\rm \mathbf{Q}~\mathbf{Q}^T.
\end{equation}
In order to integrate this probabilistic description within a typical constrained-optimisation formulation, we define the set \(\mathcal{D}_{\alpha}\) as the \(\alpha\)-level confidence region around the mean of \(\Delta G_r'^\circ\). Using Eq.~(\ref{standard_normal_transformation_of_drg0}), and noting that the squared norm of a standard normal random variable is known to be Chi-squared distributed, we can represent this set as:
\begin{equation} \label{confidence_constraint}
\mathcal{D}_{\alpha}=\{\mathbf{x} \in \Realn{N} \hspace{0.1cm}|\hspace{0.1cm}\mathbf{x}=\Delta \bar{G}_r'^\circ+\mathbf{Q}~\mathbf{m}, \quad ||\mathbf{m}||_2^2\leq \chi^2_{q;\alpha}\}
\end{equation}
where \(\mathbf{m} \in \Realn{q}\) is a vector of free parameters and \(\chi_q(.)\) denotes the quantile function (inverse cumulative distribution function) of a chi-squared distribution with \(q\) degrees of freedom. We can thus account for the uncertainty in thermodynamic estimates by treating the free energies of reaction as decision variables, rather than known parameters, and constraining their value to belong to \(\mathcal{D}_{\alpha}\). For a given reference flux distribution \(\mathbf{v}\), this results in the  following quadratically-constrained program (QCP):
\begin{equation}
\begin{aligned}
& \underset{\mathbf{m},\v{c}, b}{\text{max}}
& & b \\
& \text{s.t.} & &  \Delta_r G'^\circ = \Delta \bar{G}_r'^\circ + \mathbf{Q} ~\mathbf{m} \\
& & &  \Delta_rG' = \Delta_r G'^\circ+\rm{RT}~\mathbf{S}^\top \mathbf{c} \\
& & &  -\text{sign}(v_i) ~ \Delta_r G'_i >b, \hspace{1cm} \text{if  } v_i \neq 0\\
& & &  ||\mathbf{m}||_2^2 \leq \chi^2_{q;\alpha} \\
& & & \mathbf{c_{\rm min}}\leq\mathbf{c}\leq \mathbf{c_{\rm max}}
\end{aligned}
\end{equation}
 where \(b\) is the min driving force (or the MDF after optimisation), \(\mathbf{c} \in \Realn{m}\) is a vector of log-metabolite concentrations,  \(\mathbf{c_{\rm min}}\);  \(\mathbf{c_{\rm max}}\in \Realn{m}\) are lower and upper bounds on these log-concentrations, \(\mathbf{S} \in \Realn{m \times N} \) is the stoichiometric matrix of the model, \({\rm R}\) is the \rev{ideal} gas constant, and \(\rm T\) is the temperature used for the computation of the free energy estimates. For our MDF calculations, we used a confidence level of \(90 \%\) and set the bounds on the concentration of all metabolites to the physiologically plausible range of (\(1 \upmu \text{M}\), \(10 \text{mM}\)). The above quadratically constrained program was formulated and solved using the GUROBI package \cite{gurobi}.

\subsection{Probabilistic thermodynamic analysis}

Probabilistic metabolic optimisation (PMO) of the model was performed using the PTA Python package \cite{gollub_probabilistic_2021}, providing the software with the curated thermodynamic estimates generated for this model. The default values available through the package for growth in M9 medium with glucose (which include measurements from \citeauthoryear{gerosa_pseudo-transition_2015}) were used as priors for the concentration of metabolites, and the growth rate was bounded from below by the reported value in \cite{gerosa_pseudo-transition_2015}. Furthermore, for this analysis, which requires that each flux in the solution have a well-defined directionality, two transhydrogenase reactions (NADH17pp and THD2pp) were allowed to operate reversibly \cite{gollub_probabilistic_2021}.

To analyse the thermodynamic state computed by PTA, we compute the flux-force efficacy \(\eta\) for each reaction as (see Figure \ref{FFE_plot}):
\begin{equation}
    \eta=\frac{\mathrm{e}^{-\frac{\Delta_rG'}{\rm RT}}-1}
              {\mathrm{e}^{-\frac{\Delta_rG'}{\rm RT}}+1}
        =\tanh\Big(-\frac{1}{2}\frac{\Delta_rG'}{\rm RT}\Big)
\end{equation}

where \(\Delta_rG'\) is the Gibbs free energy of reaction in the thermodynamic state, \({\rm R}\) is the universal gas constant and \({\rm T}\) is the temperature considered for the analysis (\(310.15~K\)). To compare the PTA-predicted flux force efficacies with experimental measurements of enzyme abundance, we selected reactions that carried at least \(2.5 \%\) of the glucose uptake flux and pooled them into two groups, corresponding to \(\eta>0.5\) (\(72\) reactions) and \(\eta<0.5\) (\(31\) reactions). $\eta = 0.5$ corresponds to $\Delta_rG' \approx -1.1~{\rm RT} = -2.8$ kJ/mol.
\begin{SCfigure}[0.8]
\begin{tikzpicture}
    \begin{axis}[
        width=7cm, height=4cm,
        xlabel=Driving Force ($-\frac{\Delta_rG'}{\rm RT}$),
        ylabel=\small{Flux force efficacy ($\eta$)},
        xmin=0,
        xmax=6,
        ymin=0,
        ymax=1.1,
        ]
        \addplot [domain=0:10, samples=100, color=red, thick] 
        { (exp(x) - 1) / (exp(x) + 1) };
        \draw[black, dashed] (axis cs: 0, 0.5) -- (axis cs: 1.1, 0.5);
        \draw[black, dashed] (axis cs: 1.1, 0) -- (axis cs: 1.1, 0.5);
        \draw[black, dashed] (axis cs: 0, 1) -- (axis cs: 6, 1);
    \end{axis}
\end{tikzpicture}
    \caption{The flux force efficacy (\(\eta\)) as a function of the (scaled) negative Gibbs free energy of reaction, \(-\frac{\Delta_rG'}{\rm RT}\). The efficacy of the flux force corresponds to the ratio between the net flux (forward minus backward flux) and the total flux (forward plus backward flux) of a reaction, which approaches \(1\) for reactions operating far from chemical equilibrium (\(\Delta_rG' \ll 0\)).\\[3mm]}  \label{FFE_plot}
\end{SCfigure}
\section{Author contributions}
A.B.-E.\ conceived the project and oversaw its initial stage. M.C and H.H.\ assembled and curated the model. M.C.\ wrote software, performed simulations and analysed results. W.L., E.N. and H.H.\ oversaw the project and contributed to the analysis and interpretation of results. M.C., H.H., W.L., and E.N.\ wrote the manuscript.

\section{Competing interests}
The authors declare that they comply with the PCI rule of having no financial conflicts of interest with respect to the content of the article.

\section{Code availability}
The model, together with all relevant data, is available on Github at \url{https://github.com/marco-corrao/iCH360}. The code and files required to reproduce all analyses in this manuscript are available on Github at~\url{https://github.com/marco-corrao/iCH360_paper} and on Zenodo at~\url{https://doi.org/10.5281/zenodo.11092781}. \rev{Additional analyses and comparisons with other models are also available in these repositories.}

\printbibliography
\newpage

\begin{appendices}

\section{Supplementary Information}
\subsection{Changes to reactions from the model \textit{i}ML1515}
\label{GEM_corrections}

After assembling the model as a subset of reactions from the genome-scale model \textit{i}ML1515, a few minor corrections were applied to some of the reactions based on evidence gathered from the literature. Note, however, that these changes did not result from an exhaustive process of review of the parent model. The corrections include:
\begin{itemize}
	\item In \textit{i}CH360, the membrane-bound transhydrogenase reaction (THD2pp) only translocates 1 proton across the periplasmic membrane, as opposed to the 2 protons translocated in the \textit{i}ML1515 reaction \cite{bizouarn_nucleotide_2005}
	\item The gene-protein rule (GPR) of gene \textit{glxK} (b0514) was reassigned to GLYCK2, which produces 2-phosphoglycerate. Meanwhile, the reaction GLYCK in from \textit{i}ML1515 (which produces 3-phosphoglycerate) was removed \cite{zelcbuch_vivo_2015,bartsch_only_2008}
	\item The NAPH-dependent homoserine dehydrogenase reaction (HSDy) was made irreversible towards the homoserine production direction \cite{he_optimized_2020}. To avoid having a reaction irreversible in the backward direction, substrate and products of the reactions were flipped.
 \item The succinate transport reaction SUCCt1pp was made irreversible in the export direction, enforcing the use of the more thermodynamically favourable SUCCt2\_2pp (which translocates two protons instead of one) for succinate import. To avoid having a reaction irreversible in the backward direction, substrate and products of SUCCt1pp were flipped.
\end{itemize}

\subsection{Computation of an equivalent biomass reaction}
\label{equivalent_biomass_computation}
In constraint-based models of metabolism, the biomass reaction summarises the production of all molecules that are not explicitly described by the model. These are typically macromolecules such as proteins or polynucleotides. Which metabolites are drained from the model network towards these other parts of metabolism, and in what proportions, depends on what compounds are described by the model. Hence, starting from an existing model, taking only a subset of the reactions but leaving the biomass reaction unchanged would lead to inconsistencies.

To construct a biomass reaction for \textit{i}CH360 that corresponds equivalently to the biomass reaction in \textit{i}ML1515, the following method was used. First, we collected all the pathways required for the production of the components present in the \textit{i}ML1515 biomass reaction (with the exception of a small number of compounds present with very small stoichiometry, which were excluded from this analysis for simplicity), but not in our model. These additional pathways (available in the repository supporting this manuscript) were manually curated based on available literature and database annotation to ensure they represent the most biologically-relevant bioproduction route for each biomass component. By adding these pathways to \textit{i}CH360, we obtained an extended model (\textit{i}CH360\textsubscript{ext}), which was able to predict growth directly through the original biomass reaction.

With this extended model at hand, we can continue as follows. Let \(N\) and \(M\) denote the number of reactions and metabolites, respectively,  in \textit{i}CH360. Further, let \(N_{\rm ext} > N\) denote  the number of reactions in the extended model \textit{i}CH360\textsubscript{ext}. A reference flux distribution \(\v{v}^*_{\rm ref} \in \Realn{N_{\rm ext}}\) is computed on the extended model through FBA. We can express this flux vector as:
\begin{equation}
    \v{v}^*_{\rm ref}=
    \begin{pmatrix}
        \v{v}^* \\
        \v{v}^*_+ \\
        v_{\rm BM}^*
    \end{pmatrix}
\end{equation}
where \(\v{v}^*\) is the subset of its fluxes corresponding to reactions present in \textit{i}CH360, \(\v{v}^*_+\) is the subset of fluxes corresponding to the reactions added to create the extended model  \textit{i}CH360\textsubscript{ext}, and \(v_{\rm BM}^*\) is the flux through the \textit{i}ML1515 biomass reaction. If the fluxes \(\v{v}^*\) were to be imposed on \textit{i}CH360 (which, at this stage, does not yet contain a biomass reaction), a number of metabolites would necessarily remain unbalanced, that is:
\begin{equation} \label {eq_biomass_unbalance}
 \M{S}~ \v{v}^* \neq \v{0}
\end{equation}
where \(\M{S} \in \Realn{M \times N}\) denotes the stoichiometric matrix of \textit{i}CH360. \rev{We now seek to define an "equivalent" biomass reaction that, when added to the \textit{i}CH360 network, will drain or produce metabolites so as to balance equation (\ref{eq_biomass_unbalance}).  Let \(\v{r_{\rm eq}} \in \Realn{M}\) denote the stoichiometry of such biomass reaction.  We compute \(\v{r_{\rm eq}}\) as the additional column of \(\M{S}\) required to balance the system while achieving the same biomass flux of the reference distribution. That is, \(\v{r_{\rm eq}}\) satisfies:}
\begin{equation}
    \Big(\M{S} \; \v{r_{\rm eq}}\Big)
    \begin{pmatrix}
        \v{v}^* \\
        v_{\rm BM}^*
    \end{pmatrix} = \v{0}
\end{equation}
which can be solved as:
\begin{equation}
    \v{r_{\rm eq}}=-\frac{1}{v_{\rm BM}^*} \M{S}\v{v}^*
\end{equation}
The stoichiometry of this equivalent biomass reaction depends on the choice of reference flux distribution \(\v{v}^*_{\rm ref}\) and, generally, may be different for different growth conditions. This results from the fact that, depending on the condition, the extended model may produce the same biomass component through different pathways, which would then be converted by our procedure into different equivalent costs of precursors in the sub-model. Nevertheless, the additional biosynthesis pathways we used to form the extended model do not allow for alternative routes to biomass, hence we found the equivalent biomass reaction to be, in this case, unique across conditions.
\subsection{Computation of attributes using the knowledge graph}\label{graph_based_computations}

Using the knowledge graph supporting the stoichiometric model of \textit{i}CH360 (Section \ref{annotation_graph} in the man text), a number of useful properties can be computed based on simple operations. Here, we outline how such an approach was used to i) compute the molecular mass of all enzyme complexes in the model, based on known masses for all polypeptides and ii) construct the boolean rules (GPRs) linking all reactions and proteins  in the graph to the model genes. For a description of the different types of nodes and edges mentioned in this section, see Supplementary Tables \ref{node_description_table} and \ref{edge_description_table}.

\subsubsection{Computation of molecular masses for all protein nodes}

In order to compute the molecular masses of all protein nodes in the graph, the protein nodes corresponding to polypeptides were first annotated with their molecular masses, readily available on the EcoCyc database. Then the molecular masses of all other protein nodes are estimated recursively as follows.
Let \(\mathcal{I}\) denotes the index set of all protein nodes and \(\mathcal{C}(i) \in \mathcal{I}\) the index set of protein components of node \(i\), i.e.~all nodes connected to node \(i\) by a subunit composition relationship. The molecular mass of any protein node \(i\), \(M_i\), is computed as:
\begin{equation}
    M_i=
    \begin{cases}
        \bar{M}_i \qquad &\text{if node \(i\) is a polypeptide} \\
        \sum_{k \in \mathcal{C}(i)} w_{ik}M_k \qquad &\text{otherwise} \\
    \end{cases}
\end{equation}
where \(\bar{M}_i\) denotes the (known) molecular mass of polypeptide node \(i\) and \(w_{ik}\) denotes the weight of the edge between \(i\) and \(k\). \np

\subsubsection{Computation of gene-protein-reaction rules}
\rev{Boolean gene-protein-reaction (GPR) rules are a widely used tool defining a map between a genotype (set of active genes) and a phenotype (set of active reactions) in a metabolic model. Conventionally, this is achieved by assigning to each reaction a boolean expression given in terms of genes in the model. In this section, we show how such expressions were computed for the reactions in \textit{i}CH360 using the knowledge graph.

Starting from the leaves of the graph (genes), we construct, for each node, a boolean expression describing its state (active/inactive, corresponding to a boolean \textit{true} / \textit{false}) in terms of its children. The exact form of this boolean expression  depends on the type of the node (reaction, protein, or logical) and the type of edges connecting it to its neighbours (Figure \ref{GPR_computation_schematics}). Particularly:
\begin{itemize}
    \item A polypeptide node is active if its associated gene is also active (Figure \ref{GPR_computation_schematics}A, left).
    \item A multimeric protein is active if all of its subunits (the child nodes connected to it by a "subunit composition" edge) are active (Figure \ref{GPR_computation_schematics}A, middle).
    \item A modified protein is active if its unmodified form (the child node connected to it by a "protein modification" edge) and its modification requirements (the child nodes connected to it by a "protein modification requirement" edge) are active (Figure \ref{GPR_computation_schematics}A, right).
    \item A logical AND (logical OR) node is active if all (any) of its child nodes are active (Figure \ref{GPR_computation_schematics}B).
    \item Finally, a reaction node is active if any of its catalysing isoenzymes (the child nodes connected to it via a "catalysis" edge) are active and, at the same time, all of its catalytic requirements (the child nodes connected to it via a "non-catalytic requirement" edge) are also active (Figure \ref{GPR_computation_schematics}C).
\end{itemize}
Using these definitions, the boolean expression describing the state of a reaction can be written, ultimately, solely in terms of genes, enabling computation of conventional GPR rules and their incorporation in the standard metabolic model.}
\label{graph_node_edges_detailing}

\subsection{Estimation of enzyme complex abundances}\label{complex_abundance_estimation}

In order to estimate the abundances of all enzymes in the model from proteomics data, we use the model graph (Main Text, Section \ref{model_graph}) to construct a stoichiometric map between enzymes and polypeptides. This map takes the form of a matrix \(\M{E} \in \Realn{n \times m}\), where \(n\) is the number of enzymes in the model and \(m \geq n\) the number of polypeptides, such that \(\M{E}_{ij}\) denotes the stoichiometry of polypeptide \(j\) in enzyme \(i\). Some polypeptides may be part of additional enzyme complexes that are not part of the model. \rev{Using the available annotation to the EcoCyc database, we identified \(7\) such polypeptides, mapping to \(9\) out-of-model complexes. If these out-of-model complexes were not taken into account, the abundance of model enzymes to which these polypeptides map would be overestimated.} Hence, we constructed a matrix \(\M{\hat{E}}\) by augmenting \(\M{E}\) with additional rows corresponding to the identified out-of-model complexes. 

With this mapping at hand, we assume that polypeptide abundances \(\v{p}\) are related to enzyme abundances \(\v{e}\) (including the required additional complexes not accounted for in the model) by
\begin{equation}
    \v{p}=\M{\hat{E}}^\top \v{e} 
\end{equation}
Hence, given a vector of experimental measurements of polypeptide abundances \(\bar{\v{p}}\), we estimate enzyme abundances by solving the nonnegative least square (NNLS) problem:
\begin{align}
    &\min_{\v{e}} \quad||\bar{\v{p}}-\M{\bar{E}}^\top \v{e}||_2^2 \\
    &\rm{s.t} \qquad  \v{e}\geq 0 \nonumber
\end{align}

\subsection{Adjustment of turnover numbers based on proteomics measurements across conditions}\label{kcat_adjustment_SI}

In this section, we outline the procedure used to adjust the turnover numbers in EC-\textit{i}CH360 by fitting proteomic measurements across conditions (see Section \ref{EC-iCH360} in the main text). Briefly, our aim is to adjust the turnover numbers that parametrise the model so that enzyme allocation predictions obtained through the enzyme-constrained formulation of FBA (see Methods in the main text) match more closely experimental measurements of enzyme abundances. By \textit{simultaneously} fitting experimental measurements across many growth conditions, we improve the robustness of the fitting procedure to experimental error and generate a condition-independent set of "typical" apparent turnover numbers that predict average trends of enzyme allocation across conditions. The output of our procedure is a set of typical enzyme efficiencies, one for each enzyme, estimated from proteomic data across conditions, as well as a set of condition-specific scaling factors that account for differences in total measured enzyme abundances between conditions. In section \ref{kcat_adjustment_SI-preliminaries} we rigorously define these parameters and state the main assumption underlying our heuristic. In sections \ref{kcat_adjustment_SI-reference_distribution}, \ref{kcat_adjustment_SI-adjustment}, and \ref{turnover_reparametrisation} we formulate a two-steps optimisation problem whose solution, upon a suitable reparameterisation, yields data-fitted estimates of the desired parameters.

\subsubsection{Preliminaries} \label{kcat_adjustment_SI-preliminaries}

Consider an enzyme $i$ in a given metabolic state $j$. This enzyme  catalyses a metabolic flux $v_{ij}$ (in (mol/gDW)/h) given by  
\begin{equation}
     v_{ij} =\kappa_{ij} ~c^{\rm enz}_{ij}
\end{equation} 
where $c^{\rm enz}_{ij}$ is the enzyme concentration  and $\kappa_{ij}$ is the enzyme efficiency (or ``apparent turnover number $k_{\rm app}$''). The efficiency $\kappa_{ij}$ is a positive rate (here in 1/h, but usually reported in 1/s). Since, by definition, it must be lower than the enzyme's turnover number $k_{\rm{cat}, i}$, we write it as 
\begin{equation}\label{apparent_kcat_definition}
    \kappa_{ij} = \sigma_{ij}~ k_{\rm{cat}, i}
\end{equation}
where $\sigma_{ij}\in [0,1]$ is a unitless ``capacity usage'' factor.
In enzyme-constrained models, enzymes are often expressed by their mass abundance $e_i=M_i~c^{\rm enz}_i$ (in g/gDW) instead of enzyme concentrations, where $M_{i}$ is the enzyme molecular mass (in g/mol), so we can write the flux as 
\begin{equation}
    v_{ij} = \frac{1}{a_{ij}} ~e_{ij}
\end{equation}
where  $a_{ij}$ is the enzyme cost per catalysed flux, given by the molecular mass $M_i$ divided by the enzyme efficiency $\kappa_{ij}$.

In principle, the capacity usage factors $\sigma_{ij}$ (and therefore efficiencies) of enzymes may freely vary between growth conditions. However, as a heuristic, we here assume that they can be approximated as the product of two factors: an enzyme-specific term and a condition-specific one, that is:
\begin{equation}\label{sigma_ij_approximation}
    \sigma_{ij} \approx \sigma_{i} \cdot \tau_j
\end{equation}
Here, the enzyme-specific factor $\sigma_{i}$ denotes the "typical" capacity usage factor of our enzyme in the range of conditions studied. The condition-specific term, $\tau_j$, is a unitless scaling factor that simultaneously increases or decreases the efficiencies of all enzymes depending on the cell's growth conditions. By convention, we assume that the values of $\tau_j$ are centered around \(1\). Substituting (\ref{sigma_ij_approximation}) in (\ref{apparent_kcat_definition}), we obtain an equivalent expression (under our assumptions) for the apparent turnover number \(\kappa_{ij}\):
\begin{equation} \label{kappa_ij_heuristic}
    \kappa_{ij}=\underbrace{\sigma_{i}~ k_{\rm{cat}, i}}_{\kappa_i}~\tau_j
\end{equation}
where \(\kappa_i\) is a ''typical apparent turnover number'' that is condition-independent. Practically, the above assumption allows us to simplify the problem by reducing the number of parameters to be fitted from \(I \cdot J\) to \(I+J\), where \(I\) and \(J\) denote the total number of enzymes and conditions considered, respectively. 
Our heuristic assumption corresponds to the idea that "high-quality carbon sources" allow a cell to establish metabolic states in which enzyme efficiencies are generally high, allowing for large fluxes per enzyme abundance in all the reactions, and therefore for high cell growth rates. Probably, this heuristics would fail in some other cases, e.g.~cases in which enzymes are specifically perturbed by enzyme inhibitors.
But in fact, it turns out that our model, assuming a single "typical apparent turnover number" $k_{\rm{app}, i}$ for each enzyme, yields very good enzyme allocation predictions. This is what would be expected if our heuristic assumption were correct, and therefore supports our heuristic prediction.

We now describe how the estimates of (enzyme-specific) efficiencies $\kappa_{i}$ and (condition specific) scaling factors $\tau_j$ for our model were obtained from model simulations and proteomics data.

\subsubsection{Overview of the fitting procedure}

Fitting the typical turnover parameters to proteomic data is, in general, not simple. Due to the linear programming formulation underlying the enzyme-constrained FBA problem, optimal flux distributions (and, by direct consequence, enzyme allocation predictions) are discontinuous over the turnover parameter space, making derivative-based searches through this space problematic from a numerical perspective. Similarly, the high dimensionality of the parameter space limits the applicability of gradient-free optimisation algorithms. Hence, we restrict ourselves to the (comparatively simpler) problem of adjusting turnover parameters for a fixed set of reference flux distributions across growth conditions, constraining our search to the portion of parameter space in which these reference flux vectors remain optimal for their respective conditions. This simplified fitting procedure thus consists of two steps. In the first part of the procedure, we use an initial parameter set to compute a set of reference flux distributions (one per growth condition) using enzyme-constrained FBA. In the second part, turnover parameters are fitted based on these reference flux distributions and experimental measurements of enzyme abundances. 

\subsubsection{Obtaining a set of reference flux distributions}\label{kcat_adjustment_SI-reference_distribution}

Following the sMOMENT formulation of enzyme-constrained FBA \cite{bekiaris_automatic_2020}, we consider a metabolic network with \(N\) reactions and \(M\) metabolites where all metabolic fluxes are positive (i.e.~reversible reactions are split into forward and backwards components) and at most one enzyme is associated with each reaction \rev{(see Methods in the main text for more information about how such unique mapping between reactions and catalyzing enzymes was generated in our case)}. Hence, we assume that the enzyme cost required to sustain flux \(v_i\) for a given growth condition \(j\) is given by \(a_{ij}~v_i\), where the cost per unit flux \(a_{ij}\) is given by:
\begin{equation}\label{cost_per_unit_flux}
   a_{ij}= \begin{cases}
        \frac{M_i}{\kappa_{ij} } \quad &\text{if reaction \(i\) is enzymatic }\\
        0 \quad &\text{otherwise}
    \end{cases}
\end{equation}
Here, \(M_i\) is the molecular mass of the enzyme associated with the reaction, \(\kappa_{ij}\) is the condition-dependent enzyme efficiency, as defined in \ref{kappa_ij_heuristic}.
Given an initial guess for the value of each \(\kappa_{ij}\), we compute a reference flux distribution for the \(j\)th growth condition, \(\v{v}_j^*\) by fixing the biomass flux, \(v_{\rm BM}\) to the experimentally measured rate and minimizing the total enzyme cost:
\begin{equation}\label{ECM_problem0}
    \begin{aligned}
    \v{v}_j^*=\argmin_{\v{v}}  &\quad \v{a}_j^\top \v{v}  \\
                     \rm{s.t.} &\quad \M{S}~\v{v}=\v{0} &\qquad \qquad \rm{(a)} \\
                               &\quad \M{B}_j~\v{v}\leq \v{b}_j &\qquad \qquad \rm{(b)}\\
                               & \quad v_{\rm BM}=\bar{v}_{\mathrm{BM},j}&\qquad \qquad \rm{(c)}\\
                               & \quad \v{v}\geq \v{0}
\end{aligned}
\end{equation}
Here, the objective \(\v{a}_j^\top \v{v}\) is the total enzyme cost for the \(j\)th growth condition, \(\bar{v}_{\mathrm{BM},j}\) is the experimentally measured growth rate for the condition, \(\M{S} \in \Realn{\rm{M} \times \rm{N}}\) is the stoichiometric matrix of the network and \(\M{B}_j \in \Realn{\rm{P} \times \rm{N}}\) and \(\v{b}_j \in \Realn{\rm{P}}\) are a matrix and a vector, respectively, encoding any desired upper bound (or positive lower bound) on the fluxes for the growth condition. Noting that constraint \ref{ECM_problem0}c can be equivalently cast as a double inequality, we rewrite the problem in the more general form:
\begin{equation}\label{ECM_problem}
    \begin{aligned}
    \v{v}_j^*=\argmin_{\v{v}}  &\quad \v{a}_j^\top \v{v}  \\
                     \rm{s.t.} &\quad \M{S}~\v{v}=\v{0} &\qquad \qquad \rm{(a)} \\
                               &\quad \hat{\M{B}}_j~\v{v}\leq \hat{\v{b}}_j &\qquad \qquad \rm{(b)}\\
                               & \quad \v{v}\geq \v{0}
\end{aligned}
\end{equation}
where the biomass flux is assumed to be the last component of the flux vector and the augmented matrices
\begin{equation}
    \hat{\M{B}}_j \equiv \begin{pmatrix}
                            \M{B_j}\\
                            [0, \cdots, \hphantom{-}1]\\
                            [0, \cdots, -1]
                         \end{pmatrix}\qquad
    \hat{\v{b}}_j \equiv \begin{pmatrix}
                            \v{b}_j\\
                            \bar{v}_{\mathrm{BM},k}\\
                            -\bar{v}_{\mathrm{BM},k}
                         \end{pmatrix}                        
\end{equation}
were introduced. 

In order to solve the the linear program \ref{ECM_problem0}, we shall consider an initial guess for the turnover parameters and assume that, for each condition, all enzymes operate at the same saturation level, so that \(\sigma_{ij}\equiv\bar{\sigma}_j\). Since the optimal flux distribution obtained as a solution of problem  (\ref{ECM_problem}) is unchanged by the choice of \(\bar{\sigma}_j\) (as this merely amounts to a scaling of the objective function), we can simply set \(\bar{\sigma}_j=1\) (that is, set \(\kappa_{ij}=k_{\rm cat,i}\)) for all conditions at this stage.

Solving (\ref{ECM_problem}) for each of the \(J\) growth conditions available in the experimental dataset, we obtain a set of optimal flux distributions \( \mathcal{V^*}=\{\v{v}_1^*, \cdots, \v{v}_J^*\}\), which we will use as a reference in the next step.

\subsubsection{Estimating typical enzyme efficiencies for the reference set of flux distributions}
\label{kcat_adjustment_SI-adjustment}

We now turn to fitting the relevant parameters against experimental measurements of enzyme abundances. For this purpose, we shall express the efficiency of each enzyme-condition pair as:
\begin{equation}\label{kappaij_fitting}
    \kappa_{ij}=\theta_j~k_i
\end{equation}
where \(\theta_j\) is a condition-specific scaling term that simultaneously scales all efficiencies for a given condition, while \(k_i\) is an "adjusted" turnover parameter that simultaneously accounts for inaccuracies in the original turnover parameter numbers as well as differences in typical saturation across enzymes (and, hence, it's not formally a turnover number). While the above parameterization differs from the one in equation (\ref{kappa_ij_heuristic}) -- the terms have a different interpretation! -- it will greatly simplify the notation of the fitting problem, as it allows us to easily distinguish between "global" adjustments (those affecting all enzymes in a condition), which we wish to pick freely, and "local" adjustments (affecting the efficiency of individual enzymes), which instead we wish to regularise. While this factorisation of saturation effects is merely a computational convenience and not necessarily biologically meaningful, we will retrieve the parameters in Eq.~(\ref{kappa_ij_heuristic}) from those in 
Eq.~(\ref{kappaij_fitting}) upon a simple reparameterisation, as we detail in Section \ref{turnover_reparametrisation}

To formulate our fitting procedure, we will  denote with \(\v{p}\) a vector of  log\textsubscript{10} adjusted turnover numbers  (i.e.~\(p_i=\log_{10}{k_i}\)) and with \(\v{s}\) a vector of condition-specific log\textsubscript{10}-scaling factor (i.e.~\(s_j=\log_{10}\theta_j\)). Further, we denote with \(\bar{\v{p}}\) the vector of original log-turnover numbers (i.e.~the one used to obtain the reference flux distributions). For the choice of reference flux distribution computed in the previous step, the abundance of the \(i\)th enzyme in condition \(j\), \(e_{ij}\), is then a function of \(\v{p}\) and \(\v{s}\):
\begin{equation}
    e_{ij}(\v{p},\v{s})=\sum_{k}a_{kj}~v^*_{kj}
\end{equation}
where the summation index \(k\) runs across all reactions catalysed by enzyme \(i\) and the flux cost of reaction \(k\) in condition \(j\), \(a_{kj}\), is computed as in (\ref{cost_per_unit_flux}). From the formulation of the linear program (\ref{ECM_problem}), there must exist a region \(\mathcal{S}_j\) of log-turnover parameter space such that \(\bar{\v{p}} \in \mathcal{S}_j\) and that, for every \(\v{p} \in \mathcal{S}_j\),  \(\v{v}_j^* \in \mathcal{V}^*\) is the optimal solution of problem (\ref{ECM_problem}) for its growth condition. Hence, in this step of the adjustment procedure, we aim to find a set of typical log efficiencies, \(\v{p}^*\) and log scaling factors, \(\v{s}^*\) by minimising the discrepancy between enzyme abundance predictions and measurements, constraining our search to this region of the parameter space \(\mathcal{S} \equiv \bigcap\limits_{j} \mathcal{S}_j\) where \textit{all} reference flux distributions are optimal for their respective growth condition: 
\begin{equation}\label{kcat_adjustment_problem0}
    \begin{aligned}
     (\v{p}^*, \v{s}^*)=\argmin_{\v{p},~\v{s}}  &\quad  \frac{1}{N_{\rm e}}\sum_{i,j}~\big[ ~l( e_{ij})-l(\bar{e}_{ij})~\big]^2 + \frac{\rho}{N_{\rm p}} \sum_i \big(~p_i -\bar{p}_i~\big)^2 \\
                     \rm{s.t.} &\quad \v{u_{\rm min}}\leq \v{p}-\bar{\v{p}} \leq \v{u_{\rm max}} \\
                               &\quad \v{p} \in \mathcal{S}
\end{aligned}
\end{equation}
where \(N_{\rm e}\) is the number of enzyme-condition pairs, \(N_{\rm p}\) is the number of turnover parameters, \(\bar{e}_{ij}\) is the experimental measurement of enzyme \(i\) in condition \(j\), \(\v{u_{\rm min}}\) and \(\v{u_{\rm max}}\) are bounds on the allowable adjustment, \(\rho> 0\) is a scalar hyperparameter, and the function \(l(\cdot )\) is defined as:
\begin{equation}
l(x)=
    \begin{cases}
        \log_{10}(x) \quad  &x>0 \\
        0       \quad &x=0
    \end{cases}
\end{equation}

The objective function of the nonlinear program (\ref{kcat_adjustment_problem0}) is a combination of two terms. The first term penalises the mean squared deviation between measurements and predictions of log-enzyme abundance. Note that the above definition of \(l(\cdot)\) implies that, for each condition, only enzymes with nonzero predicted abundance are included in this term.  The second term is a regularisation expression (whose strength is controlled by the hyperparameter \(\rho\)) penalising the mean squared deviation between the adjusted turnover parameters and the original parameter set. The latter term mainly serves two purposes: first, it ensures that, whenever a turnover parameter is "free" in the problem (which can happen if its associated reaction fluxes are \(0\) across all conditions, or if no experimental measurements are available for its associated enzyme), it will be kept at its original value; secondly, it provides a mean to tune the strength of the adjustment procedure.

In order to define the region \(\mathcal{S}\), we shall exploit the sufficient optimality conditions of LP (\ref{ECM_problem}). Introducing, for each growth condition, the vectors of dual variables, \(\vs{\lambda}_j \in \Realn{M}\) and \(\vs{\mu}_j \in \Realn{P}\), corresponding to constraints \ref{ECM_problem}b and \ref{ECM_problem}c, respectively, we note that problem (\ref{ECM_problem}) admits (for the \(j\)th growth condition) a dual problem in the form:
\begin{equation}\label{ECM_problem_dual}
    \begin{aligned}
    \max_{\vs{\lambda}_j,~\vs{\mu}_j}  &\quad -\hat{\v{b}}_j^\top \vs{\mu}_j  \\
                     \rm{s.t.} &\quad \M{S}^\top\vs{\lambda}_j~+\hat{\M{B}}_j^\top \vs{\mu}_j + \v{a}_j \geq \v{0} &\qquad \qquad \rm{(a)} \\
                               & \quad \vs{\mu}_j\geq \v{0}
\end{aligned}
\end{equation}
A well-known result in linear programming duality theory \cite{maranas_zomorrodi_2016} states that a flux distribution is the optimal solution of the primal problem (\ref{ECM_problem}), if, and only if,  the dual problem (\ref{ECM_problem_dual}) is feasible (dual feasibility) and its optimal objective coincides with the primal optimal objective, that is \(\v{a}_j^\top\v{v}_j^*=-\v{b}_j^\top \vs{\mu}_j \) (strong duality). Taken together, dual feasibility and strong duality thus define the region of optimality in turnover parameter space of each reference flux distribution.  Hence, we can integrate the above definition of \(\mathcal{S}\) within problem (\ref{kcat_adjustment_problem0}) by introducing the two vectors of dual variables (\(\vs{\lambda}_j\) and \(\vs{\mu}_j\)) for each condition as additional optimisation variables, and simultaneously enforcing the optimality for each reference distribution. By doing this, we obtain the final formulation of the nonlinear program for turnover number adjustment, which we solved to obtain the results shown in the main text:
\begin{equation}\label{kcat_adjustment_problem}
    \begin{aligned}
     (\v{p}^*, \v{s}^*)= \argmin_{\v{p},~\v{s}}  &\quad  \frac{1}{N_{\rm e}}\sum_{ij}~\big[ ~l( e_{ij})-l(\bar{e}_{ij})~\big]^2 + \frac{\rho}{N_{\rm p}} \sum_i \big(~p_i -\bar{p}_i~\big)^2 \\
                     \rm{s.t.} &\quad \v{u_{\rm min}}\leq \v{p}-\bar{\v{p}} \leq \v{u_{\rm max}} \\
                               &\quad\M{S}^\top\vs{\lambda}_j~+\hat{\M{B}}_j^\top \vs{\mu}_j + \v{a}_j \geq \v{0} &\quad j=1,\dotsc,J\\
                               &\quad \vs{\mu}_j \geq \v{0}                                     &\quad j=1,\dotsc,J\\
                               &\quad \v{a}_j^\top\v{v}_j^*=-\hat{\v{b}}_j^\top \vs{\mu}_j                    &\quad j=1,\dotsc,J
\end{aligned}
\end{equation}
\subsubsection{Conversion to apparent turnover numbers}\label{turnover_reparametrisation}
The above procedure produces (after conversion back to a linear scale) a set of adjusted turnover parameters, \(\v{k}^*\) and scalings, \(\vs{\theta}^*\). In order to retrieve the typical enzyme efficiencies and condition-specific scaling factors introduced in \ref{kappa_ij_heuristic}, we simply parametrise the solution by factorising the scaling terms \(\theta_j^*\) as 
\begin{equation}
    \theta_j^* \equiv \bar{\sigma}^*~\tau_j^*
\end{equation}
Here, \(\bar{\sigma}^*\) is the geometric mean of the scalings across conditions, which we interpret as typical enzyme saturation level across conditions, while \(\tau_j^*\) is a residual scaling factor fluctuating around \(1\) which is required to account for differences in total measured enzymes between conditions. The typical enzyme efficiencies (\(\kappa_i\)) introduced in \ref{kappa_ij_heuristic} are then simply recovered by incorporating the \(\bar{\sigma}^*\) constant into the fitted turnover parameters:
\begin{equation}
   \kappa_i\equiv \bar{\sigma}^*~k_{i}^*
\end{equation}
\subsubsection{Potential extensions}
We conclude this section by noting that the procedure described above assumes that the original parameter set -- the set of $k_{\rm cat}$ values used as proxies for apparent $k_{\rm cat}$ values in section \ref{kcat_adjustment_SI-reference_distribution} -- is sufficiently good to produce a realistic flux distribution to use as a reference for the adjustment step. If this is not the case, then a multi-start approach can be implemented, where multiple turnover parameter sets are first generated by perturbing the original parameter set, and each of them is used to generate a separate set of flux distributions. Each reference set is then provided as an input to problem (\ref{kcat_adjustment_problem}), and the solution achieving the lowest objective is chosen in the end. The perturbed parameter set may be generated either randomly (for example, by introducing log-normal noise in the original turnover parameter vector) or systematically. The latter could be achieved, for example, by identifying reactions with zero-predicted flux but high measured abundance of the associated enzyme. By systematically increasing the corresponding turnover number, one can "encourage" these reactions to be included in the reference set, potentially leading to the exploration of more relevant reference distributions than achieved by random perturbations.

Note that, for in this work, we limited ourselves to the original parameter set and thus did not explore this potential heuristic.
\newpage
\section{Supplementary Figures}
\renewcommand{\figurename}{Supplementary Figure}
\renewcommand{\thefigure} {S\arabic{figure}}
\setcounter{figure}{0}
\begin{figure}[h!]
	\centering
	\includegraphics[width=\textwidth]{all_images/aminoacid_biosynthesis_annotated.pdf}
	\caption{Biosynthesis of amino acids (blue labels) from core metabolism precursors (red labels) in \textit{i}CH360. The Escher maps (including Suppl.~Fig.~\ref{nt_biosynthesis} to \ref{c1-metabolism}) are available along with the code. R5P: ribose-5-phosphate; E4P: erythrose 4-phosphate; PEP:phosphoenolpyruvate; 3PG: 3-phosphoglycerate; AKG: apha-ketoglutarate; CBP: carboamyl phosphate.}
    \label{aa_biosynthesis}
\end{figure}

\newpage
\begin{figure}[h!]
	\centering
	\includegraphics[width=\textwidth]{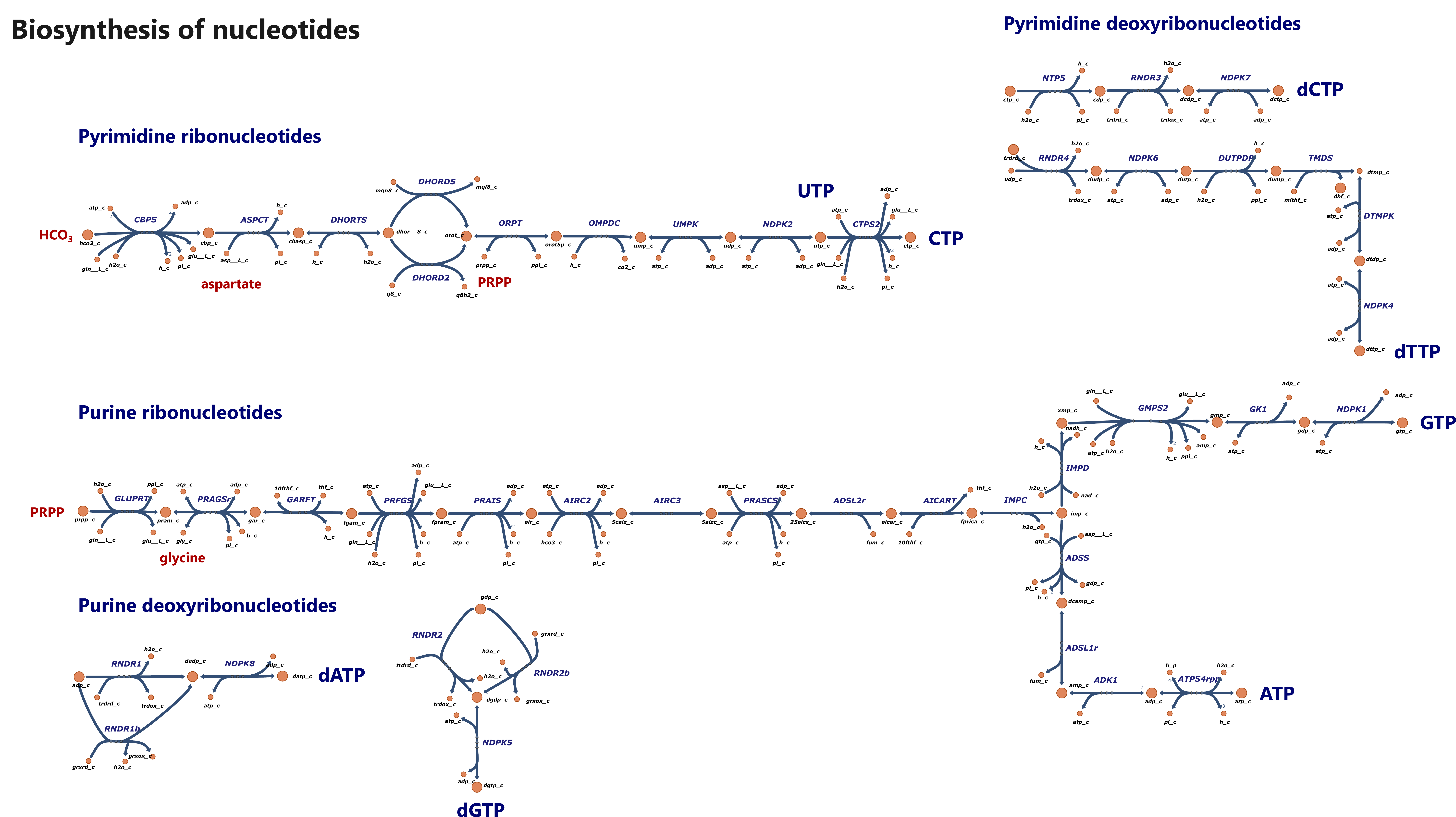}
	\caption{Biosynthesis of pyrimidine and purine (deoxy)ribonucleotides (blue labels) from core and aminoacid metabolism precursors (red labels) in \textit{i}CH360. PRPP: 5-phosphoribosyl-1-pyrophosphate.}
     \label{nt_biosynthesis}
\end{figure}

\newpage
\begin{figure}[h!]
	\centering
	\includegraphics[width=\textwidth]{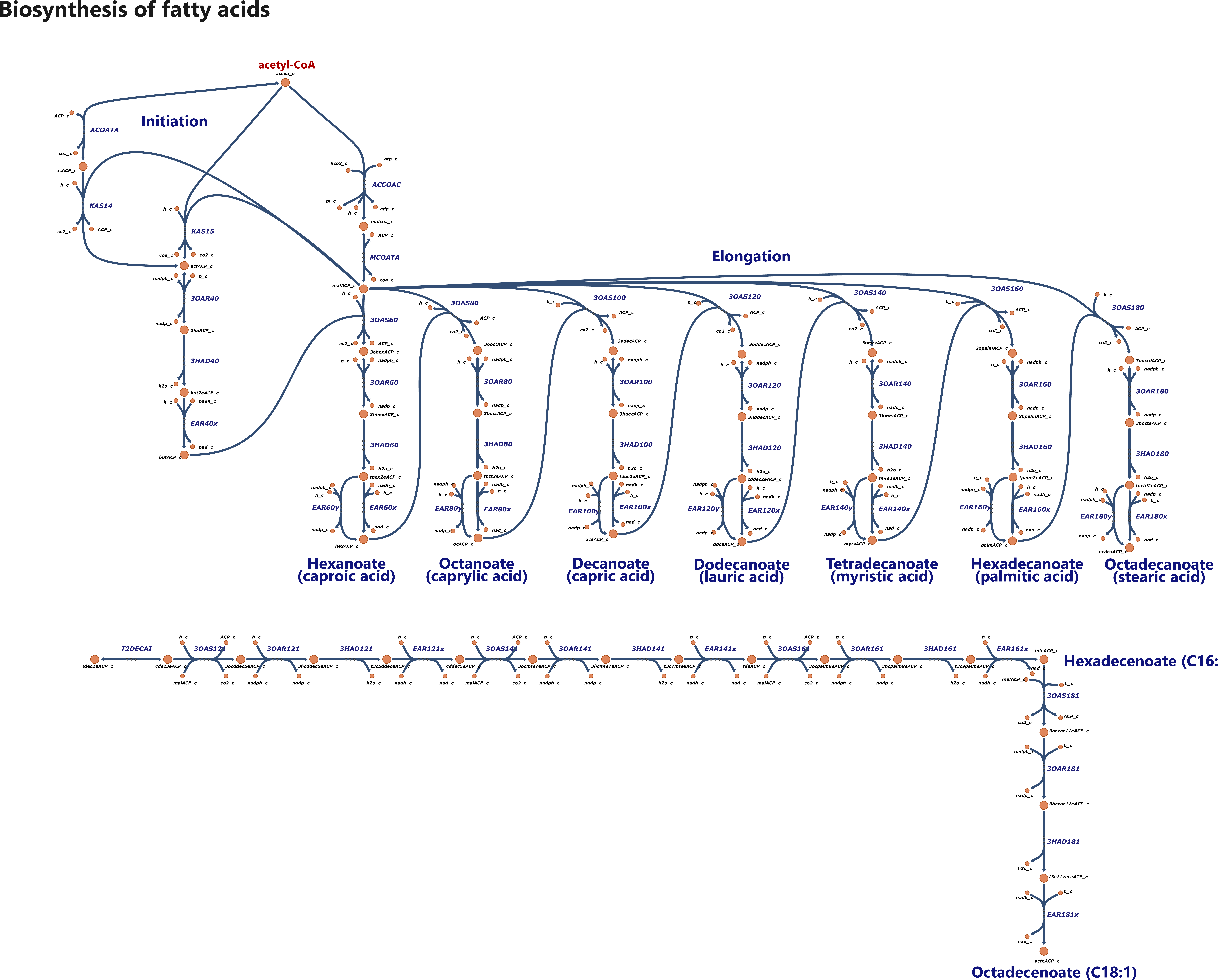}
	\caption{Biosynthesis of saturated and unsaturated fatty acids from acetyl-CoA. The map for saturated fatty acids was taken from Escher \cite{king_escher_2015}.}
     \label{fa_biosynthesis}
\end{figure}

\newpage
\begin{figure}[h!]
	\centering
	\includegraphics[width=0.6\textwidth]{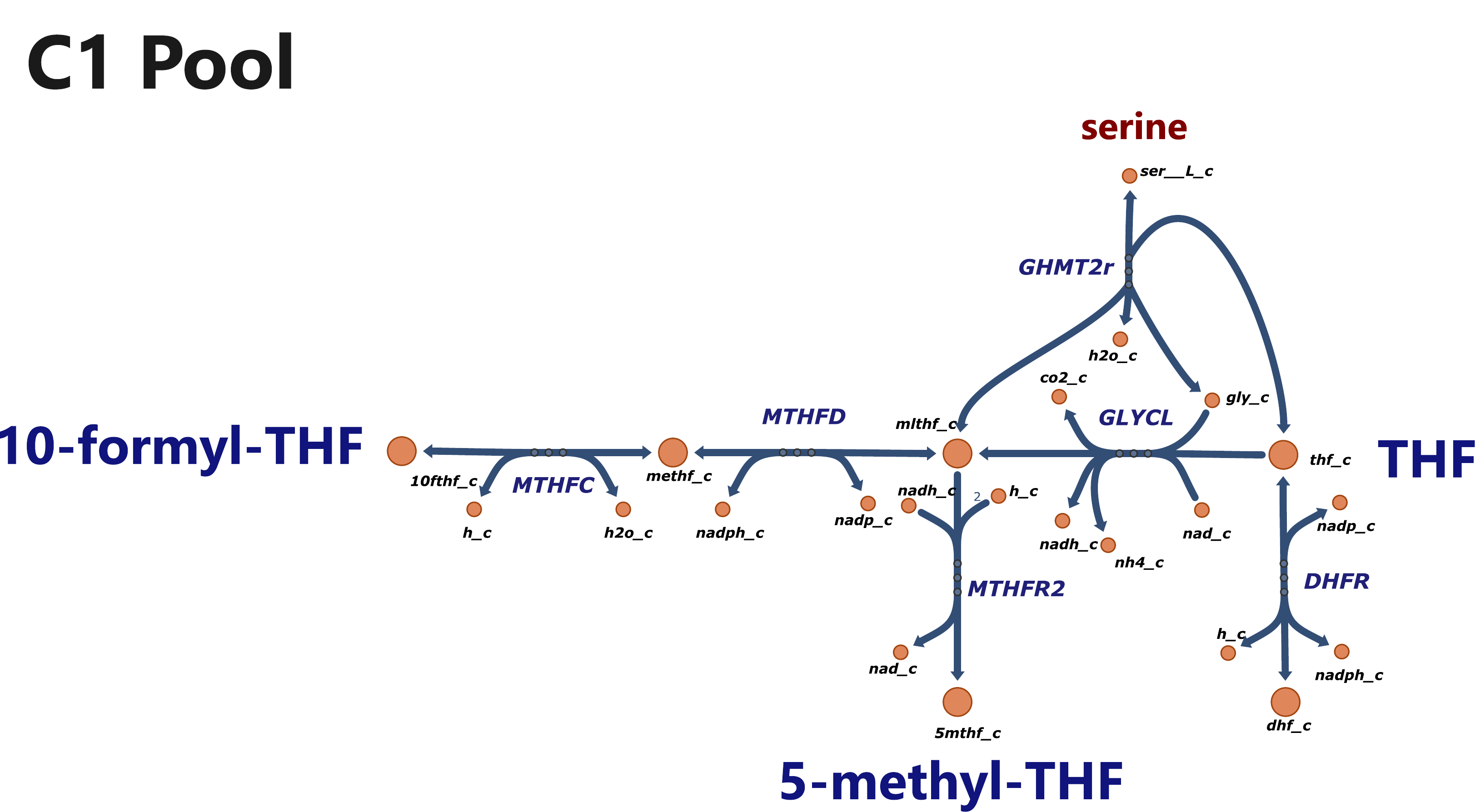}
	\caption{Metabolism of one-carbon compounds in \textit{i}CH360. THF: tetrahydrofolate. }
     \label{c1-metabolism}
\end{figure}

\newpage
\begin{figure}[h!]
	\centering
	\includegraphics[width=0.7\textwidth]{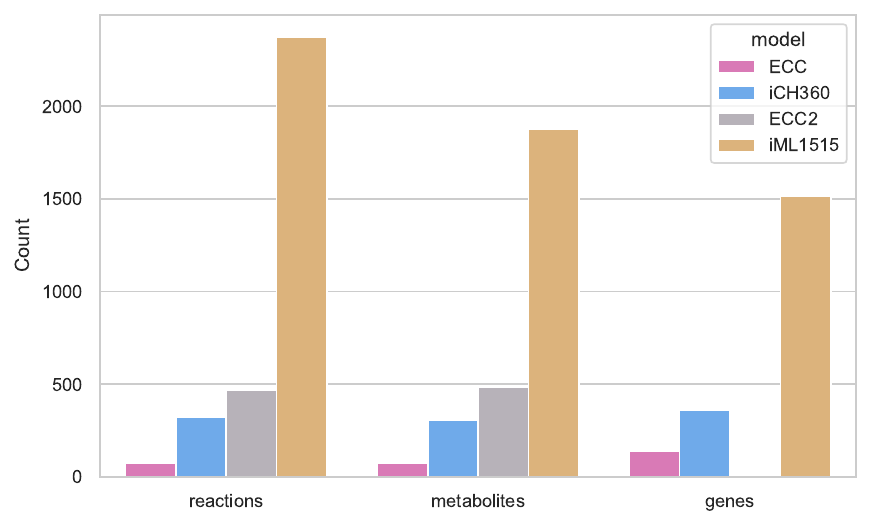}
	\caption{Comparison of model sizes between ECC, ECC2, \textit{i}CH360 and \textit{i}ML1515. To allow for a fair comparison, pseudo-reactions (e.g.~exchange reactions) were excluded from the count. Note that gene annotations were not available in the SBML model of ECC2 accompanying its publication \cite{hadicke_ecolicore2_2017}. }
	\label{model_size_comparison}
\end{figure}

\newpage
\begin{figure}[h!]
\begin{center} \includegraphics[width=0.8\textwidth]{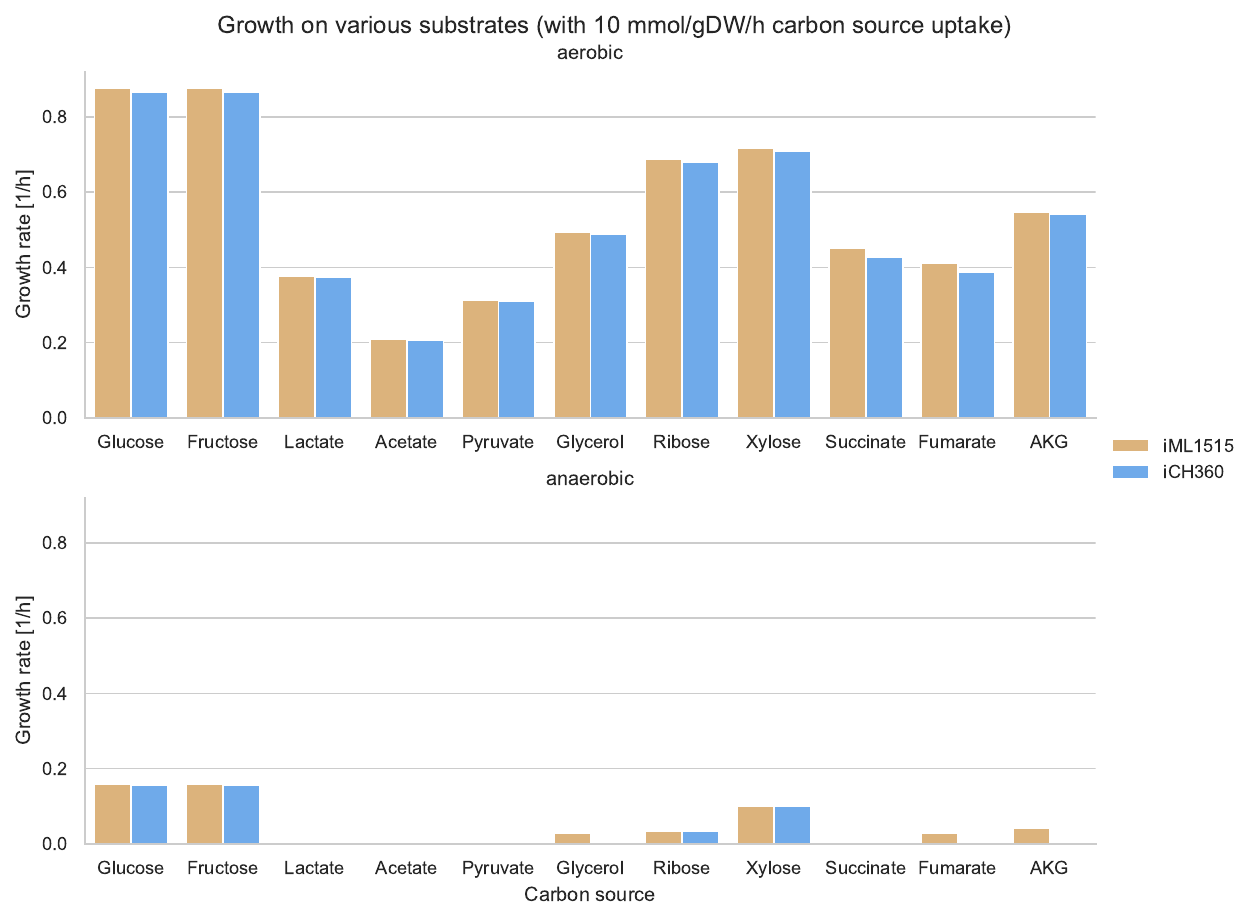}
 \end{center}
	\caption{Maximal biomass fluxes achieved by \textit{i}CH360 and its parent model \textit{i}ML1515, for aerobic and anaerobic growth across multiple carbon sources. In all cases, the substrate uptake flux was bounded to \(10\) mmol/gDW/h.}
 \label{model_comparison_growth}
\end{figure}

\newpage
\begin{figure}[h!]
\begin{center}
 \includegraphics[width=1\textwidth]{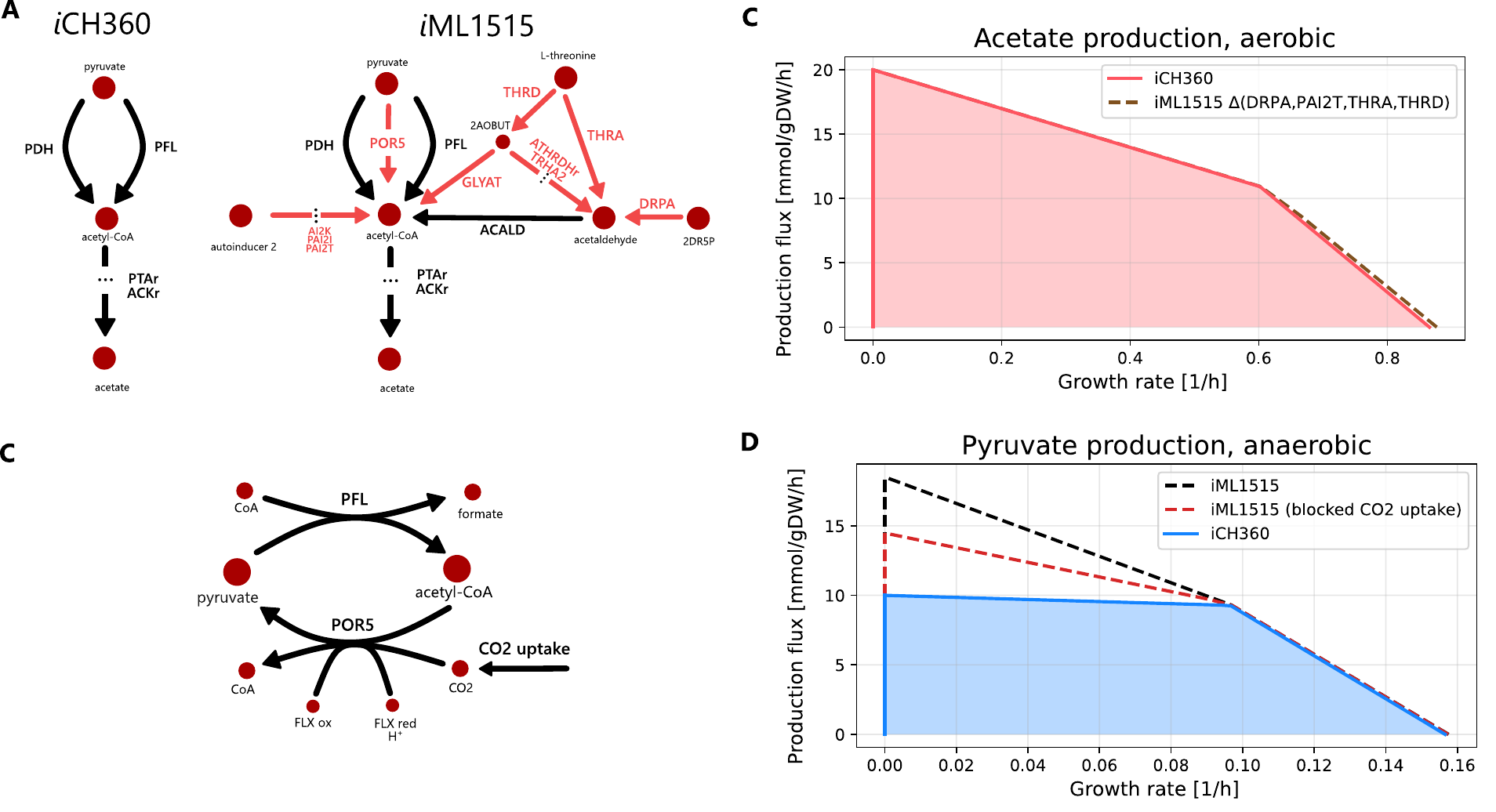}
 \end{center}
	\caption{Analysis of the differences in acetate production between \textit{i}CH360 and its genome-scale parent, \textit{i}ML1515. \textbf{A}: available metabolic routes for the production of acetate in both models. Left: \textit{i}CH360 can only produce acetyl-CoA, precursor for acetate, via the oxidation of pyruvate by either pyruvate dehydrogenase (PDH) or pyruvate-formate-lyase (PFL). Note that the latter reaction is known not to be active under aerobic conditions, but we did not block it for the purpose of this analysis. \textit{i}ML1515 can additionally produce acetate via additional pathways not present in \textit{i}CH360 (in red). Note that only the main substrate and products for each reaction are shown for clarity. 2AOBUT: L-2-Amino-3-oxobutanoate; 2DR5P: 2-Deoxy-D-ribose 5-phosphate. \textbf{B}: Blocking these degradation routes by simultaneous knockout of four reactions (DRPA, PAI2T, THRA, THRD) results in the two model sharing a virtually identical production envelope under aerobic conditions. The production envelope shown here was computed for aerobic growth using glucose as a carbon source. \textbf{C}: Under anaerobic conditions, the differences between the two model are further exacerbated by the ability of the genome-scale network to achieve higher pyruvate production yield. The genome-scale model can produce higher amounts of pyruvate by uptaking external CO\textsubscript{2} and using it as an electron sink using the POR5 reaction. FLX ox/red: oxidised/reduced flavodoxin. \textbf{D}: Blocking CO\textsubscript{2} uptake reduces the differences in pyruvate production between the two models, but does not remove them completely, implying the existence of additional mechanisms used by \text{i}ML1515 to achieve higher pyruvate yields. The production envelopes shown here was computed for anaerobic growth using glucose as a carbon source. }
 \label{aerobic_acetate_production_iCH360_vs_iML1515}
\end{figure}

\newpage

\newpage
\begin{figure}[h!]
\begin{center}
 \includegraphics[width=0.9\textwidth]{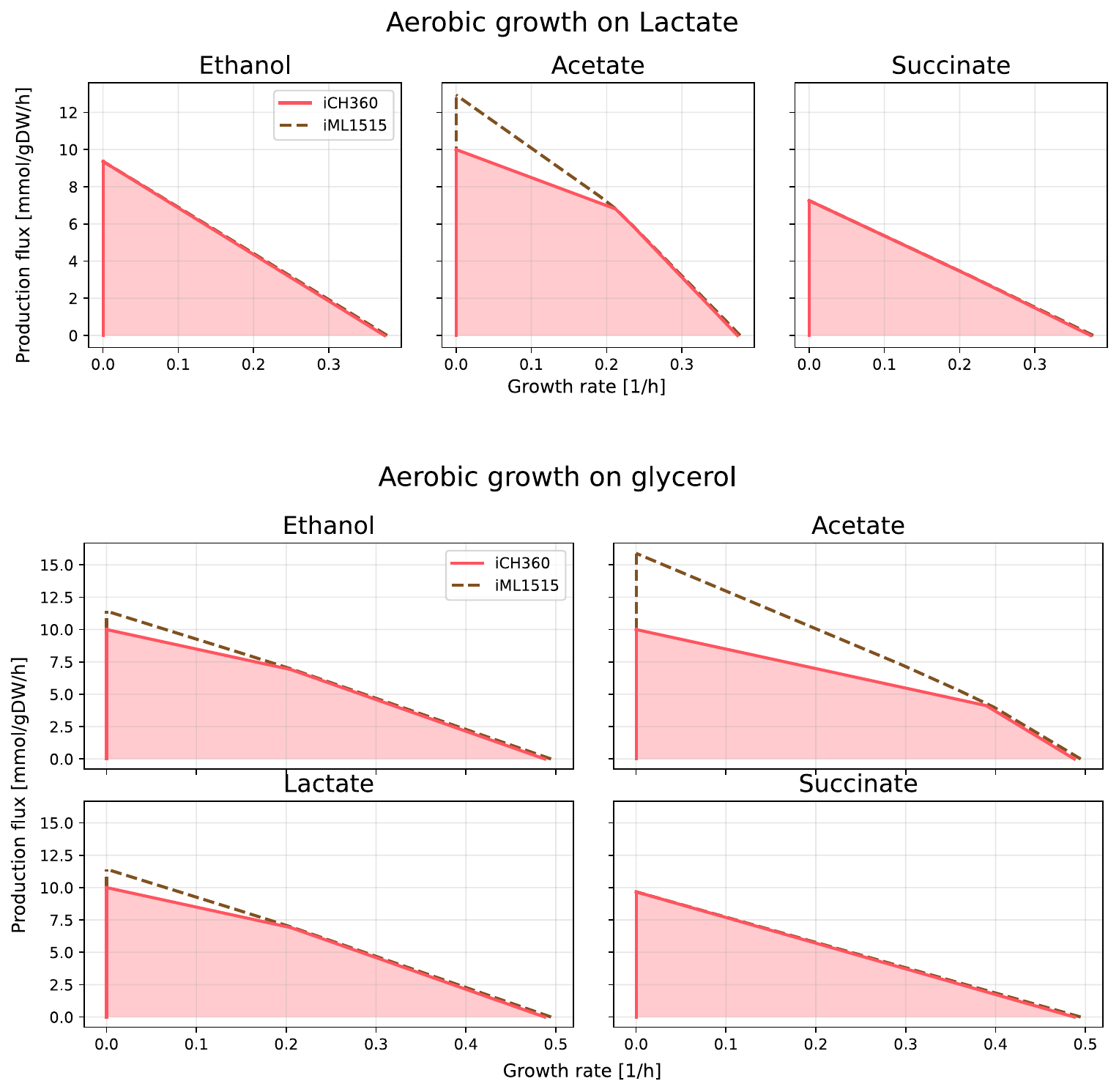}
 \end{center}
	\caption{Comparison of production envelopes between \textit{i}CH360 and its parent model \textit{i}ML1515. Top: production of ethanol, acetate and succinate during aerobic growth on lactate. Bottom: production of ethanol, acetate, lactate and succinate during aerobic growth on glycerol. Note that the dashed line representing the production envelope of \textit{i}ML1515 is sometimes hidden behind the blue line corresponding to \textit{i}CH360. }
 \label{additional_production_envelopes_lactate_glycerol}
\end{figure}

\newpage
\begin{figure}[h!]
\begin{center}
 \includegraphics[width=0.9\textwidth]{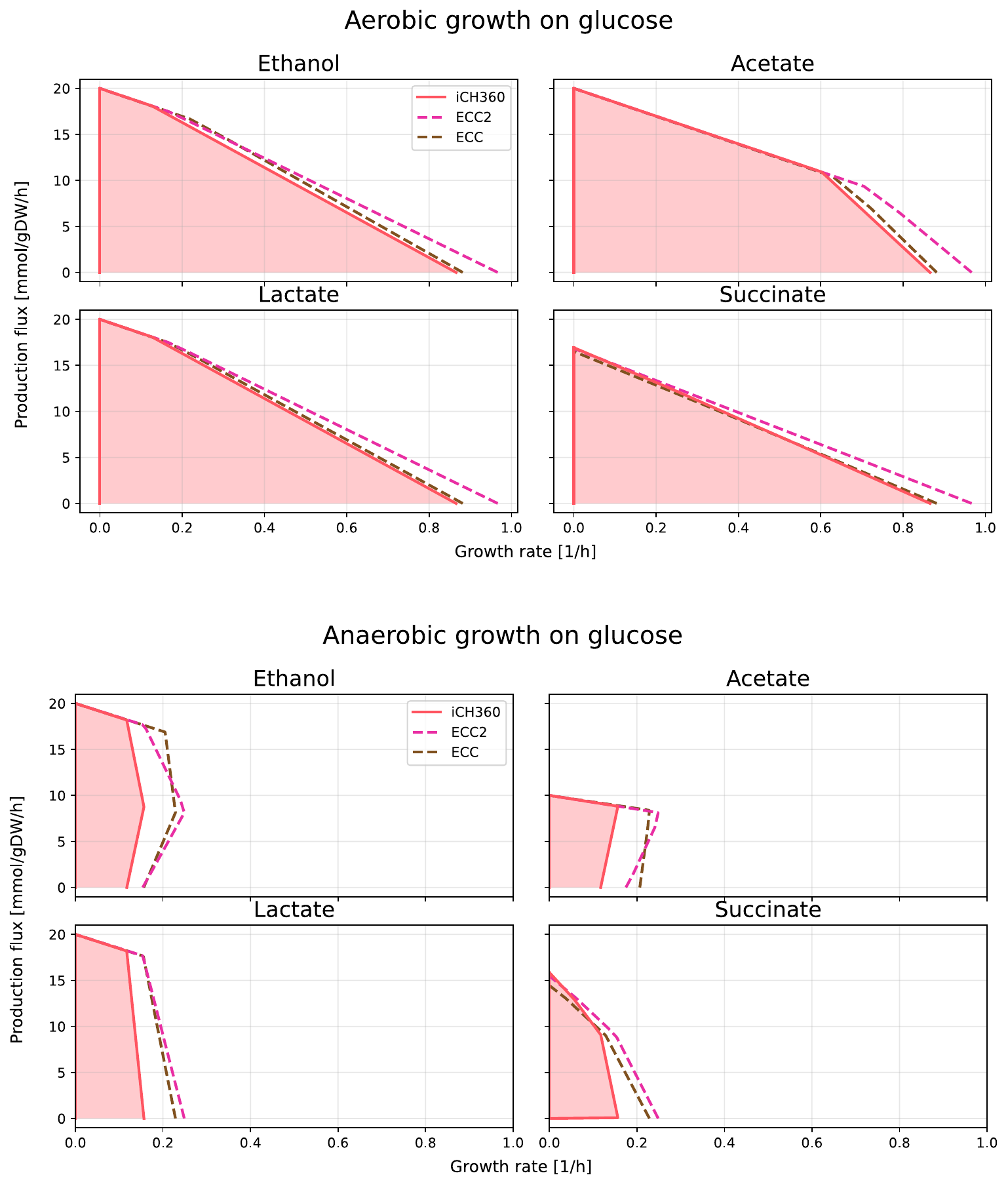}
 \end{center}
	\caption{Comparison of production envelopes between \textit{i}CH360 and other medium-scale models, namely \textit{E. coli} Core (ECC) and \textit{E. coli} Core 2 (ECC2) for growth on glucose as a sole carbon source.Top: production of ethanol, acetate and succinate under aerobic conditions. Bottom: production of ethanol, acetate, lactate and succinate under anaerobic conditions. Additional comparisons between the three models are available in the repository supporting this manuscript.}
 \label{production_envelopes_ich360_vs_ecc_ecc2_glucose}
\end{figure}
\newpage
\begin{figure}[h!]
    
	\centering	\includegraphics[width=0.4\textwidth]{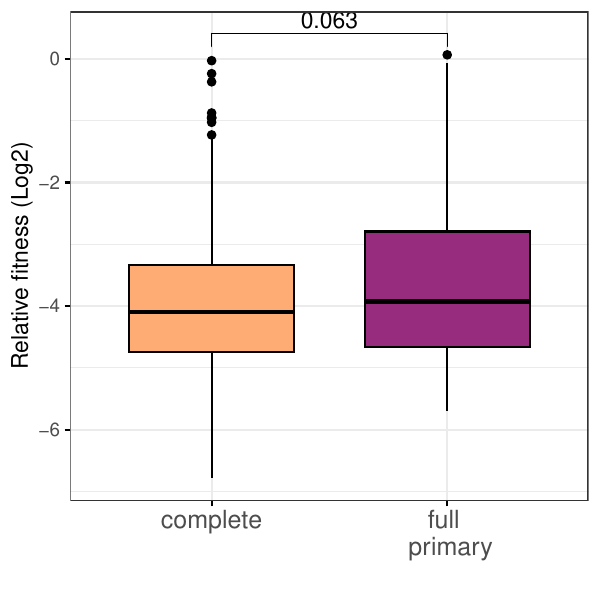}
	\caption{Fitness losses associated with disruptions of catalytic edges. There is no significant difference between the fitness effects of disruptions classified as complete disruptions (disruption of all catalytic edges for a reaction) and full primary disruption (disruption of all primary catalytic associations, but with remaining secondary ones), according to a Wilcoxon rank-sum test, \(p=0.063\).}
 \label{complete_vs_full_primary_fig}
\end{figure}
\newpage
\begin{figure}[h!]
	\centering	\includegraphics[width=1\textwidth]{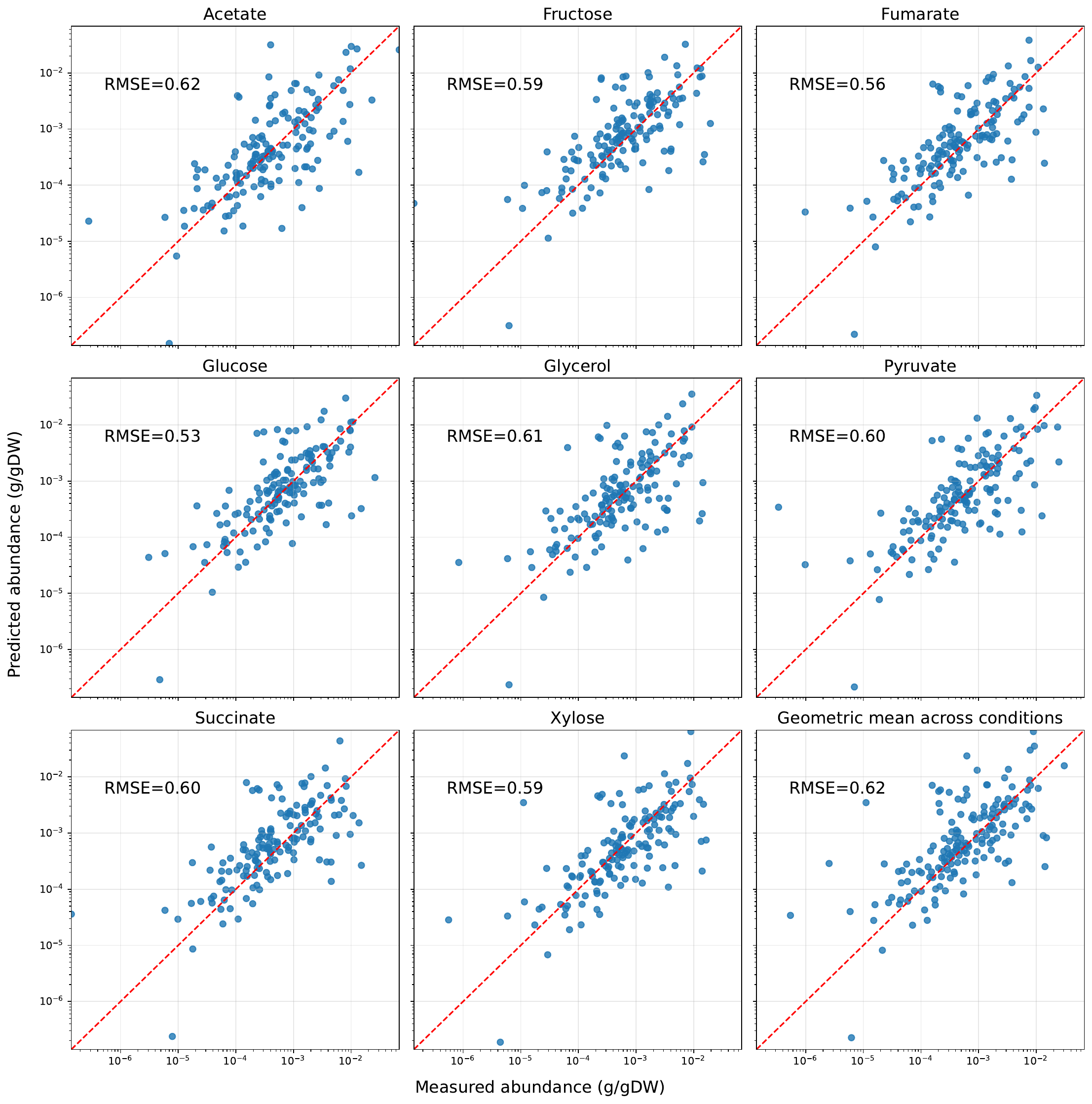}
	\caption{Predicted proteome allocation across growth conditions, obtained with the turnover parameter set from \citeauthoryear{heckmann_kinetic_2020}. The bottom-right panel shows the geometric means of measurements and predictions across conditions.}
 \label{proteome_allocation_predictions_all_conditions}
\end{figure}

\newpage
\begin{figure}[h!]
	\centering	\includegraphics[width=0.6\textwidth]{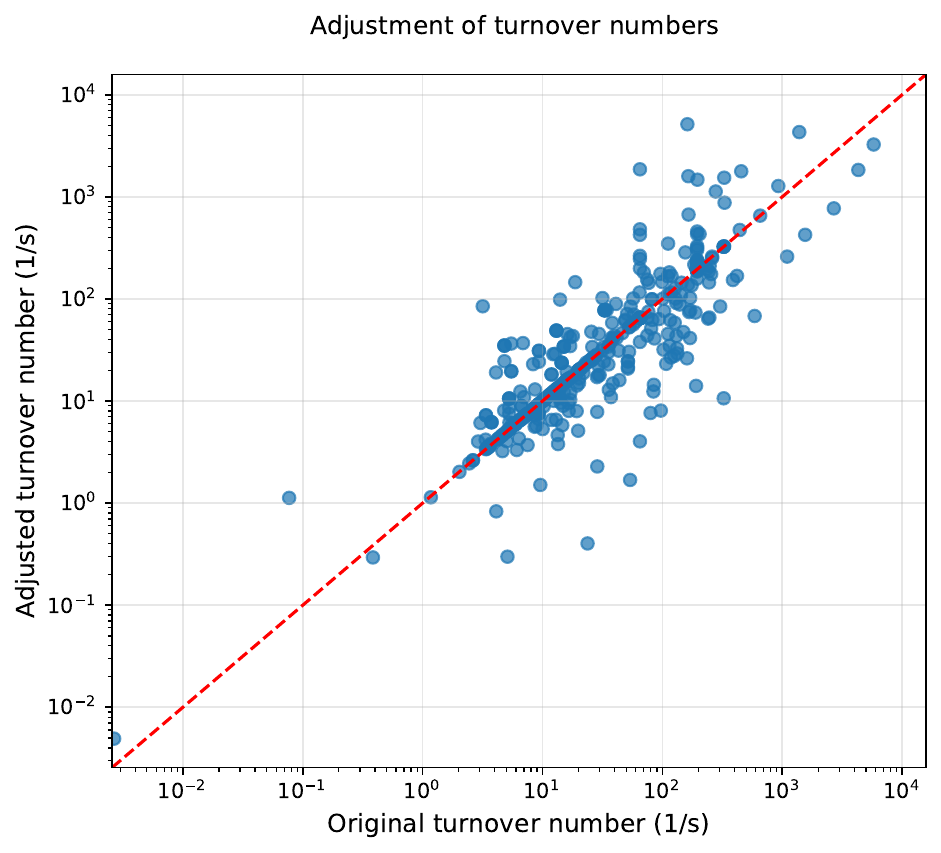}
	\caption{Turnover parameters used in EC-\textit{i}CH360 before and after the adjustment procedure (see section \ref{EC-iCH360} and Methods in the main text, as well as Supplementary Information \ref{kcat_adjustment_SI} for details). The original parameter set was directly parsed from \citeauthoryear{heckmann_kinetic_2020}.  }
 \label{original_vs_adjusted_kcats}
\end{figure}

\newpage
\begin{figure}[h!]
	\centering	\includegraphics[width=0.7\textwidth]{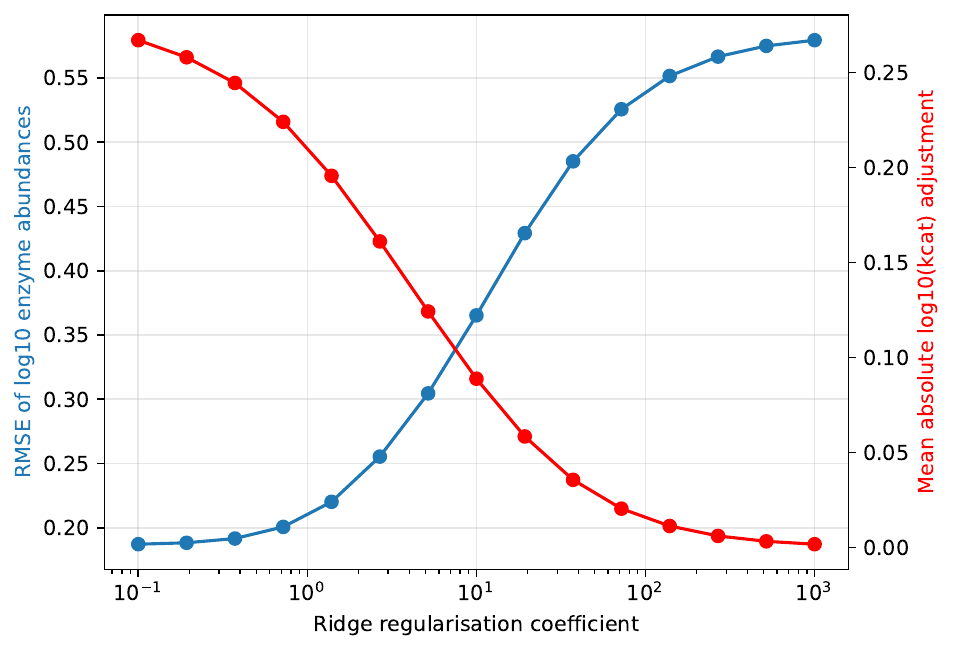}
	\caption{Impact of the magnitude of the ridge regularisation coefficient (\(\rho\) in Eq.~(\ref{kcat_adjustment_problem})) on the outcome of the adjustment procedure for turnover numbers. The blue curve represents the RMSE between measured and predicted log-enzyme abundances and decreases monotonically as less regularisation is applied to the problem. The red curve represents the mean absolute deviation between original and adjusted log-turnover numbers, which follows the opposite trend and converges to \(0\) as more regularisation is applied. }
 \label{regulatisation_parameter_sweep}
\end{figure}

\newpage

\begin{figure}[h!]
	\centering	\includegraphics[width=0.9\textwidth]{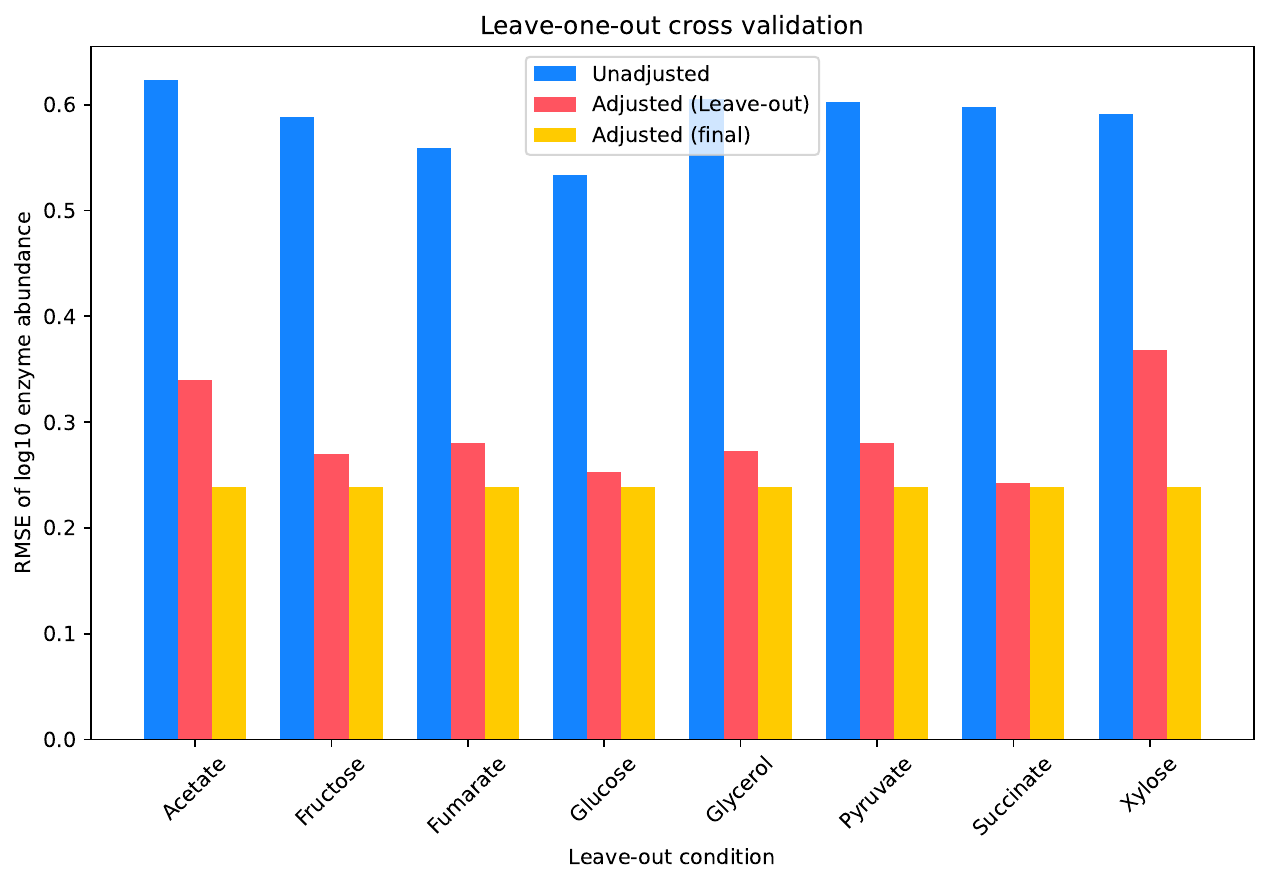}
	\caption{Leave-one-out cross-validation for the turnover parameter adjustment procedure. For each condition in the dataset, the graph shows the RMSE (computed for log\textsubscript{10}-transformed values) between measurements and predictions of enzyme abundances in that condition. Blue bars show the RMSE computed using the initial, unadjusted parameter set from \citeauthoryear{heckmann_kinetic_2020}. Red bars show the RMSE computed after parameter fitting, but excluding the condition for which the RMSE is evaluated from the training dataset. Finally, yellow bars show the RMSE computed using the final adjusted parameter set, obtained including all conditions in the training dataset. }
 \label{leave_one_out_cross_validation}
\end{figure}

\newpage
\begin{figure}[h!]
	\centering	\includegraphics[width=0.8\textwidth]{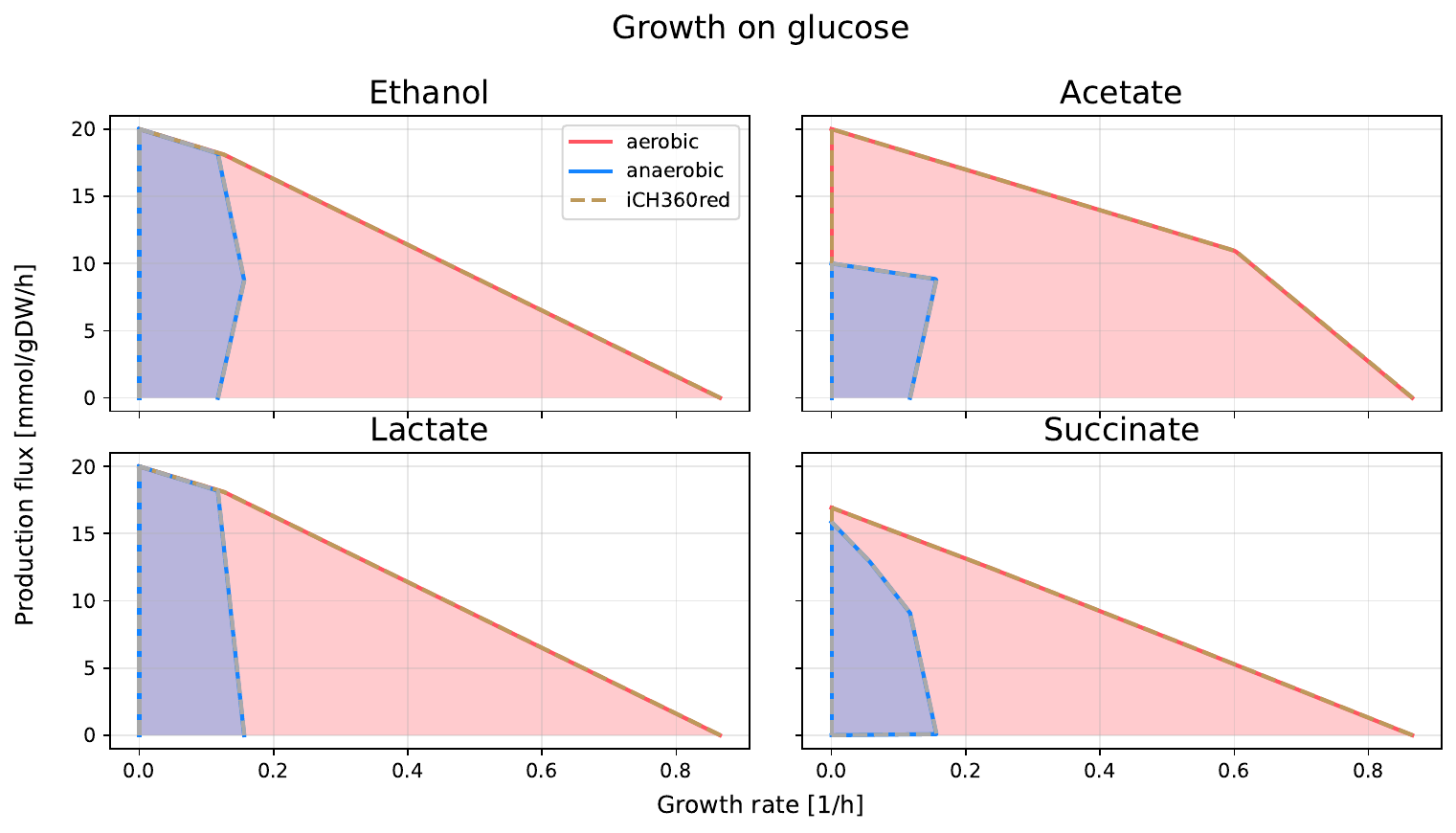}
	\caption{Comparison of production envelopes for growth on glucose between \textit{i}CH360 and the reduced variant \textit{i}CH360\textsubscript{red} used for elementary flux mode enumeration and analysis.  For the products and growth conditions shown, the two models have virtually identical solution spaces.}
\label{ich360_vs_ich360red_production_envelopes_glucose}
\end{figure}

\newpage
\begin{figure}[h!]
	\centering	\includegraphics[width=1\textwidth]{all_images/max_yield_efm_fluxes.pdf}
	\caption{Metabolic flux distribution corresponding to the maximum yield elementary flux mode (see main text, section \ref{efm_enumeration}). The mode is purely respiratory, with no excretion of typical fermentation by-products such as acetate, ethanol, or lactate. The graphics was produced in Escher \cite{king_escher_2015}.}
\label{max_yield_efm_escher_map}
\end{figure}

\newpage
\begin{figure}[h!]
	\centering	\includegraphics[width=1\textwidth]{all_images/max_growth_efm_fluxes.pdf}
	\caption{Metabolic flux distribution corresponding to the maximum growth elementary flux mode (see main text, section \ref{efm_enumeration}). The mode shows a mixed respiratory/fermentative phenotype, with significant acetate excretion. The graphics was produced in Escher \cite{king_escher_2015}.}
\label{max_growth_efm_escher_map}
\end{figure}

\newpage
\begin{figure}[h!]
	\centering	\includegraphics[width=0.7\textwidth]{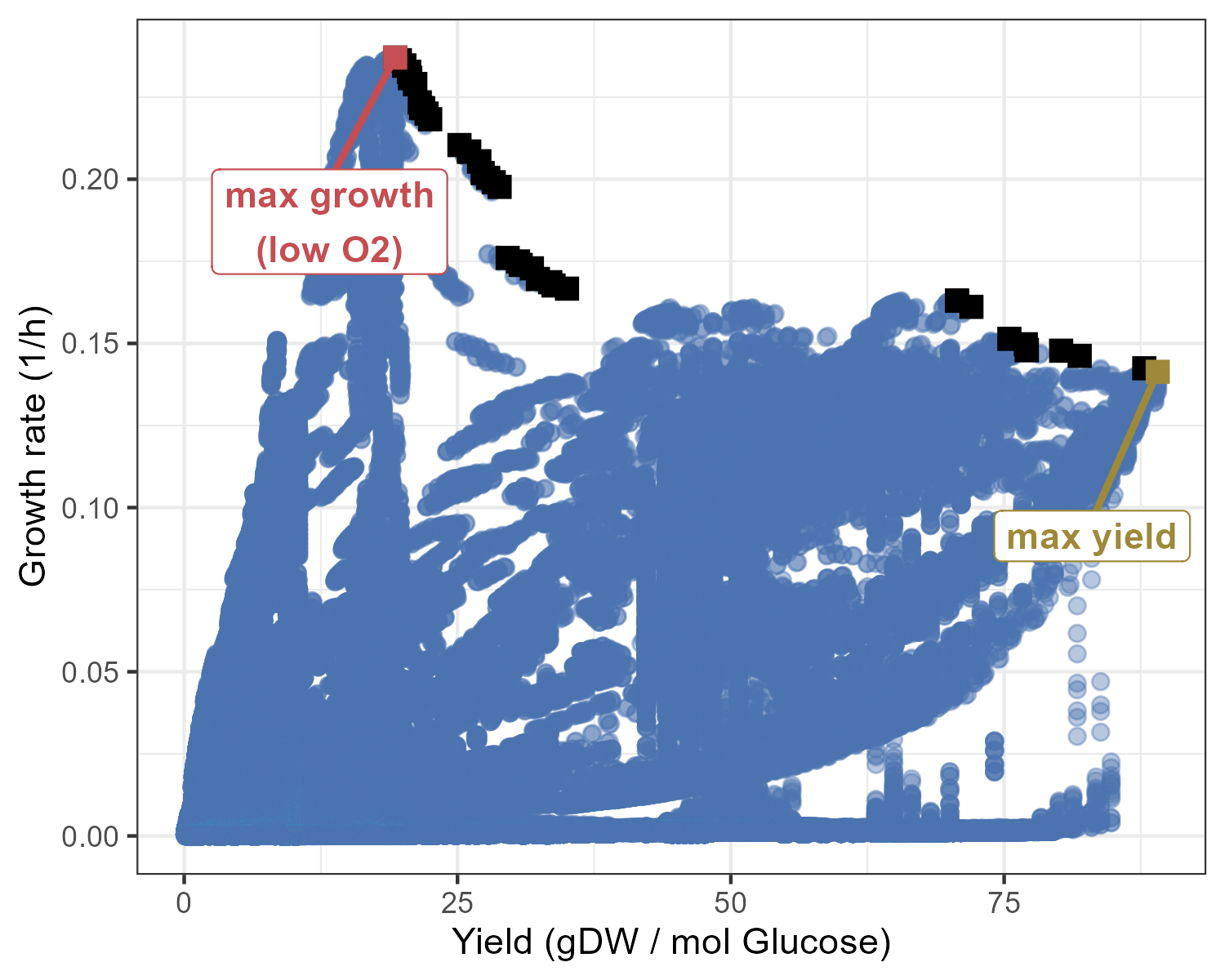}
	\caption{EFM growth-yield trade-off front under simulated low oxygen conditions. The figure resembles Figure \ref{EFM_yield_cost_tradeoff} in the main text, but was obtained by setting a 1000-fold higher cost for the cytochrome reactions. The higher enzyme cost is used to mimic the growth condition in which the oxygen-consuming reaction must operate at very low enzyme saturation, thus requiring higher enzyme-mass investment per unit flux. Black squares denote Pareto optimal EFMs. The Pareto front is much broader than the one obtained with the original enzyme cost for the oxygen-consuming reactions. }
\label{pareto_front_low_o2}
\end{figure}
\newpage
\begin{figure}[h!]
	\centering	\includegraphics[width=1\textwidth]{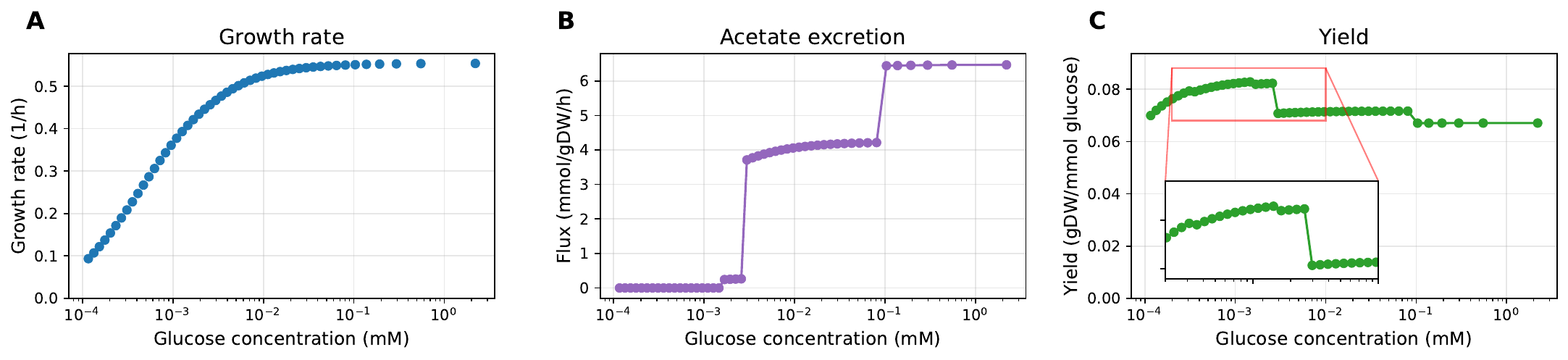}
	\caption{satFBA results (corresponding to Figure \ref{satFBA_results} in the main text) obtained by enforcing a lower bound on the ATP maintenance flux equal to \(6.86\) mmol/gDW/h (taken directly from the parent model \textit{i}ML1515). In this case, the optimal solution is no longer an EFM. This is evident from the yield profile (C), which is not piecewise constant with respect to the external glucose concentration. }
\label{satFBA_with_ATPM}
\end{figure}
\newpage

\begin{figure}[h!]
	\centering	\includegraphics[width=0.8\textwidth]{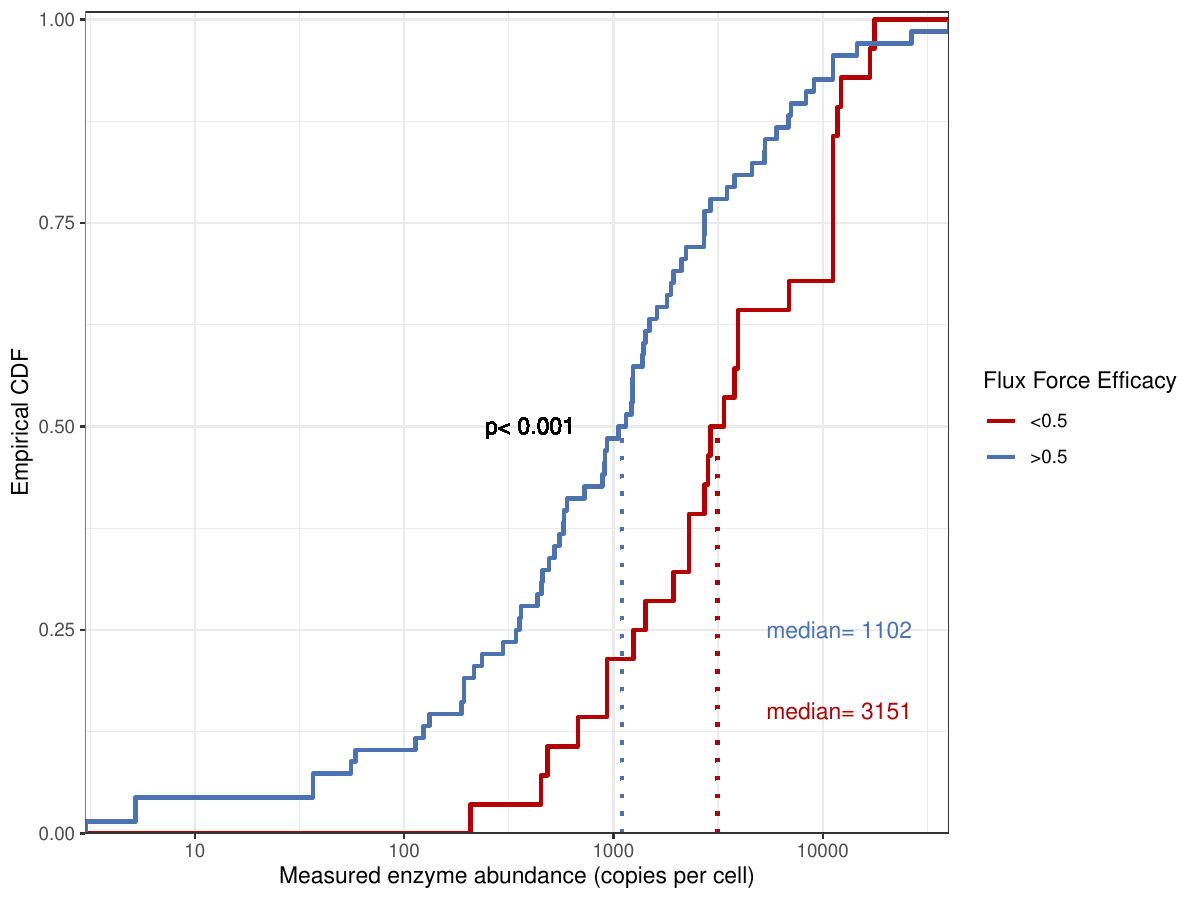}
	\caption{Empirical cumulative distribution functions (CDF) of measured enzyme abundances for reactions with predicted flux force efficacies above or below \(50 \%\) (blue and red curves, respectively). The two functions are significantly different (\(p<0.001\), two-sided Wilcox rank-sum test). The low-efficacy group shows an approximately \(3\)-fold higher median enzyme abundance than the high efficacy group. }
\label{FFE_vs_enzyme_abundance_comparison}
\end{figure}

\newpage
\begin{figure}[h!]
	\centering	\includegraphics[width=1\textwidth]{all_images/ich360_compressed_map.pdf}
	\caption{Compressed metabolic map of \textit{i}CH360, wherein linear pathways longer than two reactions were lumped into single effective reactions. The map was produced in Escher \cite{king_escher_2015} and can be used to visualise fluxes, metabolite concentration, and gene expression data.}
\label{compressed_escher_map}
\end{figure}
\newpage
\begin{figure}[h!]
    \centering
    \includegraphics[width=0.9\linewidth]{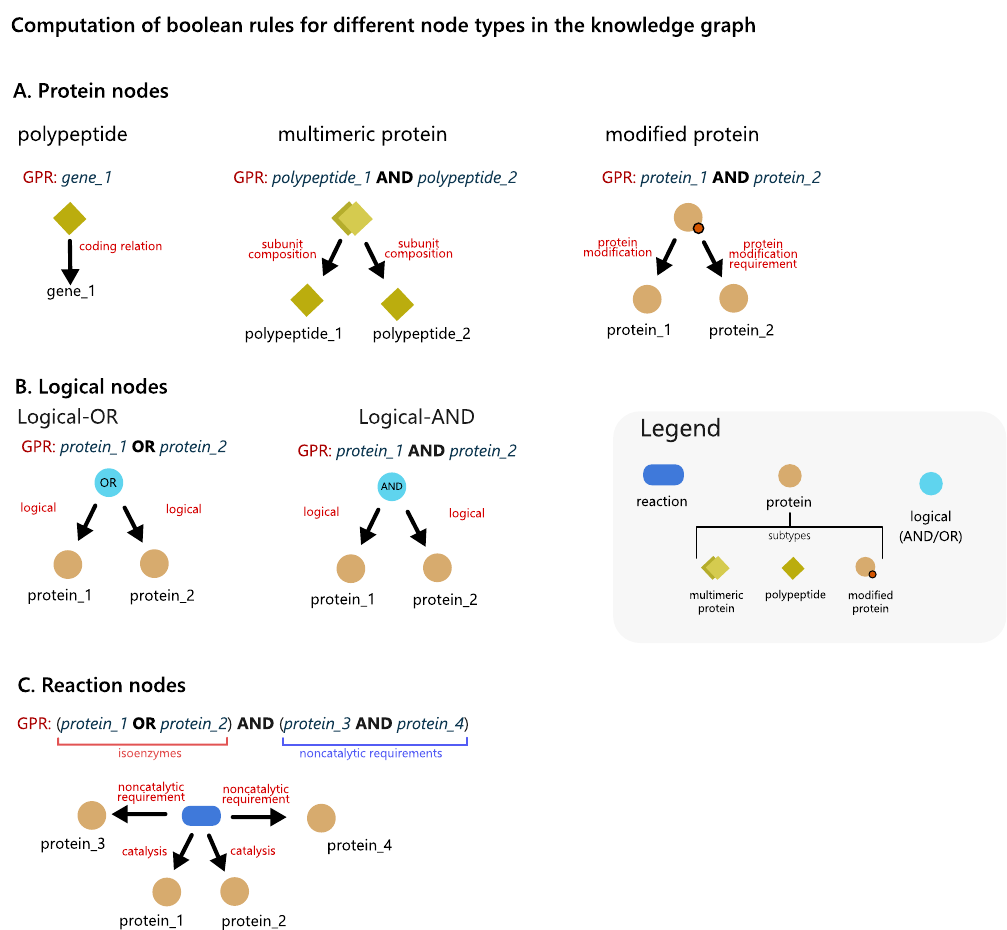}
    \caption{Computation of boolean gene-protein-reaction (GPR) rules with the knowledge graph. For each node in the graph (the top node in each diagram), a boolean expression can be constructed to describe the state of a node (active or inactive) in terms of the state of its child nodes. The exact form of this boolean expression depends both on the node type (e.g. protein, reaction, or logical) and on the type of edges that connect it with its neighbours (e.g. catalysis, subunit composition, etc.). Using these rules, the state of a reaction can be expressed, ultimately, solely in terms of genes in the model, as per convention in standard metabolic models. These GPR rules define a map between a genotype (set of active genes) to a phenotype (set of active reactions, and can be used to simulate \textit{in silico} the effect of given gene knockouts.}
    \label{GPR_computation_schematics}
\end{figure}

\newpage
\section{Supplementary Tables}
\renewcommand{\tablename}{Supplementary Table}
\setcounter{table}{0}
\begin{table}[h!]
\fontsize{10pt}{10pt}\selectfont
	 \caption{Some anecdotal examples of unrealistic predictions obtained when running FBA analyses on large-scale models (here, we considered the genome-scale parent of \textit{i}CH360, \textit{i}ML1515). We stress that these behaviours are not the result of errors in the large model. Rather, they result from applying simple methods with few constraints, such as FBA, to a large model with many degrees of freedom. Note that the examples shown here were not obtained through a thorough or systematic investigation of the model prediction abilities. }\ \\
     
  \begin{tabularx}{\textwidth}{
        |>{\RaggedRight\hsize=0.25\hsize}X
        >{\hsize=0.75\hsize}X|}
		\hline
        Example
        &Description\\
        \hline
        Production of fatty acids
        & In the (p)FBA solution computed on \textit{i}ML1515 using glucose as a carbon source and growth as an objective, the canonical fatty acid production pathway in \textit{E. coli} (see Supplementary Figure \ref{fa_biosynthesis}) is completely unused. Instead, the genome-scale model runs beta-oxidation in reverse to produce fatty acids. \\
        Anaerobic pyruvate production
        & Under anaerobic conditions, \textit{i}ML1515 can uptake external CO\textsubscript{2} and use it as a sink for glycolytic electrons, reducing acetyl-CoA produced by PFL back into pyruvate (see Main text, section \ref{sec:ProductionEnvelopes}). This behaviour, which allows the model to channel additional carbon towards pyruvate, is thermodynamically unrealistic under ambient CO\textsubscript{2} conditions. \\
        Pyruvate auxotrophic strain knock-outs prediction
        & To construct a pyruvate auxotrophic strain, only a few knock-outs in the central metabolism are necessary (MaeA, MaeB, pcK \cite{yu_augmenting_2018}). However, \textit{i}ML1515 uses reactions in amino acid degradations bypassing these knock-outs. \\
        
            Acetyl-CoA auxotrophic strain knock-outs prediction
        & Similarly, knock-out of 4 genes (aceEF, pflB, poxB), disrupting PDH, PFL and POX reactions, respectively, results in an acetyl-CoA auxotrophic strain that is unable to grow on glucose as the sole carbon source \cite{yu_augmenting_2018}. These knockouts are bypassed \textit{i}ML1515, which can use a number of additional pathways to produce acetyl-CoA (see Main text, section \ref{sec:ProductionEnvelopes}). \\
        \hline
	\end{tabularx}
     \label{examples_of_unrealsistic_GEM_behaviour}
\end{table}
\newpage
\begin{table}[h!]
\fontsize{10pt}{10pt}\selectfont
	 \caption{Description of node types in the knowledge graph supporting the stoichiometric model. The IDs used in the "Example" column refer to node identifiers in the graph.}\ \\
     
  \begin{tabularx}{\textwidth}{
        |>{\RaggedRight\hsize=0.25\hsize}X
		>{\RaggedRight\hsize=0.3\hsize}X
		>{\hsize=1\hsize}X
        >{\hsize=0.45\hsize}X|}
		\hline
		Node type 
        &Node subtype 
        &Description 
        &Example\\
        \hline
		reaction
        & 
        & A mass balanced chemical or biochemical reaction 	 
        & {bigg:GAPD}\\
		\hline
		\multirow{3}{=}{protein}
        & polypeptide 
        & A single polypeptide coded by a gene
		&GAPDH-A-MONOMER\\
		\cline{2-4}
        & multimer
        & A complex formed by stoichiometric binding of different polypeptides and/or other complexes 
        &GAPDH-A-CPLX\\\cline{2-4}

        & modified protein
        & A polypeptide or multimer which underwent post-translational modification 
        &LIPOYL-GCVH\\\cline{2-4}
        \hline
        gene 
        &
		& A gene
        &b1779\\
		\hline
        metabolite 
        &
		& An organic or inorganic molecule
        &L-ASPARTATE\\
		\hline
          {logical-AND/ \newline logical-OR}
          &
	   &Used as intermediates node to encode arbitrary logical rules in GPRs generated thorough the graph (e.g.~one of two proteins being required by another node)
        &THIOREDOXINS \\\hline
	\end{tabularx}
     \label{node_description_table}
\end{table}
\newpage

\begin{table}[h!]
\fontsize{8pt}{8pt}\selectfont
	\caption{Classification and properties of edge types (and biological meaning of their associated weights, when applicable) in the \textit{i}CH360 graph data structure supporting the stoichiometric model.}\ \\
    
	\begin{tabularx}{\textwidth}{
			|>{\hsize=0.1\hsize}X
            >{\hsize=0.08\hsize}X
            >{\hsize=0.08\hsize}X
            >{\hsize=0.08\hsize}X
			>{\RaggedRight\hsize=0.36\hsize}X
            >{\hsize=0.3\hsize}X|}
		\hline
		Edge type & Parent node type(s) &Child node type(s)&Subtype &Description & Example \\
        \hline
        \multirow{2}{=}{catalysis}
        &\multirow{2}{=}{reaction}
        &\multirow{2}{=}{protein}
            &{primary}
            &The default catalytic relationship between a reaction and an enzyme.
            &{\scriptsize bigg:PFK~\(\rightarrow\)~6PFK-1-CPX}\\ 
            \cline{4-6}
            &
            &
            &secondary
            &A catalytic relationship between a reaction and an enzyme, where the enzyme had been shown in literature (based on \textit{in vitro} or \textit{in vivo} evidence) to account for only minor catalytic activity for the reaction when compared to another (primary) isoenzyme. Notes and references to the relevant literature for the secondary annotation are included as edge metadata.
            &{\scriptsize bigg:PFK~\(\rightarrow\)~6PFK-2-CPX}\\
            \cline{4-6}
            &
            &
            &inactive
            &Indicates that the child protein is an enzyme for the parent reaction, but it's inactive in the K-12 strain due e.g.~to a frameshift mutation. \rev{There are only two such edges in the model, and they both involve the same enzyme (Acetohydroxyacid synthase II) encoded by the \textit{ilvG} and {ilvM} genes}
            &{\scriptsize bigg:ACHBS\(\hspace{0.05cm}\rightarrow\hspace{0.05cm}\)ACETOLACTSYNII-CPLX}\\
            \cline{4-6}
        \hline
        non-catalytic requirement
        &reaction
        &protein
        &
        &Indicates that the child protein is required by the parent reaction, although not as a catalyst. Typical examples include proteins used as cofactors (e.g.~glutaredoxins) in the reaction or featuring as prosthetic groups for a metabolite involved in the reaction (e.g.~Acyl-Carrier-protein)
        &{\scriptsize bigg:ACOATA\(\hspace{0.05cm}\rightarrow\hspace{0.05cm}\)ACP-MONOMER}
        \\
        \hline
         \multirow{2}{=}{subunit \mbox{composition}}
        & \multirow{2}{=}{protein}
        &\multirow{2}{=}{protein}
            &requirement
            &Indicates that the child node is a subunit of the parent node and is required for the correct functioning of the complex. The weight of the edge indicates the stoichiometry of the subunit in the complex.
            &{\scriptsize FABA-CPLX\(\hspace{0.05cm}\xrightarrow{2}\hspace{0.05cm}\)FABA-MONOMER}\\
            \cline{4-6}
            &
            &
            &accessory
            &Indicates that the child protein is an accessory subunit of the parent protein, meaning it can be part of the complex (potentially enhancing or modulating its function), but it's not strictly required for the complex to perform its physiological function. The weight of the edge indicates the stoichiometry of the subunit in the complex.
            &{\scriptsize ATPSYN-CPLX\(\hspace{0.05cm}\xrightarrow{1}\hspace{0.05cm}\)EG10106-MONOMER}\\
            \cline{4-6}
        \hline
        protein \mbox{modification}
        &protein
        &protein
        &
        &Indicates that the parent protein is a obtained by post-translational modification of the child protein. 
        &{\scriptsize PYRUVFORMLY-CPLX\(\hspace{0.05cm}\rightarrow\hspace{0.05cm}\)PYRUVFORMLY-INACTIVE-CPLX}
        \\
        \hline
        protein \mbox{modification} requirement
        &protein (modified-protein)
        &protein
        &
        &Indicates that the child protein is required to accomplish the post-translational modification leading to the parent protein.
        &{\scriptsize PYRUVFORMLY-CPLX\(\hspace{0.05cm}\rightarrow\hspace{0.05cm}\)\mbox{PFLACTENZ-MONOMER}}
        \\
        \hline
        coding relation
        &protein
        &gene
        &
        &Indicates that the child gene codes for the parent polypeptide
        &{\scriptsize RIBOKIN-MONOMER\(\hspace{0.05cm}\rightarrow \hspace{0.05cm}\)\mbox{b3752}	}
        \\
        \hline
        regulation
        &protein
        &metabolite, protein
        &
        &Indicates that the child metabolite or protein is a regulator for the enzyme. Information about the regulation mode (activation vs inhibition), the regulation mechanism (competitive vs allosteric) and the regulated reaction (if the enzyme catalyses multiple) is provided as edge metadata whenever available. If present, the weight of the edge denotes the activation/inhibition constant for the interaction as reported in EcoCyc, with units indicated as edge metadata. 
        &{\scriptsize SHIKIMATE\(\hspace{0.05cm}\xrightarrow{ \rm 160.0 \mu M}\hspace{0.05cm}\)\mbox{AROE-MONOMER}	}
        \\
        \hline
        putative association
        &reaction
        &protein
        &
        &Indicates that a putative association between the reaction and the protein has been proposed in literature.
        &{\scriptsize bigg:PFL\(\hspace{0.05cm}\rightarrow\hspace{0.05cm}\)EG11910-MONOMER}
        \\
        \hline
        logical
        &logical AND/OR
        &any
        &
        &Connects logical operator nodes to downstream nodes. Used to create arbitrary complex logic relations in the graph.
        &{\scriptsize  
        THIOREDOXINS \newline
        \(\rightarrow\) RED-THIOREDOXIN-MONOMER \newline
        \(\rightarrow\) RED-THIOREDOXIN2-MONOMER \newline
        \rev{(In this example, the logical edges are used to indicate that the THIOREDOXINS node (of type \textit{logical-OR}) is active when any of the two child nodes (representing two thioredoxins found in \textit{E. coli}) are active.}
            }
        \\
        \hline
        
	\end{tabularx}
	\label{edge_description_table}
\end{table}
\newpage
\begin{table}[h!]
	\caption{Manual curation of the reaction pruning process used to construct \textit{i}CH360\textsubscript{red}. Each reaction set represents a set of alternative routes for the production of the same metabolite (but using, for example, different cofactors). For each set, the most physiologically relevant option, based on available literature, was preserved in the reduced model variant. }\ \\
    
	\begin{tabularx}{\textwidth}{
			|>{\hsize=0.25\hsize}X|
			>{\hsize=0.25\hsize}X
                |>{\hsize=0.5\hsize}X|
		}
		\hline
		Reaction set & Pruned in \textit{i}CH360red & Notes\\
		\hline
		EAR(n)x, EAR(n)y *& EAR(n)y &Enzyme FabI can work with 
 both NADH/NADPH, but higher activity was found with NADH \cite{bergler_enoyl-acyl-carrier-protein_1996}\\	
 		\hline 
		ACOATA, KAS14, KAS15 &
		ACOATA, KAS14& Initiation of fatty acid biosynthesis can occur by either direct condensation of acetyl-CoA with malonyl-ACP (KAS15) or by transacylation of acetyl-CoA followed by condensation with malonyl-ACP (ACOATA + KAS15). Because the transacylacylase activity of FabH (ACOATA) has been to be significantly lower than its condensation activity (KAS15) \cite{tsay_isolation_1992}, only the former pathway is maintained in \textit{i}CH360red. \\
   		\hline 
		VALTA, VPAMTr &
		VPAMTr& These are both routes to production of valine. We keep VALTA (\textit{ilvE}) as it is the last step in the canonical valine biosynthesis route. \\
		\hline 
		RNDR1, RNDR2, RNDR1b, RNDR2b &
		RNDR1b, RNDR2b& The ribonucleoside diphosphate reductase can work both with the thioredoxin and glutaredoxin redox systems. \textit{i}CH360 retains only the thioredoxin version.\\
		\hline
		SULabcpp, SO4t2pp &
		SO4t2pp& SULabcpp is an ATP-mediated active transport of sulfate in the cell via an ATP-binding-cassette (ABC) transporter \cite{sirko_sulfate_1995}, while SO42tpp (\textit{cysZ}) is a proton symporter \cite{zhang_escherichia_2014}. We maintain the former as its impairment was shown to lead to cysteine auxotrophy \cite{sirko_sulfate_1995}.  \\
		\hline
	\end{tabularx}
	\label{metabolic_redundancies}
    * \(n \in (60,80,100,120,140,160,180,121,141,161,181)\)
\end{table}

\newpage
\begin{table}[!h]
\caption{ 
Numbers of elementary flux modes enumerated for the reduced model variant \textit{i}CH360red under different growth conditions. Numbers in brackets represent the numbers of EFMs after filtering. Filtered modes include only those supporting Biomass flux and, for aerobic conditions, those that a) have nonzero oxygen uptake and b) do not use either of three reactions (PFL, DHORD5, FRD2), which are known to be only physiologically active under anaerobic conditions }
\centering
\begin{tabular}{lll}
& \multicolumn{2}{c}{number of EFMs (filtered)} \\
\cmidrule(lr){2-3}
           & Aerobic & Anaerobic\\\midrule
Glucose    &13468719 (1035696))  &204028 (195670)) \\
Pyruvate  &1763631 (135266))    & 6949 (6480))\\
Glycerol   & 922217 (82112))  &  NA \\
Acetate    &38099 (7596) & NA\\
Lactate   & 1270315 (5897) )  & 1497 (1424))\\\bottomrule
\end{tabular}
\label{EFM_count}
\end{table}

\end{appendices}

\end{document}